\documentclass[npg,manuscript]{copernicus}

\begin{document}

\nolinenumbers

\title{Laser-Like Waves Amplification in Straits}

\Author[1,2]{Andrei}{Pushkarev}

\affil[1]{Skoltech, Bolshoy Boulevard 30, bld. 1, Moscow, 121205, Russia}
\affil[2]{Lebedev Physical Institute RAS, Leninsky 53, Moscow 119991, Russia}

\runningtitle{RUNNINGTITLE}

\runningauthor{RUNNINGAUTHOR}

\correspondence{Andrei Pushkarev (dr.push@gmail.com)}


\firstpage{1}

\maketitle

\begin{abstract}
We present the research of ocean surface wind waves excitation in non-homogeneous situations, on the example of deep water strait in presence of the constant wind, blowing perpendicular to the coastal line. The used statistical wave model is based on Hasselmann equation with high-wavenumbers wave-breaking dissipation, exact nonlinear four-waves  interaction and ZRP wind input term. At the first stage, the waves propagate in the wind direction in a step-like moving front manner, which is the combination of self-similar fetch-limited and duration-limited solutions of Hasselmann equation. The second stage begins further, when sufficient amount of wave energy is concentrated  at the incoming waves shore line. Beginning with that time, the nonlinear  wave interactions start formation of the wave groups, propagating across and against  the wind. Despite the absence of long-wave dissipation, the system asymptotically evolves into complicated quasi-stationary state, comprising of  the self-similar ``wind sea'' in the wind direction, and quasi-monochromatic waves, radiated  close to orthogonally with respect to the wind, while slightly tilting counter the wind direction with the angle increasing toward the wave turbulence origination shore line, and reaching $15^{\circ}$ close to it. The total wave energy in the asymptotic state surpasses the wave sea energy propagating along the wind by two times  due to the presence of quasi-orthogonal and counter the wind wave fields. The very similar turbulence structure was previously observed experimentally, this paper presents theoretical explanation of these results.  It is suggested  to name this laser-like radiation phenomenon by {\bf N}onlinear {\bf O}cean {\bf W}aves {\bf A}mplifier, simplified to the abbreviation \textit{\bf NOWA}.

\end{abstract}

\section{Introduction} \label{Int}

The natural events have been influencing humanity history on many occasions in the past.
The fairly recent example, related to ocean wave weather prediction, is especially illustrative during World War II allied incursion Operation Overlord  in Normandy, in June 1944, which was the biggest maritime military operation in history. Incapacity of good enough wind ocean waves weather forecasting affected the course of the invasion -- rough seas significantly contributed into human losses of 10,000, including 4,414 dead.

Since that time, surface waves weather prediction turned into big industry, using sophisticated tools, such as satellites, meteostations, booey and parallel supercomputers, running operational wave forecasting models. In our opinion, mathematical equations used in these models are not properly justified, but their infirmity is compensated by tunable parameters. This approach, apparently, works in many open seas situations, where the influence of inhomogeneity, such as coastal topography, or sharp wind turns, is not essential. 

The presence of the inhomogeneity, for example, in the form of a coastline, compicates the situation. It turns out that the wave climate forecasting in English Channel nowadays is still almost as hard as it was in 1944. The results, presented in current research, show that the wave spectra in the channels with parallel shores appear to be very different from the spectra in open seas.

The generally accepted model, used nowadays for wind waves turbulence description, is Hasselmann equation (hearafter HE), or kinetic equation for waves, also known in oceanography as energy balance equation \cite{R1,R2} :
\begin{equation}
\label{HE}
\frac{\partial \varepsilon }{\partial t} +\frac{\partial \omega _{k} }{\partial \vec{k}} \frac{\partial \varepsilon }{\partial \vec{r}} =S_{nl} +S_{in} +S_{diss}
\end{equation}
where $\varepsilon =\varepsilon (\omega_k,\theta,\vec{r},t)$ is the spectrum of wave energy,  $\omega_k =\omega (k)$ is linear wave dispersion, two-dimensional real space coordinate vector $\vec{r}=(x,y)$ defines location in real space and $t$ is time. $S_{nl}$ is complex integral function, describing nonlinear four-waves interactions, $S_{in}$ is the wind input and $S_{diss}$ is the wave energy dissipation source functions, describing wind energy input and wave energy dissipation due to wave-breakings. In the presented research we consider only the deep water situation $\omega=\sqrt{g k }$, where $k=|\vec{k}|$ is the wavenumber absolute value, and $g$ is the acceleration of gravity force . 

It is commonly distributed oceanographic community opinion, that Eq.({\ref{HE}}) is already well-studied  in the deep-water case, and subsequent productive study should be focused on the accomplishment of the description of  the situations related to very strong tornado winds, the modifications associated with the shallow water effects as wells as the bottom friction impact. We show in the current study that Eq.({\ref{HE}}), yet, even in the ``primitive'' form is so far the mathematical item, worthy additional in-depth discussion, which can uncover so far unknown physical properties. It is important to emphasize that those properties can be properly studied only using the exact formulation of nonlinear four-wave interactions source function $S_{nl}$, since current routine usage of highly computationally efficient DIA-like substitutes is dangerously misleading. As our results show below, that nonlinear interaction definition difference produces significantly altered physical results with respect to seemed to be perfectly known physical understanding of the wind-driven wave turbulence in unbounded domains .

In presence of the shorelines, Eq.({\ref{HE}}) must be supplied with boundary conditions, which properties can vary depending on the kind of coastal line. In the current research we use the simplest ``totally dissipative'' boundary conditions, implying that the wave energy incoming to the coast is completely absorbed. The effects of more sophisticated boundary conditions are the subject of future research.

The paper is organized in the following way:

\begin{itemize}

\item{} section \ref{Recent_Developments} briefly describes recent developments in self-similarity analysis of HE, which, in particular, lead to the formulation of ZRP balanced source terms, used in the presented research, and produces specific characteristics, observed in the presented numerical simulation

\item{} section \ref{HE_Bounded} discusses Cauchy problem statement for HE solution in limited domains

\item{} section \ref{Num_Sim} contains the results of numerical simulation and their interpretation 

\item{} section \ref{Experimental_Evidence} presents experimental evidence of the observed phenomena

\item{} section \ref{Concl} wraps up the results of the presented research

\end{itemize}

\section{Self-similar properties of HE} \label{Recent_Developments}

Inventions of HE source functions have been continuing for the last 60 years, both experimentally and theoretically. The vast majority of them did not pass through HE numerical testing to check if they reproduce ocean waves observations, as it would be expected in the routine approach in the field of numerical simulation of nonlinear partial differential equations. It is explainable, though, by two major difficulties. First, HE is the complex integro-differential equation with hidden symmetries, which theoretical analysis requires significant efforts in studying Weak Turbulence Theory \cite{R5},  and second,  -- the nonlinear four-wave interaction term is extremely computationally heavy even for modern parallel computers.

Recent comparative analysis of several historically developed wind input source functions  $S_{in}$ has been performed in \cite{R7,R6}, which showed their fivefold scatter. As far as concerns the wave energy dissipation source function $S_{diss}$, the situation might be even worse, since  there is no so far even an agreement on the localization of the energy sink in wave-number space.   

Due to limited volume of current publication, we cannot get into details of the related discussion, which can be found in \cite{R7}, and only mention that, as the result of the circumstances, noticed above, the modern operational models contain several dozens of tunable parameters.

The research on physically based source functions has been presented in the  publications \cite{R8,R88}, where Eq.(\ref{HE}) was analyzed for two situations: spatially inhomogeneous semi-infinite limited fetch case, bounded by the coast line:
\begin{equation}
\label{LFeq}
\frac{1}{2} \frac{\omega}{k} \cos\theta \frac{\partial \epsilon}{\partial x} = S_{nl} + S_{wind}
\end{equation}
and spatially homogeneous, or duration limited case in infinite ocean:
\begin{equation}
\label{DLeq}
\frac{\partial \epsilon}{\partial t} = S_{nl} + S_{wind}
\end{equation}
Eqs.(\ref{LFeq}) and (\ref{DLeq}) allow self-similar substitutions in the assumptions of  wind input term power dependence on the frequency:
\begin{equation}
S_{in} \sim \omega^{s+1} f(\theta)
\end{equation}
where $f(\theta)$ is unknown spreading function of the angle $\theta$. In conjunction with the experimental regression line \cite{R20}, they give the following self-similar solutions with corresponding relationships:
\begin{itemize}

\item for the duration limited case:
\begin{eqnarray}
&&\varepsilon(\omega,\theta,t) = t^{p+q} F(\omega t^q,\theta)	\label{SelfSimDL}   \\
&&9q-2p = 1												\label{MagicRelationTime}  \\
&&p =10/7, \,\,\, q=3/7, \,\,\, s=4/3							\label{IndicesDL}    \\		
&&E \sim t^p												\label{EnTimeSelfSim} \\
&& <\omega> \sim t^{-q}									\label{FrTimeSelfSim}  
\end{eqnarray}

\item for the limited fetch case:
\begin{eqnarray}
&&\varepsilon(\omega,\theta,x) = x^{p+q} F(\omega x^q,\theta)	\label{SelfSimLF}  \\
&&10q-2p = 1												\label{MagicRelationFetch} \\
&&p=1, \,\,\, q = 3/10, \,\,\, s=4/3								\label{IndicesLF}  \\
&&E \sim x^p												\label{EnFetchSelfSim}   \\
&&<\omega> \sim x^{-q}									\label{FrFetchSelfSim}
\end{eqnarray}

\end{itemize}
\noindent where $E$ and $<\omega>$ are the total wave energy and mean frequency correspondingly, $x$ is the fetch coordinate.

It was found in \cite{R88,R8} that the set of balanced ZRP wind input  and explained below ``implicit'' dissipation due to wave breakings source terms numerically yields self-similar dependencies Eqs.(\ref{SelfSimDL})-(\ref{FrFetchSelfSim}), observed as well in the field limited fetch experiments, and later analyzed in \cite{R18}. ZRP model also does not need change of tuning parameters with respect to wind speed magnitude increase from five to ten meters per second \cite{R88}. Though ZRP model can not be declared as the completely free of tuning parameters, as there are two of them in the procedure of construction of the wind input source function $S_{in}^{ZRP}$, the described approach might be considered as the advance toward HE models, having solid physical basement.

The arguments, presented here, have been considered as the justification for the usage of ZRP set of  source functions in the performed numerical simulation.

\section{Statement of the limited fetch problem in strait} \label{HE_Bounded}

\subsection{Limited and semi-infinite domains}

As it was summarized in previous section, self-similarity analysis appears to be quite productive in spatially homogeneous and semi-infinite situations, where it predicts specific power-like laws of energy and mean frequency dependencies, and allows to formulate the balanced set of ZRP wind input and ``implicit'' dissipation source functions. 

The situation can be different, however, in inhomogeneous situations, such as ocean straits. It is possible to imagine that due to positive and negative signs of energy advection in real space, the assumed self-similar picture of wave energy distribution could be modified via nonlinear four-waves interaction of its locally enhanced values in the specific locations of real space geometrical domain. 

Publications \cite{R106,R107,R108,R116} have already reported the formation of the additional wave modes to the expected wind-sea in the direction of the wind -- the quasi-monochromatic waves, propagating almost perpendicular to the permanent wind direction in deep-water straits. Those wind-orthogonal quasi-monochromatic waves exhibited up to eight-fold spectral amplitude excess with respect to the wind-sea spectral maximum at the established asymptotic quasi-stationary state in time.

Current manuscript is focused on the in-depth study of this effect, and the efforts on clarifying its physics. In the next sections we are beginning our journey considering HE as the mathematical physics object, requiring proper formulation of the initial and boundary condition, or Cauchy problem.

Further in this section, we consider the geometrical domain in $x$, $k_x$, $k_y$ independent variables, which can be construed as the ``pipeline'' in the mixed Real-Fourier domain. Such pipeline representation happens to be quite helpful  in  graphic imagination of wave energy advection from one coast line to another, and in the opposite direction as well as the boundary and initial conditions formulation.

\subsection{Boundary and initial conditions for limited domains: Cauchy problem formulation} \label{CauchyProblem}

\begin{figure}
\centering
\includegraphics[width = 1.2\linewidth]{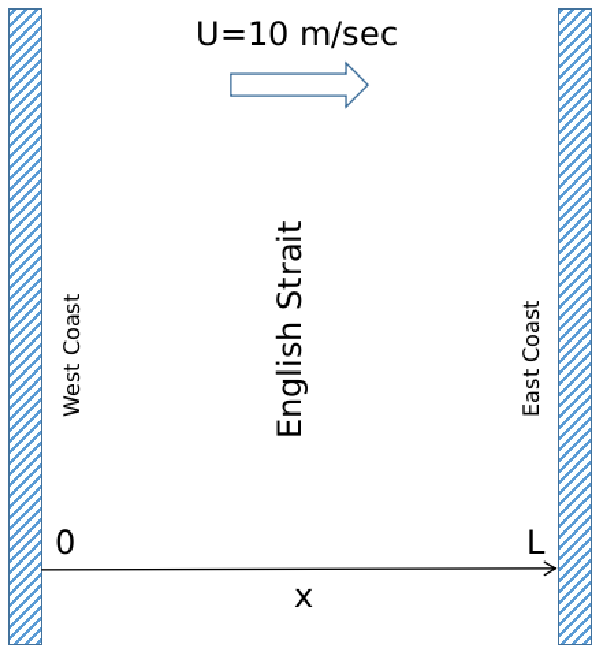}
\caption{The schematic presentation of the simulation domain in real space. $\vec{U}$ is the constant wind velocity in the direction of the fetch axis $\vec{x}$.} 
\label{Geometry}
\end{figure}

Eq.(\ref{HE}) is considered in Real-space domain, presented on Fig.\ref{Geometry}, which is vertically infinitely stretching strait of the width L. The left and right straight boundaries  will be called thereafter the ``west'' and the ``east" coasts. It is assumed that the permanent wind speed vector $\vec{U}$ is directed orthogonally to the boundaries, and therefore, the associated physical evolution will be homogeneous in vertical direction, and can be parameterized by the Real-space axis $\vec{x}$, directed perpendicularly to the shore lines.

The symmetries, associated with the described geometrical domain lead to the corresponding non-stationary HE:
\begin{eqnarray}
\frac{\partial \varepsilon}{\partial t} + \frac{1}{2} \frac{\omega}{k} \cos{\theta} \frac{\partial \varepsilon}{\partial x} = S_{nl}+S_{in}+S_{diss}
\label{HE2}
\end{eqnarray}
\noindent
where $\theta$ is the angle between the direction of the axis $\vec{x}$ and wave-vector $\vec{k}$; it is assumed that the directions of $\vec{x}$ and $\vec{U}$ are associated with $\theta_{wind} = 0^{\circ}$.

\begin{figure}
\centering
\includegraphics[scale=0.4]{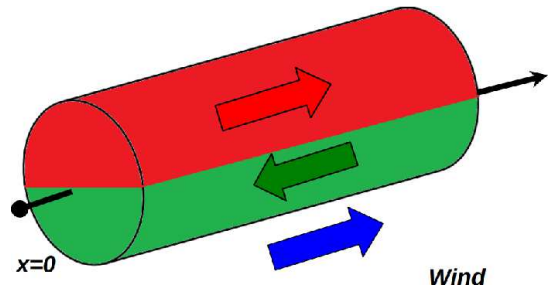}
\caption{Schematic description of the wave energy fluxes propagation along the fetch in Real and Fourier space.}
\label{Fluxes}
\end{figure}

Wave energy spectral density  $\varepsilon=\varepsilon(k_x,k_y,x,t)$  in Eq.(\ref{HE2}) is the function of four independent variables $k_x$, $k_y$, $x$ and $t$ in mixed Real-Fourier space, graphically presented on Fig.\ref{Fluxes}. One can interpret the corresponding problem configuration as the stack of Fourier spaces of wave energy spectral density $\varepsilon(k_x,k_y, x_0,t_0)$, threaded on the axis $\vec{x}$, each at the corresponding spatial coordinate $x_0$ and time $t_0$. All those wave energy Fourier spaces form the ``pipe'', divided into red and green semi-pipes. Red semi-pipe corresponds to the values $\cos\theta>0$, while green semi-pipe corresponds to $\cos\theta<0$. Therefore, the ``advective'' part of the Eq.(\ref{HE2}) 
\begin{eqnarray}
\frac{\partial \varepsilon}{\partial t} + \frac{1}{2} \frac{\omega}{k} \cos{\theta} \frac{\partial \varepsilon}{\partial x} = 0
\label{HE3}
\end{eqnarray}
\noindent
conveys the energy along the characteristics from the west to the east coast in the red (along the wind), and from the east to the west coast in the green (against the wind) semi-pipes, due to different signs of the $\cos\theta$ in the term $\frac{1}{2}\frac{\omega}{k}\cos{\theta}$.

Advection of the wave energy in the mixed Real-Fourier is, however, only the part of the dynamics described by Eq.(\ref{HE2}). Another process is the nonlinear four-wave interaction.
Such splitting of the dynamics by processes hints about possible numerical integration technique of the Eq.(\ref{HE2}), explained below.

We are finishing current subsection with the discussion of the boundary and initial conditions of  Eq.(\ref{HE2}), or Cauchy problem, which will accomplish formulation of the problem statement, suitable for numerical simulation.

Consider the waves having component in the direction of the axis $\vec{x}$ and propagating, therefore, in the red cylinder. We will assume that they have zero amplitude at west end of this semi-cylinder, i.e. the west coast does not generate any waves:
\begin{eqnarray}
\varepsilon(\omega,\theta,x,t)|_{x=0} &=& 0, \,\,\,  -\pi/2 < \theta < \pi/2
\end{eqnarray}
As far as concerns the green cylinder, the situation is mirror-symmetrical to the red cylinder, and it is assumed that the waves, having wave-number components against the axis $\vec{x}$, have zero amplitude at the east end of that semi-cylinder, i.e. the east coast does not produce any waves either:
\begin{eqnarray}
\varepsilon(\omega,\theta,x,t)|_{x=L} = 0,  \,\,\,  &&  \pi/2 < \theta \le \pi         \\
										     &&  -\pi < \theta \le -\pi/2
\end{eqnarray}
There are still 2 undefined boundary conditions left: the east boundary condition for the red semi-pipe and the west boundary condition for the green semi-pipe. While it is possible to define partial reflection of the waves from them, we will assume their perfect transparency, or the free advection for the incoming waves. One can call, therefore, such boundary condition ``fully dissipative'', or perfectly absorbing the wave energy. They can be observed both in natural beaches and laboratory waves tanks as slightly tilted with respect to the horizon pebble shores, perfectly absorbing wave energy.

The last thing left to accomplish Cauchy problem is the definition of the initial conditions, which were chosen in the form of extremely low-amplitude linear waves in the red semi-cylinder:
\begin{eqnarray}
\varepsilon(\omega,\theta,x,t)|_{t=0} = 10^{-6},  \,\,\,\,\,\,  -\pi/2 < \theta < \pi/2
\label{InCond}
\end{eqnarray}
and no initial waves in the green cylinder, i.e. zero level wave energy for the waves, running against the wind in the green semi-cylinder:
\begin{eqnarray}
\varepsilon(\omega,\theta,x,t)|_{t=0} = 0,  \,\,\,\,\,\,   &&  \pi/2 < \theta \le \pi         \\
										     &&  -\pi < \theta \le -\pi/2
\end{eqnarray}
at the initial time. That choice of the initial conditions was making initial nonlinear four-waves interactions negligible.

\section{Numerical modeling} \label{Num_Sim}

\subsection{Discretization approach}

Eq.(\ref{HE2}) is the complex nonlinear integro-differential equation with partial derivatives, and it is tempting to solve it using sophisticated high-accuracy techniques, such as implicit numerical schemes, resolved by iterations. As it was tested, implicit numerical schemes show relative accuracy improvement with some time step enhancement, but necessity of making about 4-5 iterations per time advance does not justify obtained time step gain. Therefore, the necessity to perform time-consuming Fourier-space integration for every advance in time, makes the explicit numerical low-order approximation schemes the viable economical alternative for time integration.

It is known that in situations with several sophisticated source functions, the method of decomposition over them is quite effective \cite{R104}. In this ideology, Eq.(\ref{HE2}) has been resolved with respect to time derivative in the left hand side, and all other terms -- advection and source functions --- have been considered as different physical processes, referred to the right hand side. 
 
The advection source term was solved on square template with the help of unconditionally stable numerical scheme with the second order approximation in space and time \cite{R103}. Exact quadruplet nonlinear interaction source function $S_{nl}$ part was defined from current distribution of spectral energy density $\varepsilon$ by Webb-Resio-Tracy algorithm \cite{R69,R68}, the wave energy dissipation source function due to wave-breakings was considered ``implicitly'' through $ \sim \omega^{-5}$ tail, which was replacing the dynamical part of the spectrum from the specific fixed frequency on every time step, and ZRP wind input source function part was resolved analytically due to its linear form.

\subsection{Definition of wind energy forcing and wave energy dissipation source terms}

The self-similarity analysis, carried in \cite{R8,R7,R88} produced so-called ZRP wind input source function, which we present here for reader convenience:
\begin{eqnarray}
\label{ZRP}
&&S_{in}(\omega,\theta) = \gamma(\omega,\theta)\cdot \varepsilon(\omega,\theta) \\
&&\gamma(\omega,\theta ) = \left\{\begin{array}{l} {0.05 \frac{\rho _{air} }{\rho _{water} } \omega \left(\frac{\omega }{\omega _{0} } \right)^{4/3} q(\theta )  {\rm \; \; for\; } f_{min} \leq f  \leq f_{d}, \; \; \omega =2 \pi f} \\ {0{\rm \; \; otherwise}} \end{array}\right. \label{ZRP1} \\
&&q(\theta ) = \left\{\begin{array}{l} {\cos^2 \theta {\rm \; \; for\; -}\pi {\rm /2}\le \theta \le \pi {\rm /2}} \\ {0{\rm \; \; otherwise}} \end{array}\right. \label{ZRP2} \\
&&\omega _{0} = \frac{g}{U }, \,\,\,  \frac{\rho _{air} }{\rho _{water} } =1.3\cdot 10^{-3} \label{ZRP3}
\end{eqnarray} 
\noindent
where $U$ denotes the amplitude of wind speed at $10$ meters elevation above the ocean surface, air density is $\rho_{air}$ and water density is $\rho_{water}$ , the frequencies $f_{min}$ and
$f_d$ are defined below.

\subsection{Parameters of the numerical simulation}

The numerical modeling of  Eq.(\ref{HE2}) has been carried out for the given below parameters: 

\begin{itemize}

\item{} strait width  $L=40$ km, reminiscent of English channel (La-Manche);
\item{} 40 equidistant points in Real space along the fetch
\item{} Fourier space domain $f_{low} < f < f_{high}$, resolved by 72 logarithmically positioned discrete frequencies with $f_{low}=0.025$ Hz and $f_{high}=2.0$ Hz
\item{} angular domain $-\pi < \theta \le \pi$ with 36 discrete angular directions, given the wind angle $\theta_{wind}=0$
\item{} wind velocity $U=10$ m/sec at 10 meters height above the ocean surface

\end{itemize}

The wave energy dissipation source function  was defined in the ``implicit'' manner   \cite{R8,R7,R88} through the replacement of the portion of the dynamical spectrum  for $f>f_d $ by Phillips tail $ \sim \omega^{-5}$. The frequency $f_d$ plays the dual role of cut-off frequency for the growing with frequency wind-forcing source function, as well as the starting frequency of the ``implicit'' dissipation. The value $f_d = 1.1$ Hz has been estimated from \cite{R20}.


\subsection{Numerical tests}

As it was mentioned above, the presence of the advection term in Eq.(\ref{HE2}) can modify the expected dynamics due to nonlinear interaction between wave fields, contained inside upper red and lower green pipelines. It is important in this relation to test the correctness of numerical implementation of different terms in Eq.(\ref{HE2}). 

The easiest tests are related to wind energy input and wave energy dissipation source functions, which are trivial to perform due to their linear nature, and while it was checked that they are correctly implemented, we omit corresponding discussion.

As far as concerns the nonlinear part of the interaction, it was calculated by WRT approach \cite{R69,R68}, which has been proved to reproduce theoretical analysis and field experimental observation in series of works \cite{R47,R46,R86,R48,R50,R72,R63,R7,R88,R101}. 

\begin{figure}
\centering
\includegraphics[scale=0.6]{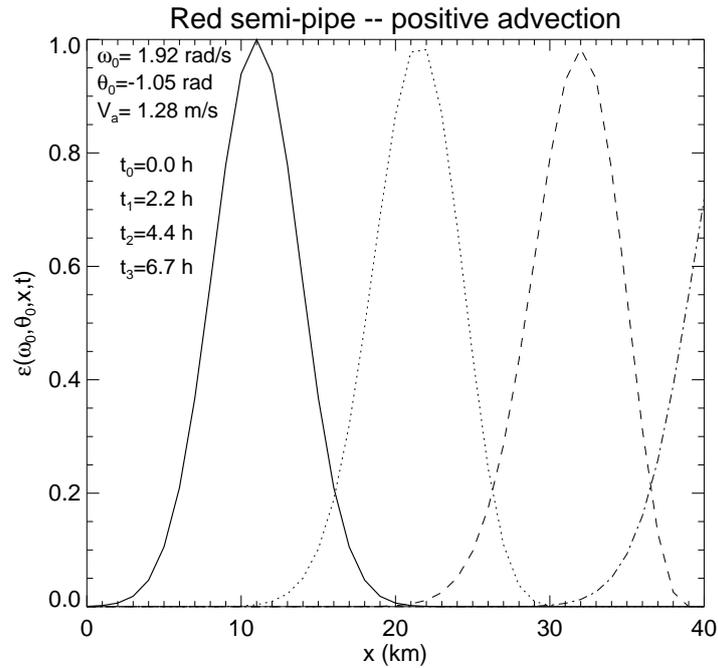}
\caption{Numerical test of the advection of the initial condition $\varepsilon(\omega,\theta,x,t)|_{t=0}= \varepsilon_0(x) \delta(\omega-\omega_0)\delta(\theta-\theta_0)$,  in the frame of Eq.(\ref{HE3}). Here $\varepsilon_0(x)$ is  bell-like smooth function. One can see its motion to the right with the constant positive velocity $V_a = \frac{1}{2}\frac{g}{\omega_0}\cos\theta_0$, where $\omega_0 = 1.92$  rad/s and $\theta_0 = 1.05$ rad,  as well as eventual disappearance at the east shoreline in accordance with ``totally dissipative'' boundary conditions.  Time correspondence to the curves is: initial condition $t_0=0$ h  -- solid line; $t_1=2.2$ h - dotted line; $t_2=4.4$ h -- dashed line; $t_3=6.7$ h -- dash-dotted line.}
\label{AdvectionRed}
\end{figure}

\begin{figure}
\centering
\includegraphics[scale=0.6]{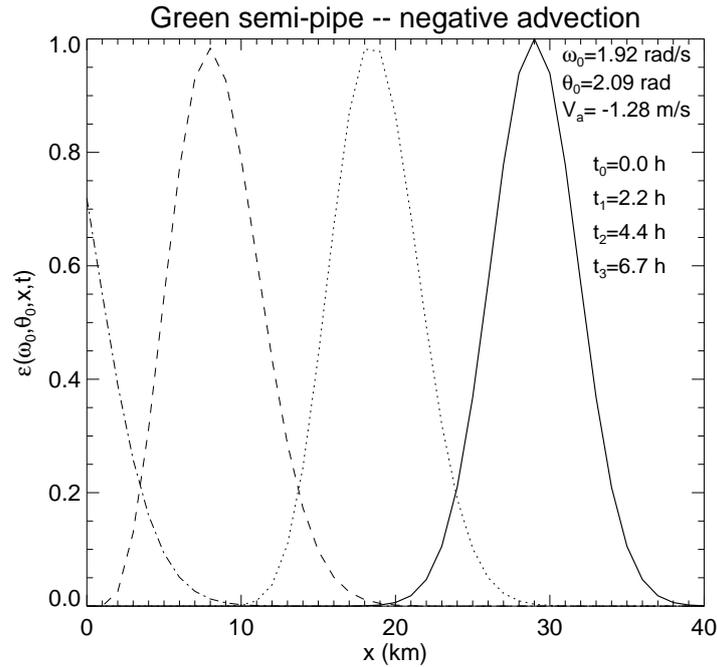}
\caption{Numerical test of the advection of the initial condition $\varepsilon(\omega,\theta,x,t)|_{t=0}= n_0(x) \delta(\omega-\omega_0)\delta(\theta-\theta_0)$ (solid line),  in the frame of Eq.(\ref{HE3}). Here $\varepsilon_0(x)$  is bell-like smooth function. One can see its motion to the left with the constant negative velocity  $V_a = \frac{1}{2}\frac{g}{\omega_0}\cos\theta_0$, where $\omega_0 = 1.92 $ rad/sec and $\theta_0 = 2.09$ rad,  and eventual disappearance at the west  shoreline in accordance with ``totally dissipative'' boundary conditions. Time correspondence to the curves is: initial condition $t_0=0$ h  -- solid line; $t_1=2.2$ h -- dotted line; $t_2=4.4$ h -- dashed line; $t_3=6.7$ h -- dash-dotted line.}
\label{AdvectionGreen}
\end{figure}

The advection part of HE Eq.(\ref{HE3}) has been tested with respect to the smooth localized initial condition propagation in the red and green semi-pipes, and showed reasonable correspondence with the analytical predictions of their sign-dependent advection along the characteristics without significant shape change, till they disappear at the corresponding boundaries, see Fig.\ref{AdvectionRed} and Fig.\ref{AdvectionGreen}.

The above mentioned separate tests of wind energy input, wave energy dissipation, nonlinear interaction as well as advection provide reasonable confidence in the functionality of the numerical model as the whole.

\subsection{Results of numerical simulation} \label{NumRes}

As it was already discussed in the subsection \ref{CauchyProblem}, the numerical modeling has been started from white noise small amplitude waves, propagating in the wind direction. 

\begin{figure}
\noindent\center\includegraphics[scale=0.4, angle=90]{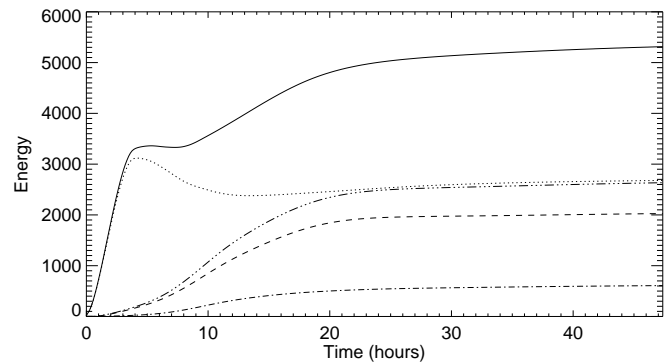} 
\caption{Total wave energy of the fetch $\int\limits_0^{2\pi}\int\limits_{f_{min}}^{f_{max}}\int\limits_{0}^{L}\varepsilon(f,\theta,x,t) df d\theta dx$ as the function of time $t$: solid line. Its components: wave energy against the wind direction (green pipe) - dashed line; in the wind direction (red pipe) - dotted line; orthogonal to the wind direction: dash-dotted line; not in the wind direction - dash-triple-dotted line.}
\label{TotalEnergyOnTime}
\end{figure}

The dependence of total wave energy on time is presented on Fig.\ref{TotalEnergyOnTime}.
The evolution consists of the stages: relatively rapid initial increase, corresponding to  $t<$ 5 hours, and following protracted growth to the asymptotically permanent value.  Given the existence of two wave energy dissipation possibilities -- one through the dissipative shores  and another through the wave-breakings -- we come to the conclusion that the wind energy supply quite unexpectedly comes into the balance with these two wave energy dissipation sinks,  apparently the first one being the dominating.

The analysis of the total wave energy on Fig.\ref{TotalEnergyOnTime} shows that it splits asymptotically in time in the following proportions: 51 $\%$ - in the wind direction, 38 $\%$ - opposite to the wind direction and 11 $\%$ - perpendicular to the wind.  That fact indicates that total wave energy in quasi-stationary state is divided in approximately equal proportions for waves running in the wind direction, and not in the wind direction (i.e. opposite or perpendicular to it). Existence of the waves not in the wind direction can be explained solely by nonlinear four-waves interaction, because according to the  Eq.(\ref{ZRP2}), wind forced waves can appear only in the range of angles $-\pi/2 < \theta < \pi/2$.  Nonlinearity of the waves is, thought,  necessary, but not sufficient condition of this quasi-isotropization of the wave spectrum: the usage of the exact form of $S_{nl}$ instead of also nonlinear DIA-like substitutes, indeed, is the real cause of the observed wave energy back-scatter.

\subsection{Spatio-temporal dynamics of the wave energy spectra}

While the total wave energy behavior as the function of time on Fig.\ref{TotalEnergyOnTime} naively assumes asymptotic isotropization of the wave energy spectra, the more detailed analysis shows  their strongly  anisotropic multi-modal spatio-temporal dynamics.

The wave energy spectra for time $t=2$ hours, presented on Fig.\ref{Spectrum3D2h}, \ref{Polar2h}, has the form of solitary single-maximum hump. Its maximum value, drawn along the fetch coordinate, closely resembles threshold shape, consisting of linear and flat parts.

\begin{figure}
\noindent\center\includegraphics[scale=0.6]{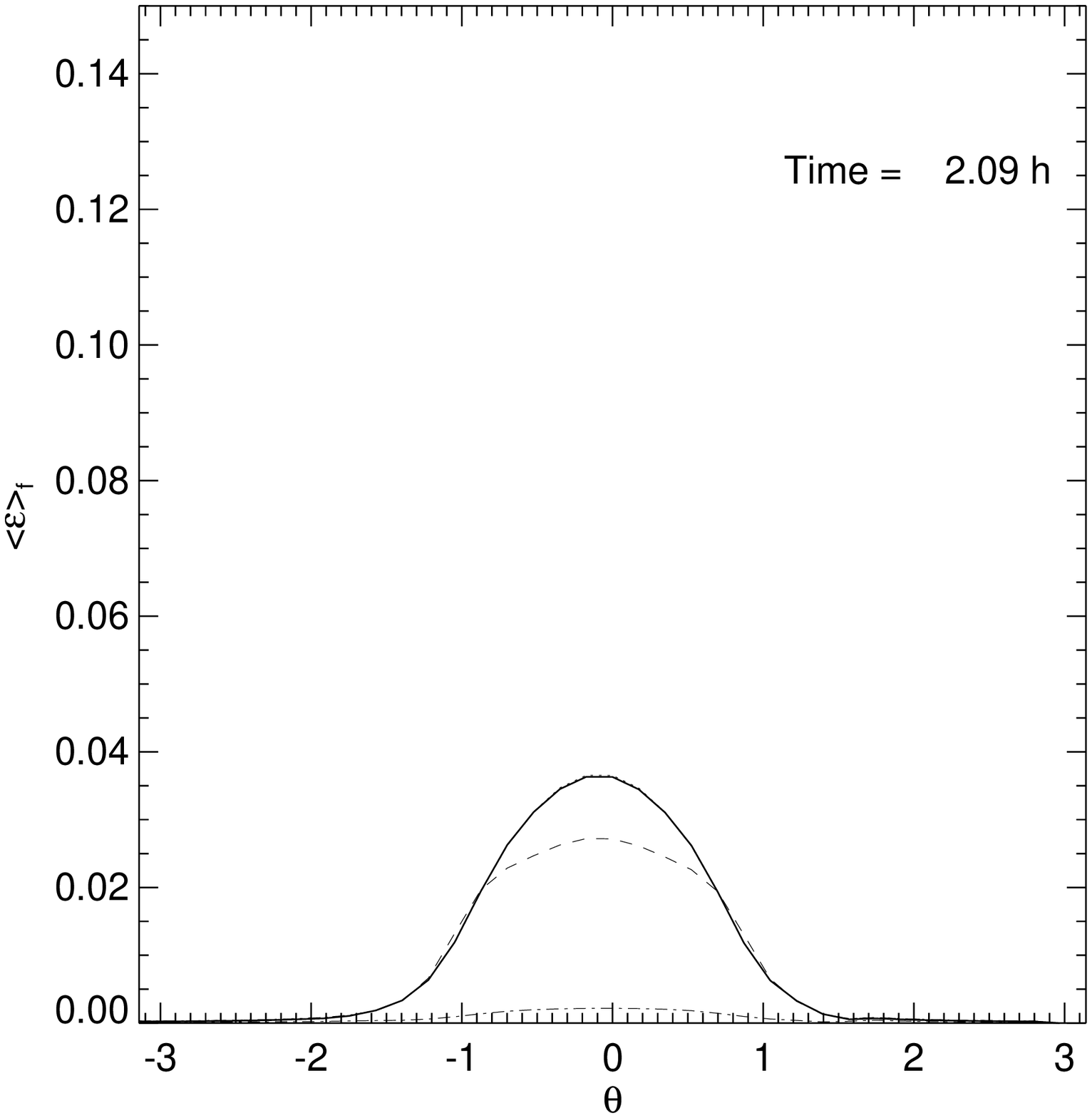} 
\caption{Frequency averaged spectra $<\varepsilon>_f = \int\limits_{f_{low}}^{f_{high}} \varepsilon (f,\theta,x,t) df$, as the function of the angle $\theta$ at fetch coordinates: $x=2$ km - dash-dotted line;  $x=14$ km - dashed line; $x=26$ km - dotted line; $x=38$ km - solid line for time $t=$ 2 h.} \label{FreqAvSpectra2h}
\end{figure}

The frequency-averaged spectra, presented on Fig.\ref{FreqAvSpectra2h}, show smooth unimodal shape, localized in the angular spread of approximately $60^{\circ}$. Fig.\ref{AngAvNormSpectra2} and Fig.\ref{AngAvSpectra2} present normal and decimal logarithm forms of the angle averaged and along the wind spectra for 4 different locations of the fetch. Both also exhibit unimodal spectral structure with the slope shape close to KZ (Kolmogorov-Zakharov \cite{R4}) $\sim\omega^{-4}$ dependence.

For time $t=6$ hours, which is close to the characteristic time of crossing the strait by the waves, corresponding to the spectral peak wavenumber, due to the advection, and estimated as $t_{adv} = L/V_{adv}$, where $L$ is the channel width and $V_{adv}$ is the characteristic advection speed of spectral peak, see Fig.\ref{Spectrum3D6h} and Fig.\ref{Polar6h}, the spectral hump starts to develop side lobes, exceeding the amplitude of the central hump. The intensity of nonlinear interaction between the wave fields in the red and green pipelines is apparently relatively high at this stage of the system evolution. One can interpret this effect as the start of nonlinear interaction between red and green semi-pipes at that time, corresponding to the maximum of the wave energy concentration in the red semi-pipe, closer to the east coast. 

Fig.\ref{AngAvNormSpectra6} and Fig.\ref{AngAvSpectra6} present normal and decimal logarithm plots of the angle averaged and uni-directional along the wind spectra in 4 different locations of the fetch for $t=$ 6 hours. The uni-directional and angle-averaged spectra still exhibit qualitative similarity relative to the wave system state at $t=4$ hours, with close to KZ  $\sim\omega^{-4}$ dependence of high-wavenumbers spectral tail, but already start to develop the divergence of angle-averaged and uni-directional spectra, caused by multi-modality development: one can observe additional spectral mode appearance closer to the west coast.  

\begin{figure}
\noindent\center\includegraphics[scale=0.6]{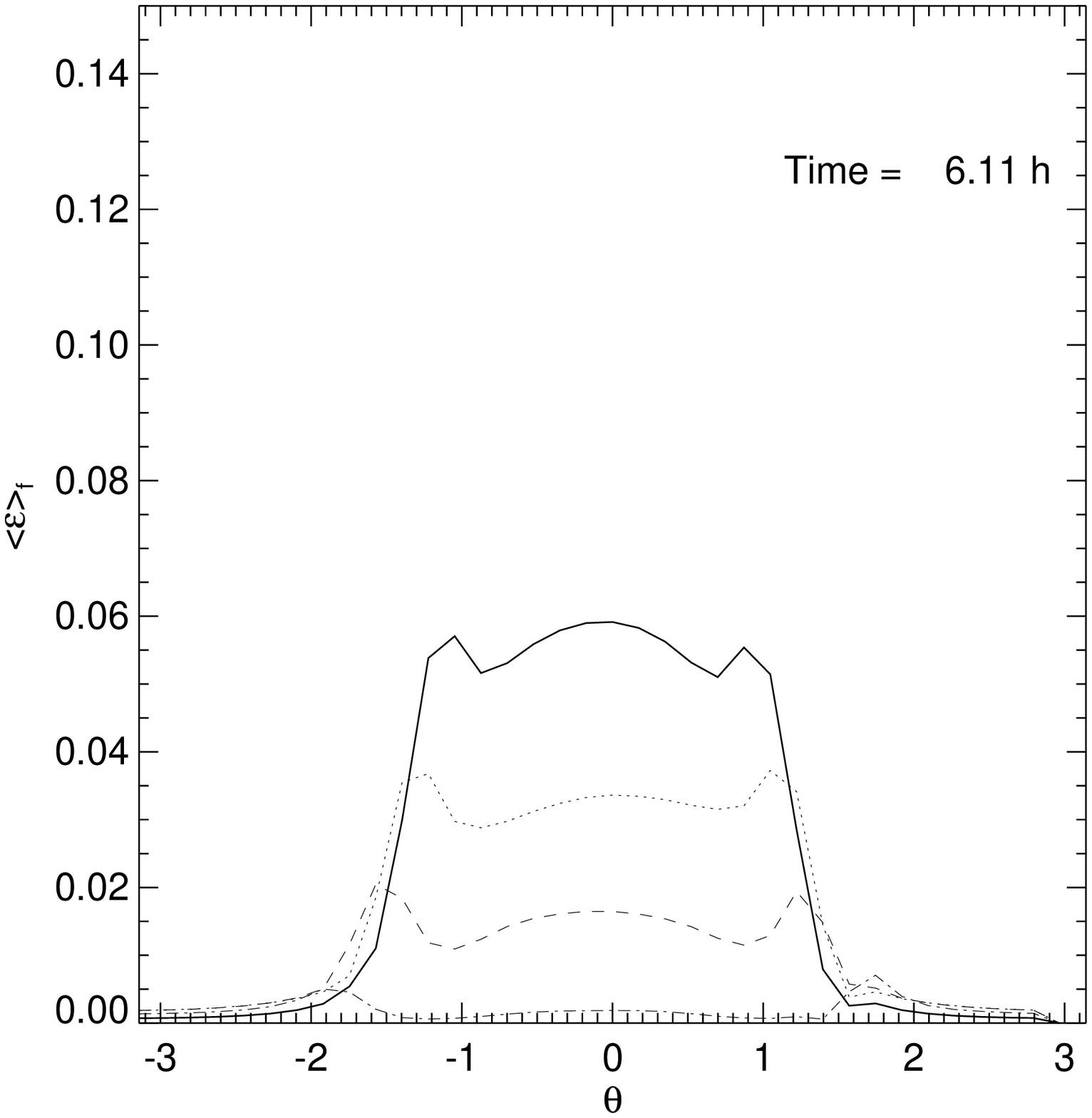} 
\caption{Frequency averaged spectra $<\varepsilon>_f = \int\limits_{f_{low}}^{f_{high}} \varepsilon (f,\theta,x,t) df$, as the function of the angle $\theta$ at fetch coordinates: $x=2$ km - dash-dotted line;  $x=14$ km - dashed line; $x=26$ km - dotted line; $x=38$ km - solid line for time $t=$ 6 h.} \label{FreqAvSpectra6h}
\end{figure}

The frequency-averaged spectra, presented on Fig.\ref{FreqAvSpectra6h} at the same fetch locations, unequivocally exhibit the tendency to multi-modality development, still being localized in the same angle spread of approximately $60^{\circ}$. 

For time $t=40$ hours, corresponding to the wave system quasi-stationary asymptotic state, the wave energy spectrum is fairly sophisticated, see Fig.\ref{Spectrum3D40h}-\ref{Polar40h}: alongside observed before wind  hump, increasing from the west to the east coast line, the powerful side peaks are detectable, which exhibit the waves, moving in the directions, almost perpendicular to the wind.

\begin{figure}
\noindent\center\includegraphics[scale=0.6]{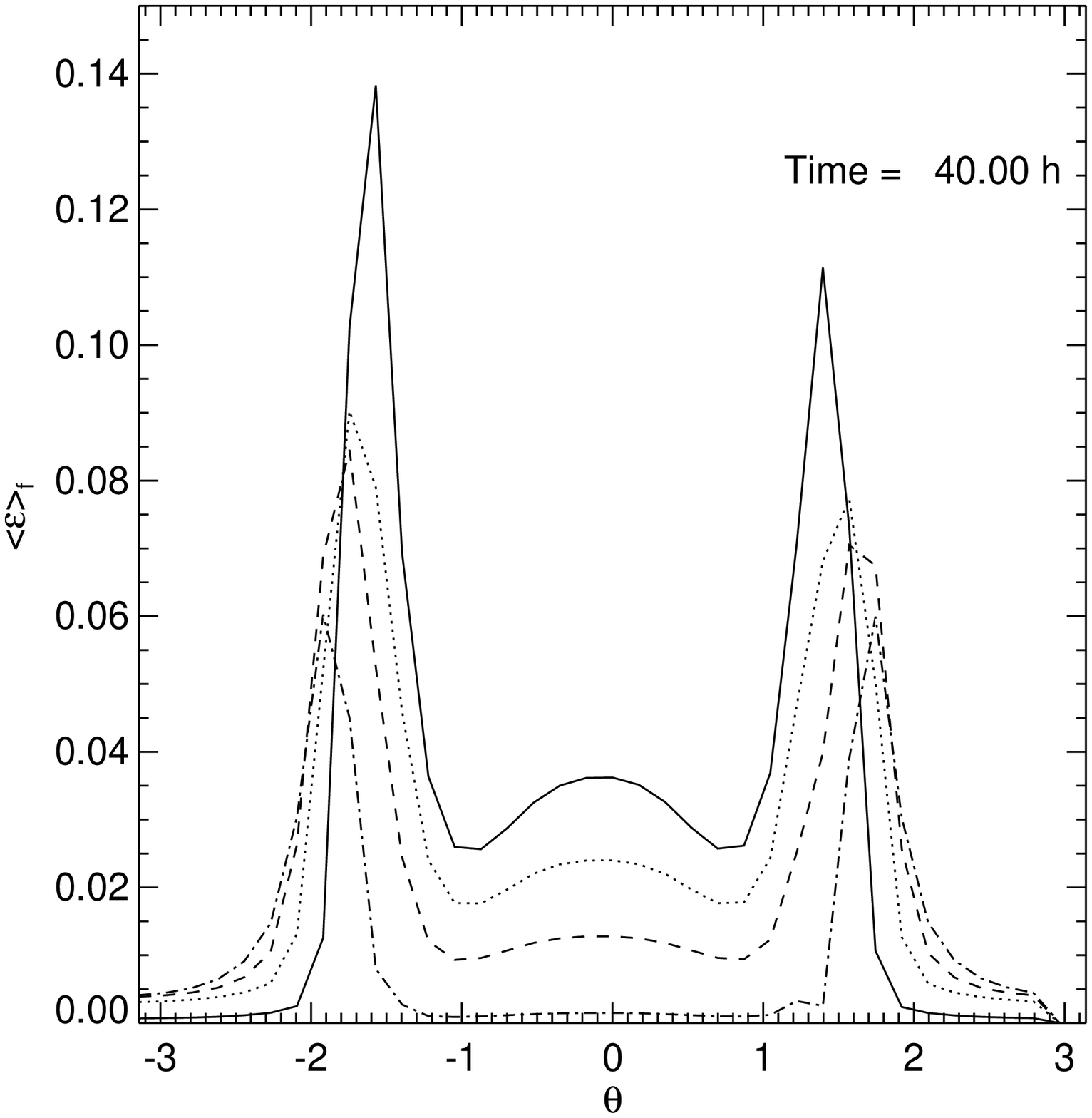} 
\caption{Frequency averaged spectra $<\varepsilon>_f = \int\limits_{f_{low}}^{f_{high}} \varepsilon (f,\theta,x,t) df$, as the function of the angle $\theta$ at fetch coordinates: $x=2$ km - dash-dotted line;  $x=14$ km - dashed line; $x=26$ km - dotted line; $x=38$ km - solid line for time $t=$ 40 h.} \label{FreqAvSpectra40h}
\end{figure}

Fig.\ref{AngAvNormSpectra40} and Fig.\ref{AngAvSpectra40} present the normal and decimal logarithm plots of the angle averaged and uni-directional along the wind spectra in 4 different locations of the fetch for $t=40$ hours. The uni-directional and angle-averaged spectra are qualitatively different at this time from what has already been observed earlier along the fetch: the spectrum along the wind is unimodal and presents the wind sea, while the angle averaged spectrum is multi-modal. This circumstance is important for experimental data interpretation, since angular averaged data present mixed information of two different wave ensembles, which should not be misinterpreted as an artifact. The frequency-averaged spectra, presented on Fig.\ref{FreqAvSpectra40h}, also confirm strikingly multi-modal shape for this time, consisting of wind sea and two side peaks, corresponding to the quasi-orthogonal to the wind modes.

\subsection{NOWA mechanism of waves radiation perpendicular to the wind}

The above observations unequivocally demonstrates the different level of the complexity of wind excited surface wave turbulence in straits with respect to widely accepted perceptions.  The wave system splits into different temporal and spatial  subsystems: we observe unimodal spectral behavior with step-like front propagation for $t \leq 5$ hours, and multimodal spectral behavior thereafter, consisting of wave ensemble, propagating in the wind direction and growing along the fetch from the west to the east coast, which we call the wind sea, and quasi-monochromatic waves, radiated in the direction, close to perpendicular to the wind. This multimodal system eventually comes to quasi-stationary asymptotic state in time.

It should be noted that the angle at which quasi-monochromatic waves are radiated with respect to the wind, is not exactly equal $90^{\circ}$, which coinsides with the shorelines direction -- it is slightly tilted against the wind. The angle of this tilt increases from $0^\circ$ in the direction opposite to the wind, i.e. from the east to the west shore, reaching $15^\circ$ in the close proximity of the west shore line. The presence of this tilt might indicate an alignment of the wave system in the attempt to find the way to balance between the wind energy input and wave energy absorption at the shorelines, to reach the asymptotic quasi-equilibrium state.

It seems that there is the some similarity of the observed quasi-monochromatic radiation with laser radiation, with the role of the active media resonator played by the fetch, filled with  nonlinearly interactive medium, which is limited by the strait shorelines, playing the role of fully transparent mirrors. We emphasize that underlying physical mechanisms of that "nonlinear laser" is completely different. In the considered case, it is purely nonlinear, and we propose to call this effect of emitting of quasi-monochromatic waves, pumped by the orthogonally blowing wind, by {\bf N}onlinear {\bf O}cean {\bf W}ave {\bf A}mplifier, reduced to {\bf NOWA}. The remarkable physical property of this effect is  the ``condensation'' of quasi-monochromatic waves in the near vicinity of the separatrix of the velocity advection distribution, i.e. the locations, where the advection velocity changes its sign. The waves, associated with such locations, are ``trapped'' due  their zero advection velocity. Such condensation effect looks similar to Bose-condensation, known in different areas of modern physics.  Those waves, propagating almost parallel to the shore lines, have the least change to get absorbed at the shorelines, being compared to the other waves, which fate, roughly speaking, is destined to get eventually absorbed  at the shore lines, forming  the main energy dissipation channel.

\subsection{Wave energy distributions along the fetch}

\begin{figure}
\noindent\center\includegraphics[scale=0.4, angle=90]{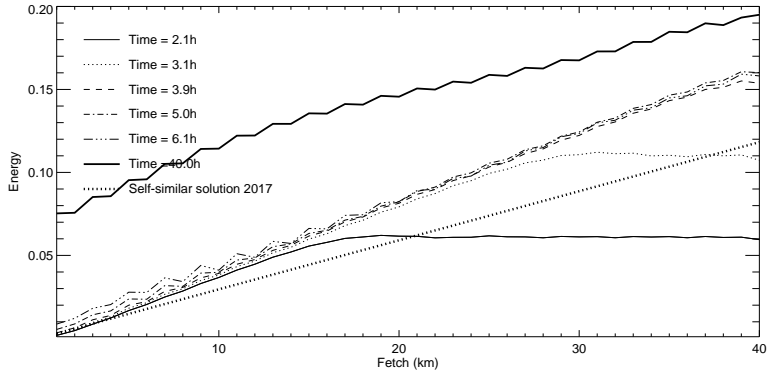} 
\caption{Total wave energy $\int\limits_{0}^{2 \pi} \int\limits_{f_{min}}^{f_{max}} \varepsilon(f,\theta,x,t) df d\theta $ distribution along the fetch for different moments of time, calculated in the angle spread $-180< \theta \le 180^\circ$. The legend ``Self-similar solution 2017'' (dotted line) is referred to the self-similar solution from \cite{R8,R88}. }
\label{TotalEnergyOnFetch}
\end{figure}

\begin{figure}
\noindent\center\includegraphics[scale=0.4, angle=90]{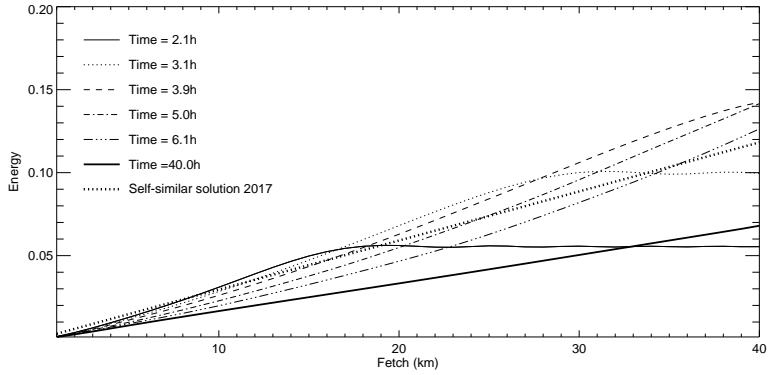} 
\caption{Total wave energy $\int\limits_{-\frac{4}{9}\pi}^{\frac{4}{9}\pi} \int\limits_{f_{min}}^{f_{max}} \varepsilon(f,\theta,x,t) df d\theta $ distribution along the fetch for different moments of time, calculated in the angle spread $-80^\circ < \theta \le 80^\circ$ with respect to the wind direction $\theta_{wind}=0^\circ$. The legend ``Self-similar solution 2017'' (dotted line) is referred to the self-similar solution from \cite{R8,R88}.}
\label{TotalEnergyCentralOnFetch}
\end{figure}

Fig.\ref{TotalEnergyOnFetch} and Fig.\ref{TotalEnergyCentralOnFetch} present total energy dependence on the fetch coordinate for different times, but calculated in the angular spreads $-180^\circ < \theta \le 180^\circ$ and $-80^\circ < \theta \le 80^\circ$ correspondingly. The choice of the angular spread for the Fig.\ref{TotalEnergyCentralOnFetch} is caused by the attempt to separate the wave sea turbulence component from the quasi-monochromatic one. One can see that both pictures show qualitatively similar behavior for $t<5$ hours, until the wave system is separated into wind sea and quasi-orthogonal to the wind waves. 

The behavior for $t<5$ hours on both pictures consists in threshold-like wave energy front propagation along the fetch in the form of the function, consisting of linear and flat part. The propagation of the front is accomplished around $t\simeq 5$ hours, forming self-similar linear function. But the later evolution is completely different. The linear function on Fig.\ref{TotalEnergyCentralOnFetch} relatively slowly reduces its slope to its asymptotic value, while its counterpart on Fig.\ref{TotalEnergyOnFetch} continues its growth, evolving to some curvilinear shape. This effect of deviation from linear shape is natural to expect, since the intensity of quasi-monochromatic waves is growing with their approach to the west coast, along with the degree of their interaction with wind sea, leading to the deformation of self-similar behavior.

\begin{figure} 
\noindent\center\includegraphics[scale=0.4, angle=90]{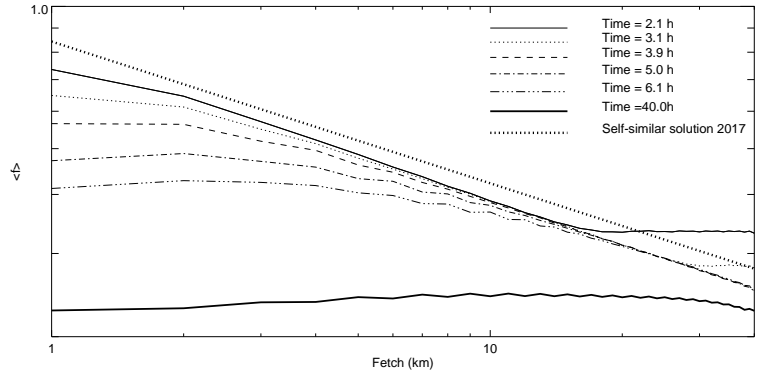} 
\caption{Decimal logarithm of the mean frequency $<f> = \frac{1}{2\pi} \frac{\int\limits_{f_{min}}^{f_{max}}\int\limits_{0}^{2\pi}\varepsilon(f,\theta,x,t) df d\theta } {\int\limits_{f_{min}}^{f_{max}}\int\limits_{0}^{2 \pi} \frac{\varepsilon(f,\theta,x,t)}{\omega} df d\theta }$ as the function of the decimal logarithm of the fetch for different moments of time calculated for angular spread $-180^\circ< \theta \le 180^\circ$. The legend ``Self-similar solution 2017'' (dotted line) is referred to the self-similar solution from \cite{R8,R88}.}
\label{MeanFreqOnFetch}
\end{figure}

Fig.\ref{MeanFreqOnFetch} shows the decimal logarithm of the mean frequency distribution as the function of the decimal logarithm of the fetch for different moments of time, calculated in the angular spread $-180^\circ < \theta \le 180^\circ$. One can observe similar to total wave energy behavior, consisting also in threshold-like function, propagating from the west to the east coast. Asymptotically in time, the mean frequency distribution along the fetch, calculated over both ensembles, comes to almost constant value as the function of the fetch coordinate.

\begin{figure} 
\noindent\center\includegraphics[scale=0.4, angle=90]{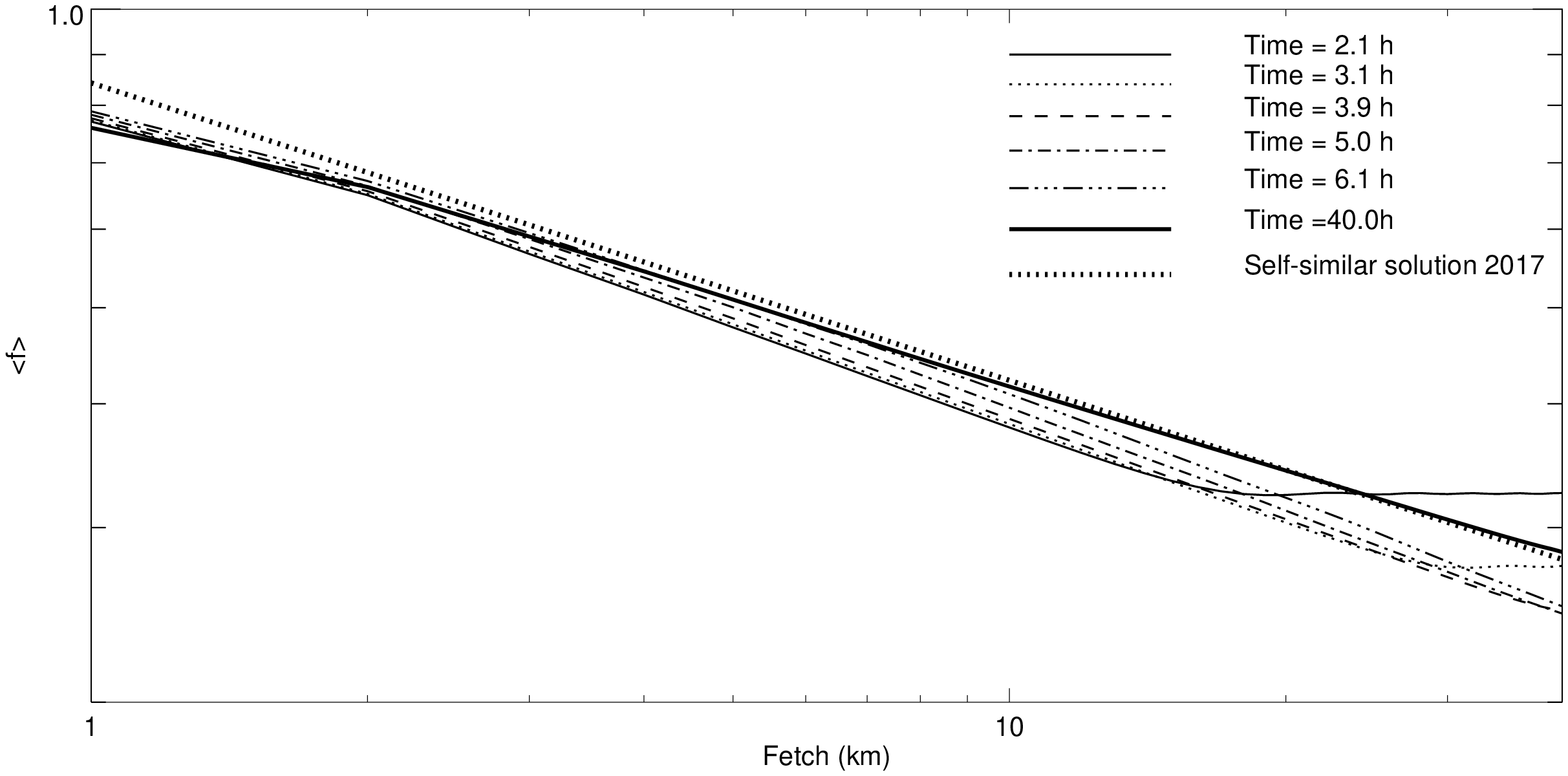} 
\caption{ Decimal logarithm of the mean frequency $<f> = \frac{1}{2\pi} \frac{\int\limits_{f_{min}}^{f_{max}}\int\limits_{-\frac{4}{9}\pi}^{\frac{4}{9}\pi}\varepsilon(f,\theta,x,t) df d\theta } {\int\limits_{f_{min}}^{f_{max}}\int\limits_{-\frac{4}{9}\pi}^{\frac{4}{9}\pi}\frac{\varepsilon(f,\theta,x,t)}{\omega} df d\theta }$ as the function of the decimal logarithm of the fetch for different moments of time calculated for angular spread $-80^\circ< \theta < +80^\circ$ with respect to the wind direction $\theta_{wind}=0^\circ$. The legend ``Self-similar solution 2017'' (dotted line) is referred to the self-similar solution from \cite{R8,R88}.}
\label{MeanFreqCentralOnFetch}
\end{figure}

Similar to the separate turbulent components analysis of the total wave energy, Fig.\ref{MeanFreqCentralOnFetch} shows decimal logarithm of the mean frequency distribution along the fetch for different times as the function of the decimal logarithm of the fetch, calculated in the angular spread $-80^\circ < \theta \le 80^\circ$, which takes into account only the wave sea effects, excluding the quasi-monochromatic waves. The threshold-like function, propagating from the west coast to the east coast is also clearly distinguishable, as in  Fig.\ref{MeanFreqOnFetch}, and is in correspondence with self-simlar laws Eq.(\ref{IndicesDL}),  (\ref{FrTimeSelfSim}) (the flat portion of the function) and Eq.(\ref{IndicesLF}),  (\ref{FrFetchSelfSim}) (the linear portion of the function). One can see that the asymptotic curve at $t=40$ hours is closer to the linear self-similar form, than the corresponding curve on Fig.\ref{MeanFreqOnFetch}, except closely adjacent to the west coast region of 3 km width. This effect is also natural, because of quasi-monochromatic waves  intensity growth from the east to the west coast line as well as their interaction intensity with the wind sea, i.e. nonlinear interaction of the wave energy in the red and green semi-pipes, leading to the deformation of the linear self-similar behavior.

\subsection{Non-stationary limited fetch solution of HE in straits}

The important effect to return to is the above mentioned threshold-like wave energy front in the form of spectral wave energy hump, propagating along the fetch in the wind direction for time $t<5$ hours, consisting of linear and flat parts and evidenced, for example, as the total energy and mean frequency dependencies on the fetch. The linear part of this function is described by limited fetch self-similar solution Eqs.(\ref{IndicesLF})-(\ref{FrFetchSelfSim}), while the flat part is described by growing in time duration-limited solution Eqs.(\ref{IndicesDL})-(\ref{FrTimeSelfSim}). The connection point of these different self-similar solutions is moving with the speed of spectral peak advection velocity.  

\begin{figure}
\noindent\center\includegraphics[scale=0.4, angle=90]{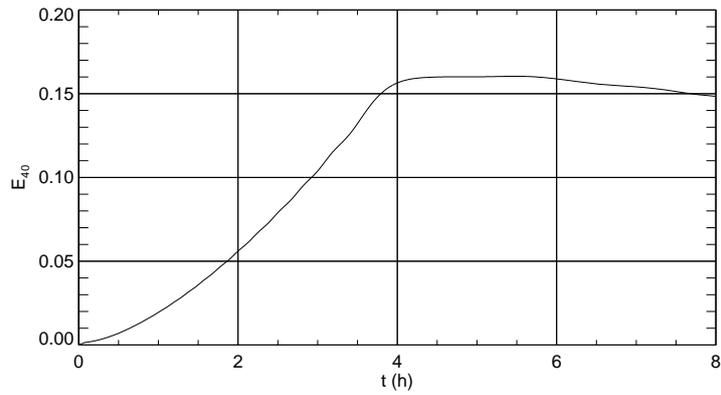} 
\caption{Energy $\left. E_{40}(t) = \int\limits_{0}^{2 \pi} \int\limits_{f_{min}}^{f_{max}} \varepsilon(f,\theta,x,t) df d\theta \right|_{x=40 \,\,\, \rm km}$ for the fetch coordinate $x=40$ km as the function of time.}
\label{EnergyVsTime40}
\end{figure}

\begin{figure} 
\noindent\center\includegraphics[scale=0.4, angle=90]{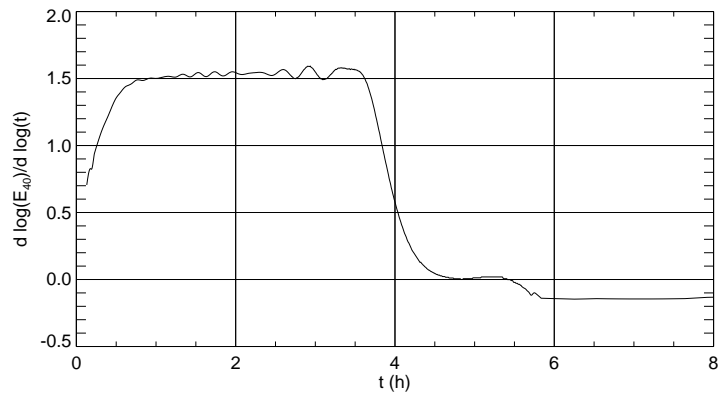} 
\caption{Energy power index $\frac{d \log{ E_{40}}(t)}{d \log{t}}$ dependence on time at the fetch coordinate $x=40$ km for the function $E_{40}$ from Fig.\ref{EnergyVsTime40}.}
\label{EnergyVsTime40Index}
\end{figure}

While the limited fetch part of this solution is quite clear to observe due to its linear total wave energy dependence on the fetch, the duration limited part is hindered. To confirm the existence of duration limited solution, governing the flat part of the wave energy propagating front, we present Fig.\ref{EnergyVsTime40}, showing energy dependence on time at the fetch coordinate $x=40$ km, i.e. at the east coast. Fig.\ref{EnergyVsTime40Index} presents local power index of this wave energy dependence on time, which is derived from Fig.\ref{EnergyVsTime40}. One can see that for the time span between 1 and 3.5 hours, indeed, the observed power index value is contained within $5\%$ deviation from the self-similar solution index $q=10/7$ of the predicted wave energy growth in duration limited case, see Eq.(\ref{IndicesDL}) .

Similar consideration can be applied to the behavior of threshold-like front propagation of mean frequency on Fig.\ref{MeanFreqCentralOnFetch}, described in the previous subsection, which we omit for the sake of brevity.

Thus, we presented here the non-stationary ``running'' threshold-like solution of HE, describing the wave system evolution for $t<5$ hours, which corresponds to the initial wave system development.

\section{Experimental evidence} \label{Experimental_Evidence}

Multi-modality of the ocean wind waves spectra in the conditions of quasi-constant wind has been reported from experimental observation for at least last three decades. Recent advances in 3D data topography acquisition technologies, including global positioning, remote sensing and wireless communication improved those measurements qualitatively and quantitatively. 

Different mechanisms, including Phillips resonance theory of wind-wave generation as well as nonlinear wave quadruplet interaction have been attracted for spectral multi-modality explanation. Unfortunately, significant volume of the acquired information has been lost for general scientific community due to rejection to publish unexpected results.

Those results include \cite{R16,R109,R110,R111}, which report:

\noindent
 {\it ``In the near shore region, the normal of the two waves fronts are almost perpendicular  to the wind vector. As fetch increases, the angle between wind and wave direction decreases, and the wave components veer toward wind as they grow. ...In fact, at very short fetches the dominant wave direction is cross-wind rather than along-wind.''}.

That observation is strikingly similar to what we report in the current publication. With the kind permission of Paul A. Hwang, we present Fig.\ref{Hwang}, showing spatio-temporal and spectral-temporal plots of the corresponding experimental observations.

One can also see the same phenomenon in publications \cite{R20} (see Fig.13) and \cite{R13} (see Fig.11-13).

\section{Conclusions} \label{Concl}

We studied numerically the development of surface wave turbulence in deep strait for permanent wind, directed orthogonally to the shore lines. It is shown that associated limited fetch growth problem is the complex process, exhibiting multimodal spectra, that splits into various spatio-temporal sub-processes. Let us wrap up the observed dynamics of the wave energy turbulence in straits:

\begin{enumerate}

\item{} The initial process of waves excitation from white noise initial conditions consists in threshold-like wave energy front  propagation in the wind direction in the form of spectral energy single hump from the west to the east coast for characteristic times, defined by the ratio of the channel width to the characteristic spectral peak advection velocity. This regime is localized in the positive advection velocity semi-pipe of the mixed Real-Fourier space, corresponding to the waves, moving in the wind direction, and has similarities with already knows wind-sea self-similar limited fetch regimes for unlimited domains.  This propagating wave energy front from the west to the east cost consists of two parts -- self-similar solution for limited fetch growth and self-similar solution for duration limited growth, and is accomplished with the formation of intermediate self-similar limited fetch asymptotic -- the perfectly linea shape of the front. The joint point of these self-similar solutions moves in real space with local spectral peak advection velocity in the direction of the wind. That describes the non-stationary solution for initial turbulence development, limited by the time span of spectral peak advection from the west to the east shoreline of the strait.

\item{} The second regime develops thereafter, when the wave energy threshold-like front has aready reached the east coast, and its start is caused by nonlinear wave energy ``throwing-off''  of already formed spectrum in the positive velocity advection region (red semi-pipe) into the negative advection region (green semi-pipe). Its main exhibition consists in the amplification of quasi-monochromatic waves, propagating quasi-orthogonally to the wind in a veer-like manner. We propose to call the described laser-like effect by \textit{Nonlinear Ocean Wave Amplifier} of the quasi-monochromatic waves in the wind-orthogonal direction, which we reduce to the acronym \textit{NOWA}.  It is apparently associated with the wave energy Bose-like condensation on zero-advection separatrix, splitting the regions of different advection direction.

\item{} We demonstrate that the surface wave turbulence in the channel is divided asymptotically to already known self-similar wind-sea and lower-frequency quasi-monochromatic waves, radiating nearly perpendicular to the wind. It is quite surprising that the wave system eventually reaches asymptotic equilibrium, or ``mature sea'' state, due to the balancing of the wave energy coming through the wind input channel by another two channels of wave energy dissipation: the main wave energy dissipation channel -- absorption at the shorelines, and the secondary wave energy dissipation channel through the wave-breaking . The described mechanism of mature sea formation yields the alternative to the dubious mature sea concept, based on physically unexplainable long wave energy dissipation, circulating in oceanographic community. The quasi-orthogonal to the wind spectral component tends to slope against the wind closer to the west coast at the angle of $15^{\circ}$ with respect to the shore line. This slanting angle is growing in the direction against the wind from the east to the west shore, creating veer-like pattern. Those slanting waves, having wave-number component opposite to the wind, demonstrate, in particular, the fact of nonlinear waves generation against the wind. 

\item{} Asymptotically, the wave energy fraction in the wind direction, is equal one orthogonal  and opposite to the wind. It is the important result of the presented numerical experiments, demonstrating the necessity of  taking into account of the nonlinear interaction in the exact form, which is responsible for backscattering of the wind-driven waves against the wind as well as quasi-monochromatic waves radiation orthogonally to the wind.

\item{} Our experiments demonstrate that quasi-monochromatic waves in asymptotic stationary state have the wave length about four time longer than the wind sea spectral peak. That's why these results holds the promise for explanation of the \textit{seiches}, presenting significant problem for moored ships in ports as well as the prediction of the amplitude and localization structure in confined basins, since the described laser-like \textit{NOWA} mechanism can explain significantly lower frequency spectral part appearance in the bounded systems.

\item{} Another lesson of the presented research emphasizes the importance of proper understanding of inhomogeneity effects, demonstrated, but not limited to, by ocean straits case, in the form of quasi-monochromatic waves, generated orthogonally to the wind. It is conceivable to imagine the open sea situation very far from the shore line, where the wind sharply changes its direction at some particular point in space and time. That condition is mathematically equivalent to the creation of the inhomogeneity just like it is done in the presence of the shorelines. We are led to infer that the effects, similar to ones observed in straits, could be also observed in the the open ocean as well, which is confirmed by experimental observations from ConocoPhillips Ecofisk platform, located about 320 km offshore in North Sea, with water depth in the area between 70 and 80 m (see Fig.11-13 in publication \cite{R13}). The expected physical picture of the open ocean turbulence will contain, therefore, the footprints of two different self-similar solutions -- the duration limited as well as the limited fetch ones, which can bee identified, for example, by calculation of the corresponding spatial and temporal indices of their power-law growth.

\end{enumerate}

The described \textit{NOWA} laser-like effect should inevitably appear in operational wave forecasting models in the case of exact nonlinear quadruplet interaction usage, and might be significantly suppressed or even absent in the case of currently used DIA-like substitutes. We are concerned that the practice of modification of the original $S_{nl}$ term for the sake of saving computational time, can significantly contribute into discrepancy between the field experimental observations and operational hindcasting as well as forecasting. It is important in this relation to carefully perform the comparison of individual hindcasting cases both for DIA as well as the exact $S_{nl}$ nonlinear term, and not to misinterpret possible multi-modal spectra appearance as numerical artifacts. 

As far as concerns our future plans, we are planning to consider the waves reflection from the coastal lines, which is expected to produce strong amplification of the \textit{NOWA}  laser-like effect.

\begin{acknowledgements} 

The presented  research has been accomplished due to the support of the grant “Wave turbulence: the theory, mathematical modeling and experiment” of the Russian Scientific Foundation No 19-72-30028.

The author gratefully acknowledge the support of this foundation.

\end{acknowledgements}

\bibliographystyle{copernicus}
\bibliography{references}

\newpage


\begin{table}
\centering
	\begin{center} 
		\begin{tabular}{c c}
			\includegraphics[width=0.4\linewidth]{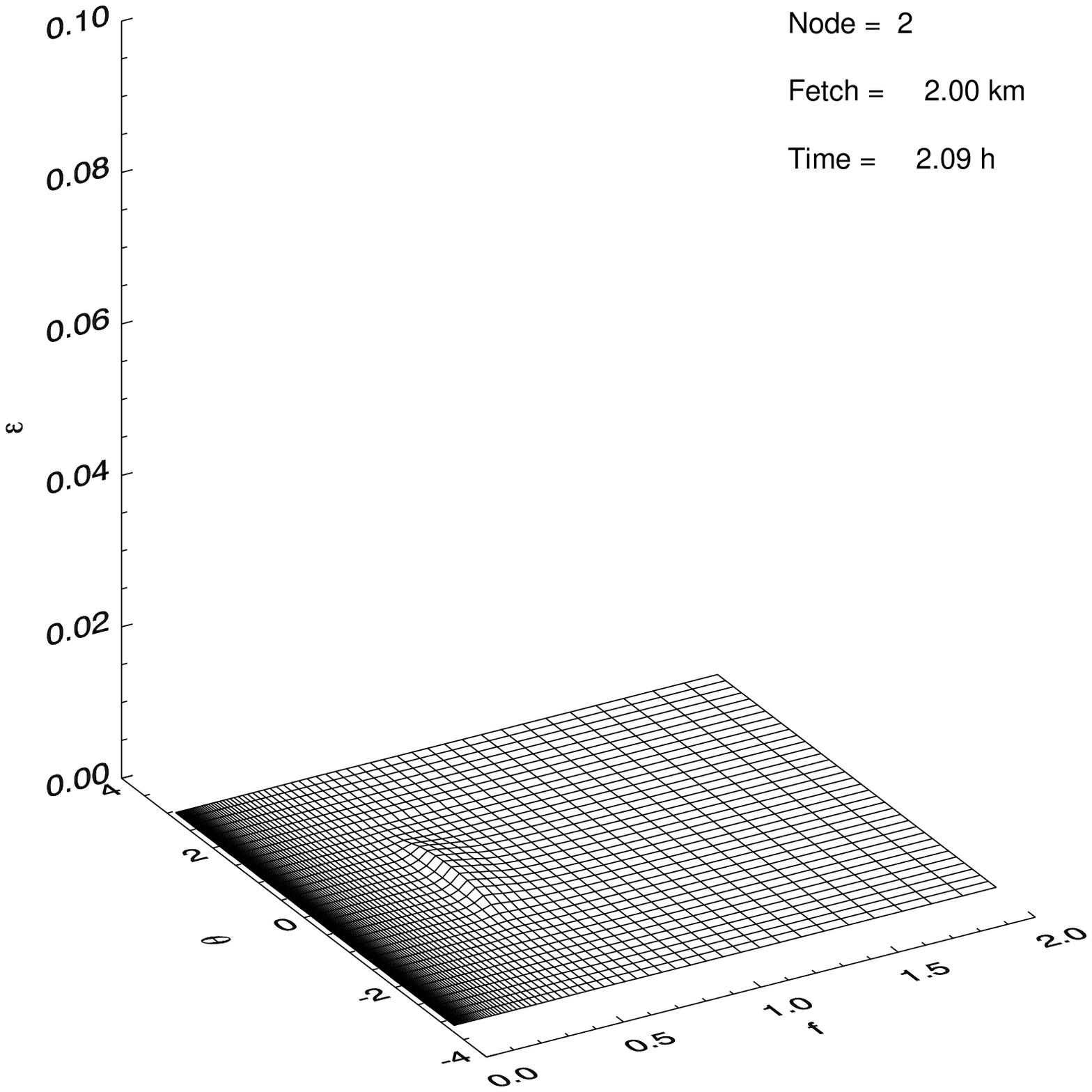} & \includegraphics[width=0.4\linewidth]{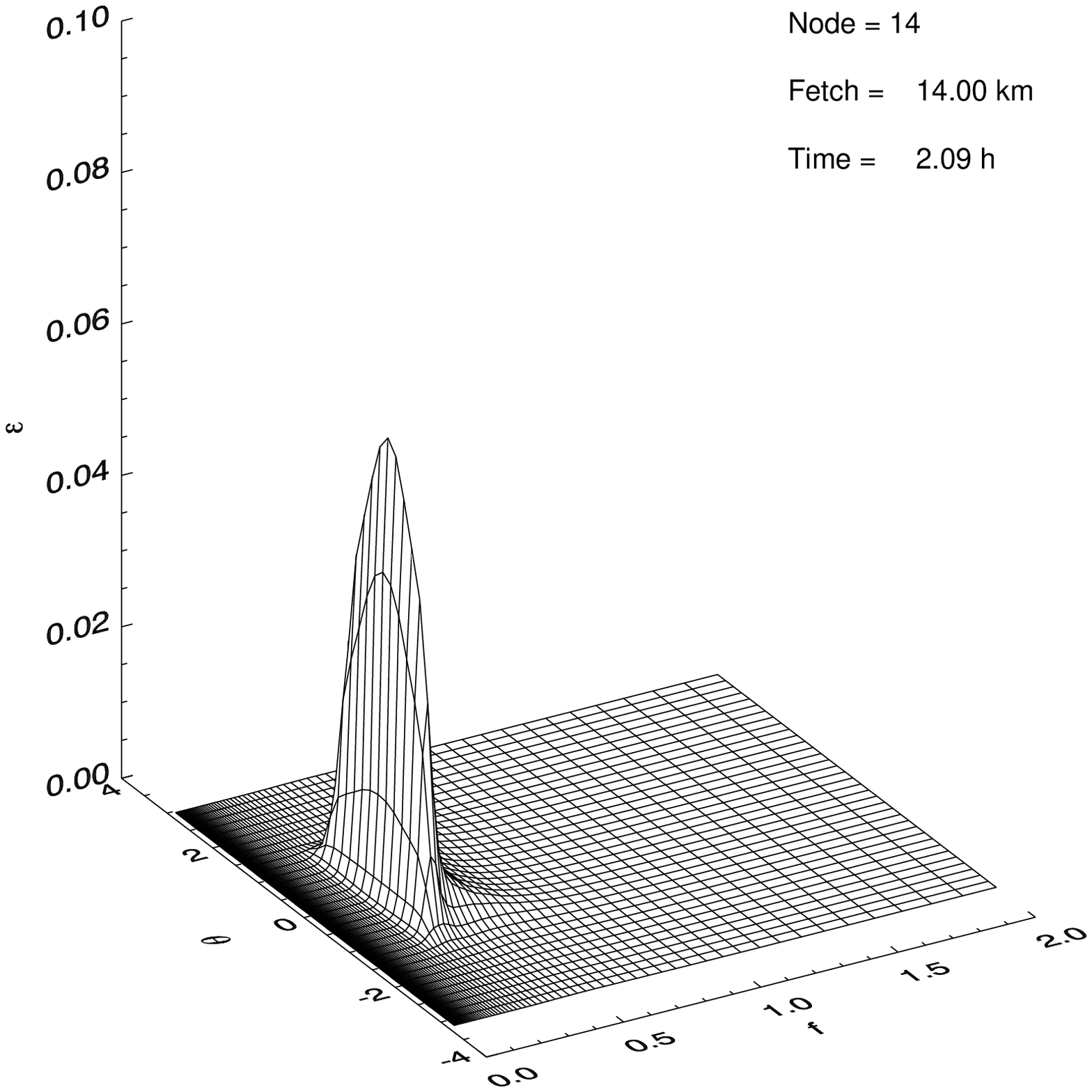} \\ [5mm]
			\includegraphics[width=0.4\linewidth]{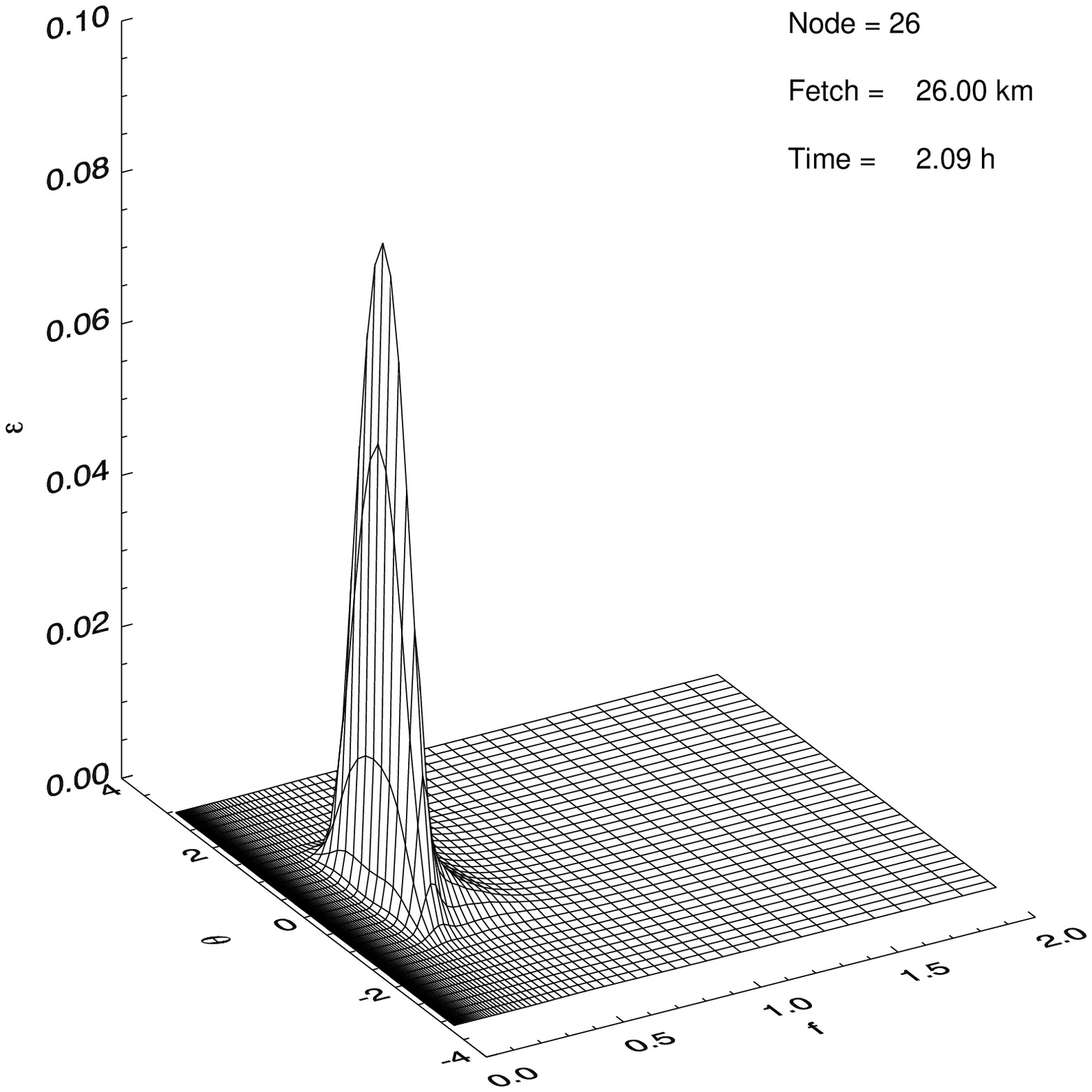} & \includegraphics[width=0.4\linewidth]{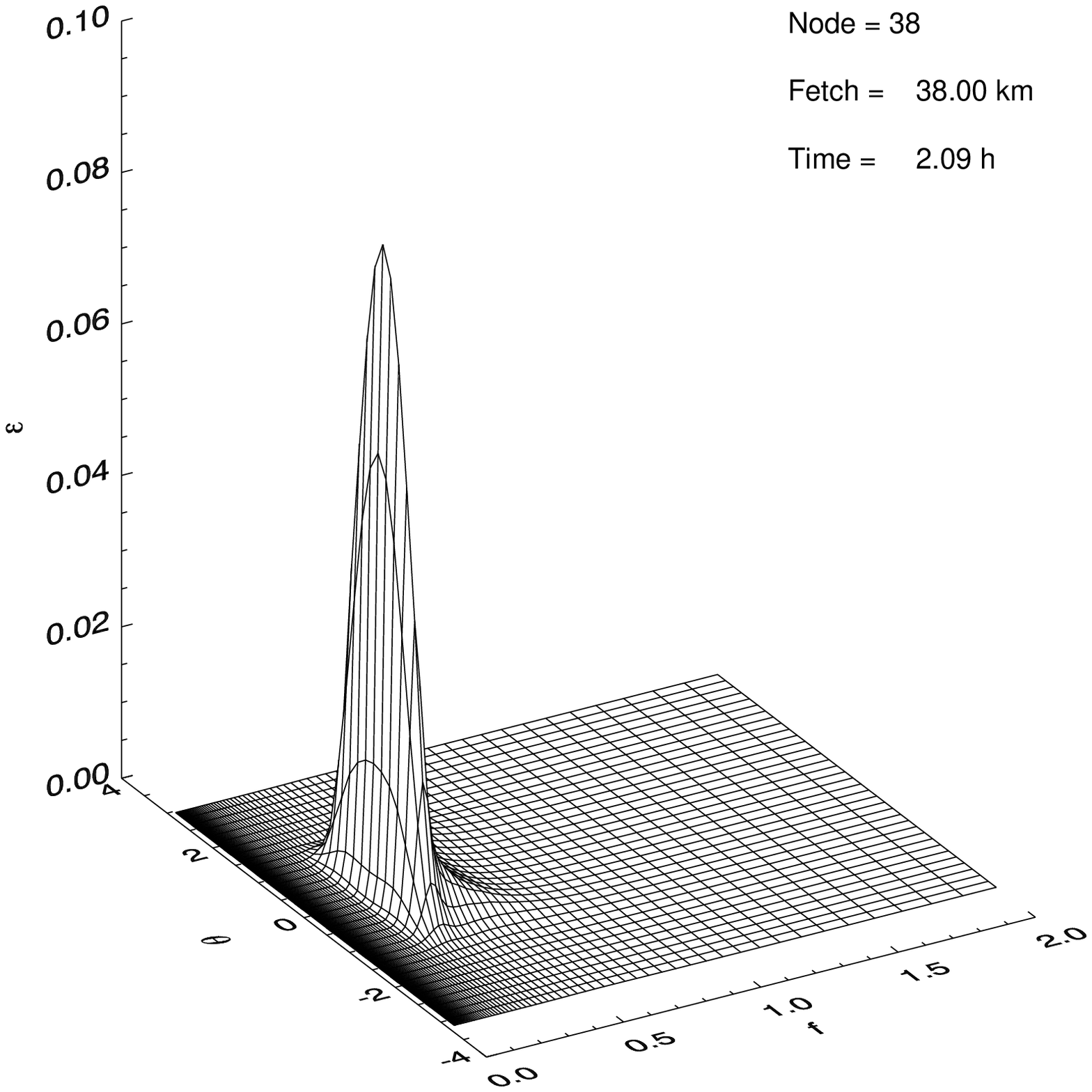} \\
		\end{tabular}
	\captionof{figure}{Energy spectrum $\varepsilon(f,\theta,x,t)$ as the function of the frequency $f$ and angle $\theta$ at the fetch coordinates $x=$ 2, 14, 26 and 38 km for time $t=2$ h.} \label{Spectrum3D2h}
	\end{center}
\end{table}

\begin{table}
\centering
	\begin{center} 
		\begin{tabular}{c c}
			\includegraphics[width=0.4\linewidth]{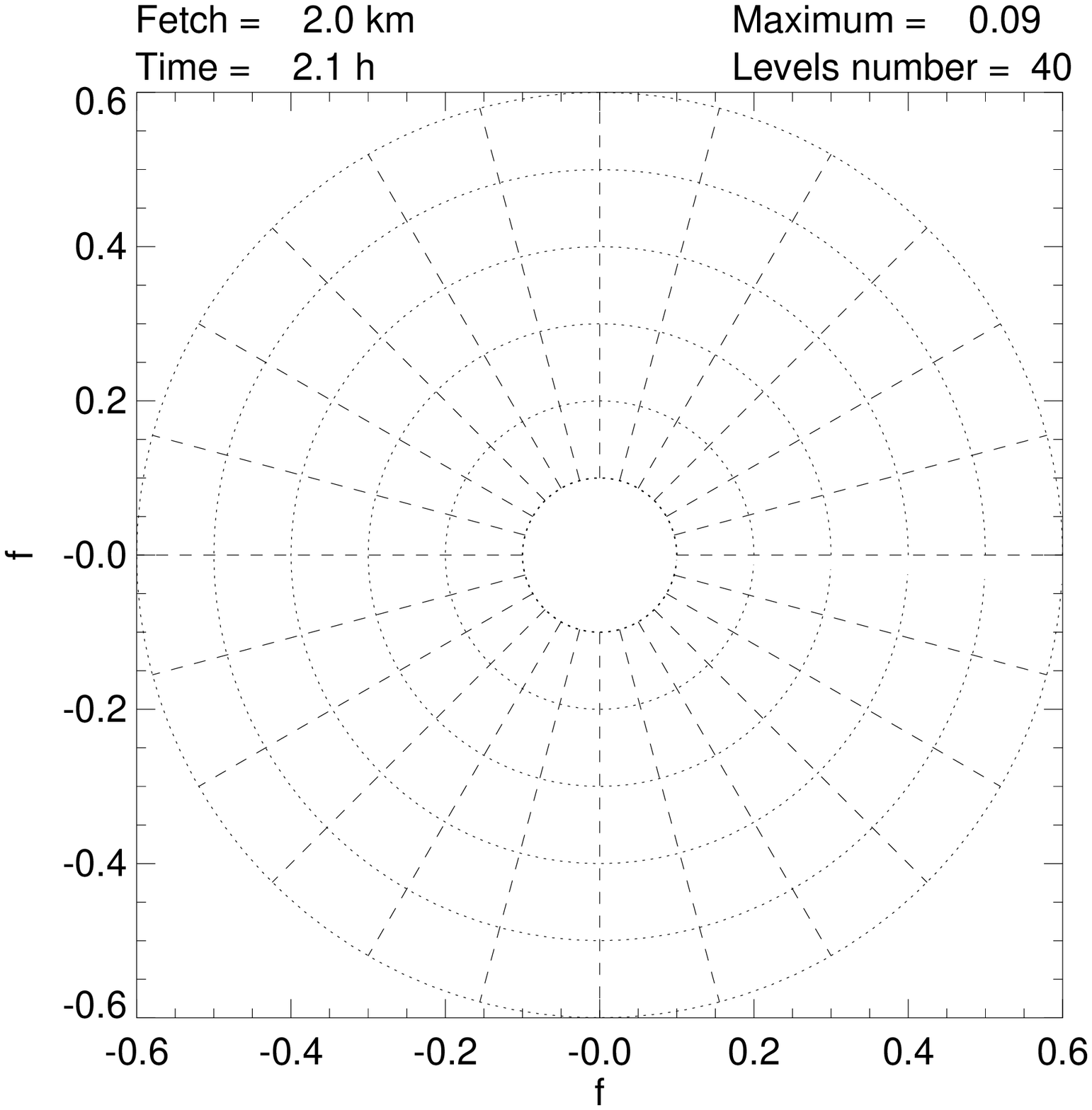} & \includegraphics[width=0.4\linewidth]{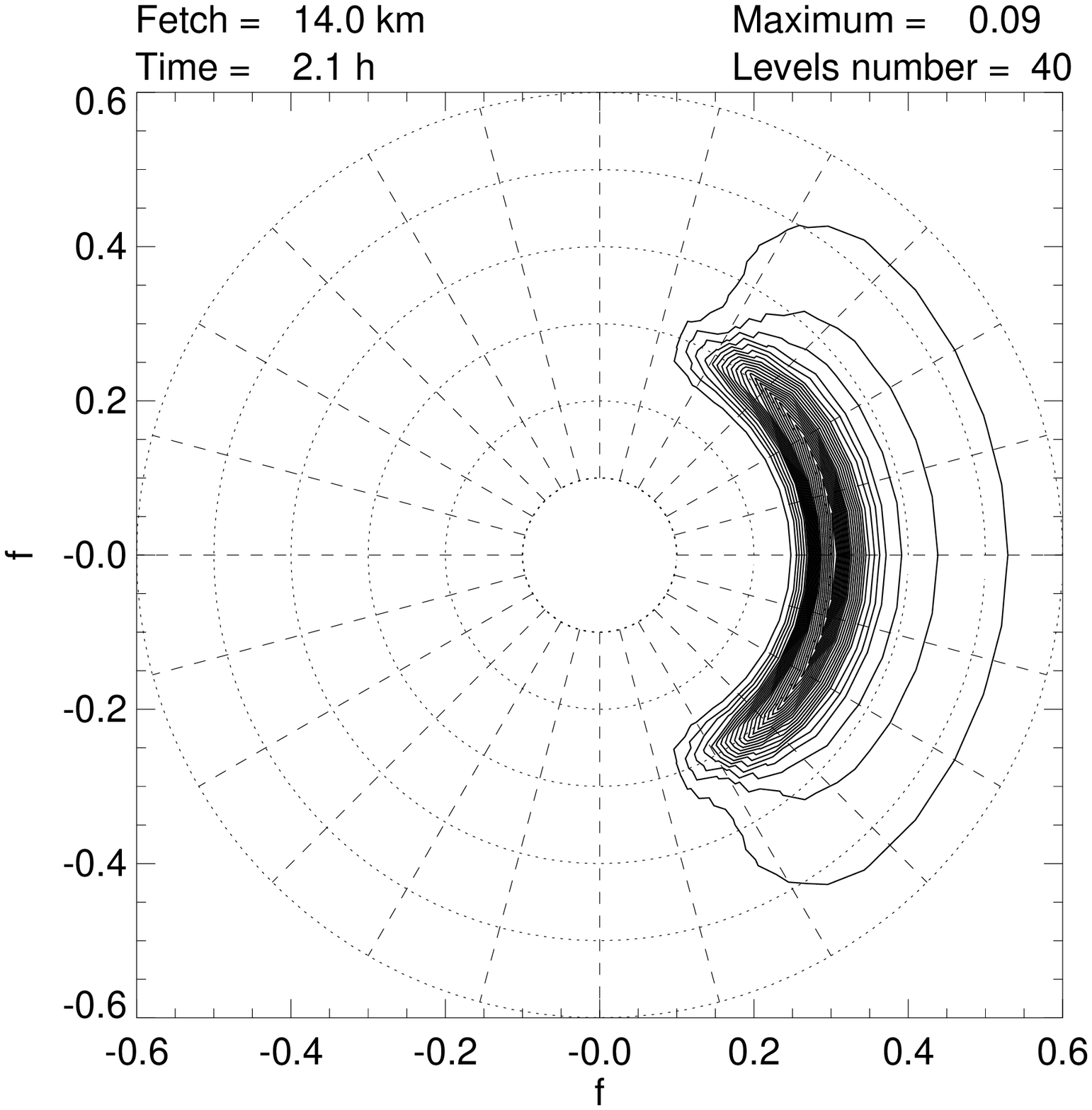}   \\ 
[5mm]			

\includegraphics[width=0.4\linewidth]{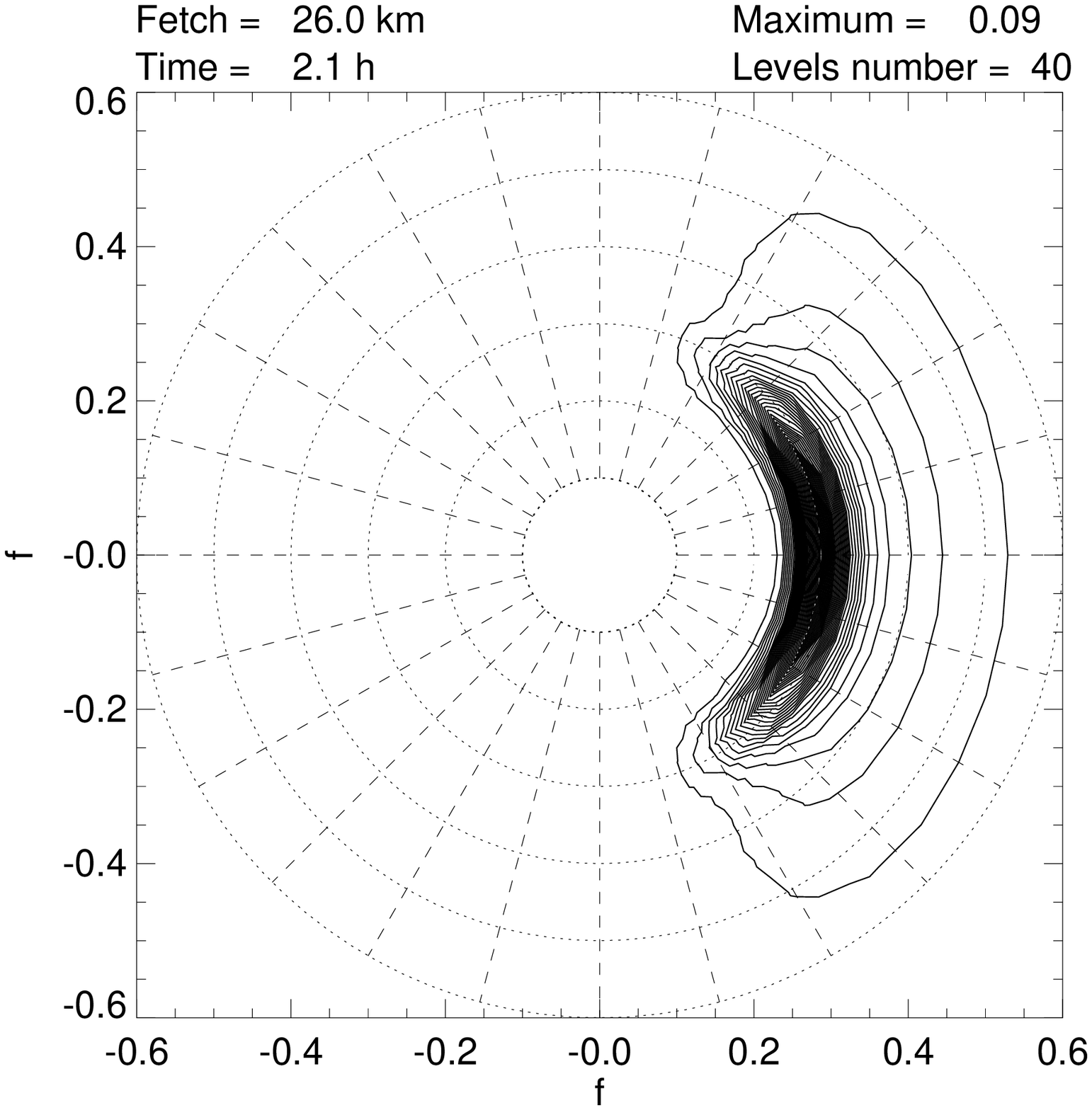} & \includegraphics[width=0.4\linewidth]{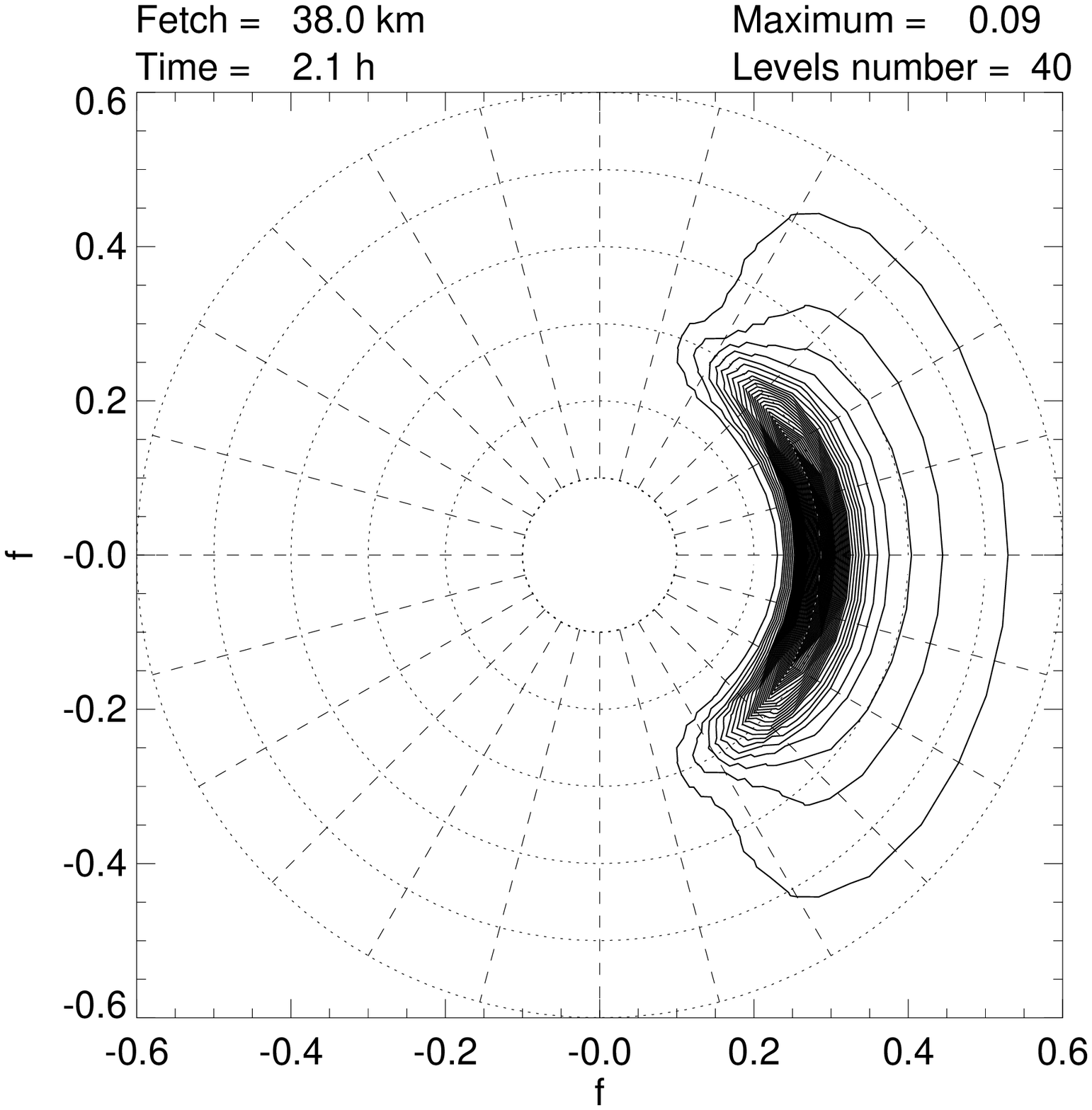} \\
		\end{tabular}
	\captionof{figure}{Energy spectrum $\varepsilon(f,\theta,x,t)$ as the function of the frequency $f$ and angle $\theta$ at the fetch coordinates $x=$ 2, 14, 26 and 38 km for time $t=2$ h.} \label{Polar2h}
	\end{center}
\end{table}

\begin{table}
	\centering
		\begin{center} 
			\begin{tabular}{c c}
			
	\includegraphics[width=0.4\linewidth]{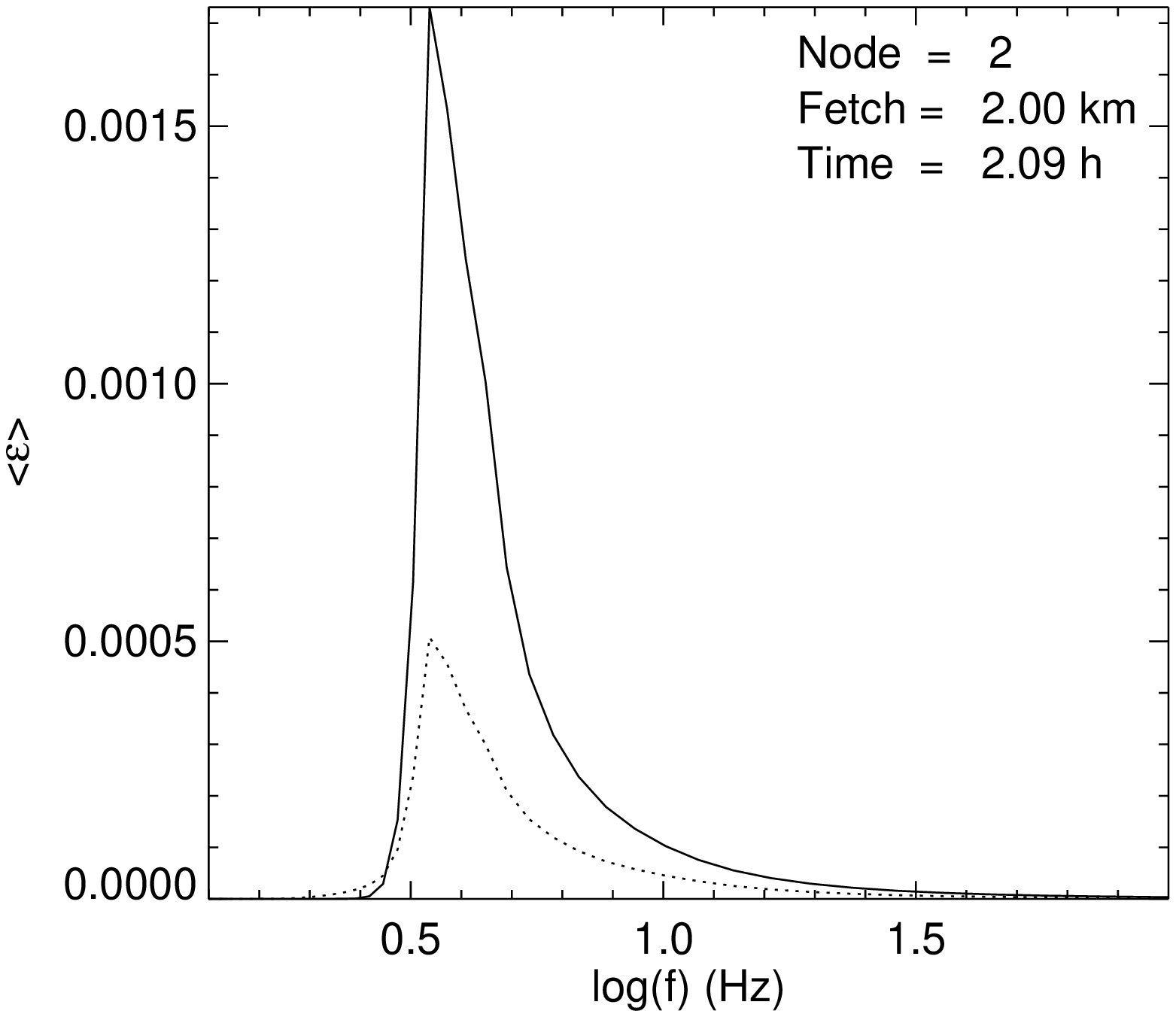} & \includegraphics[width=0.4\linewidth]{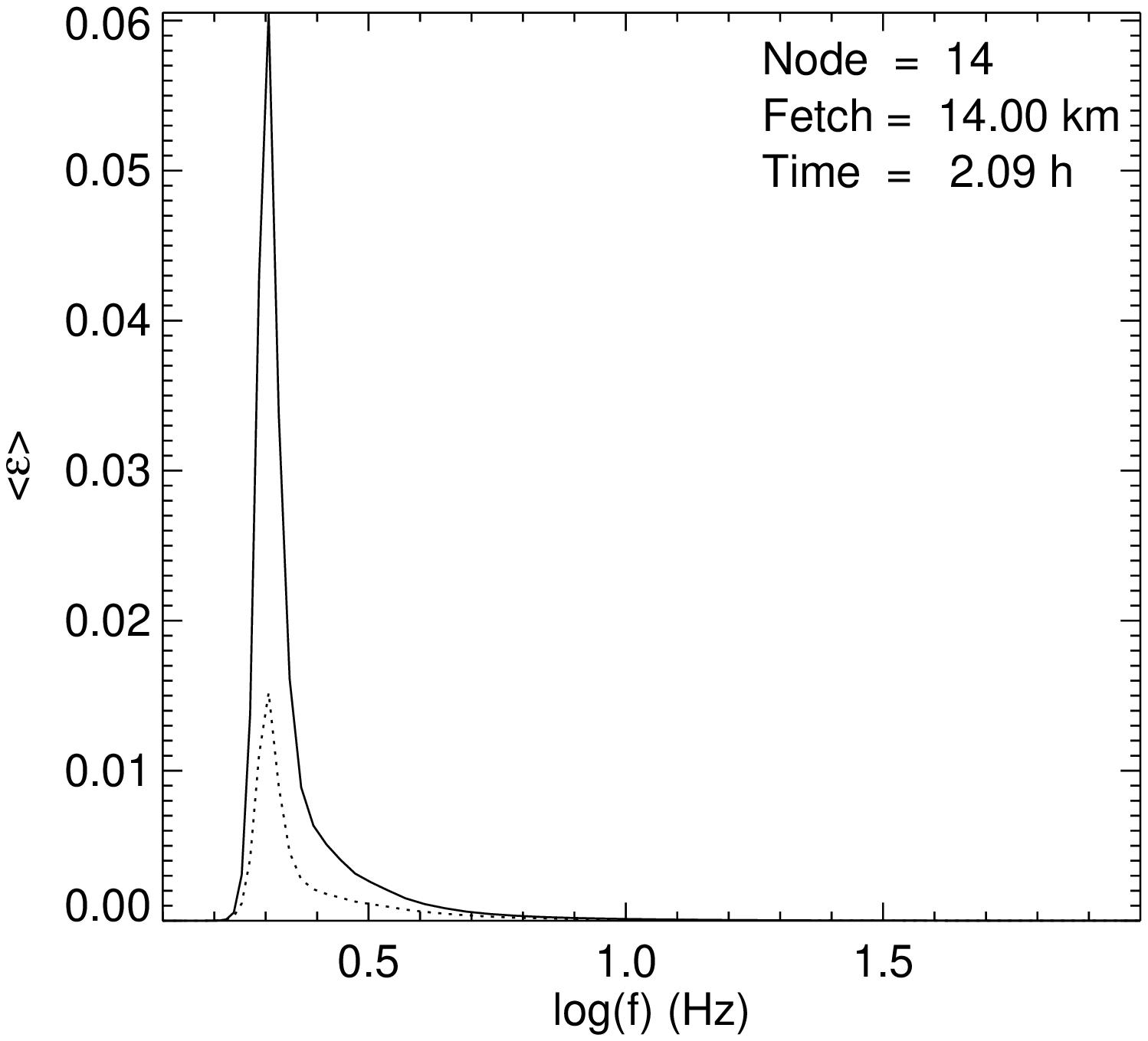} \\
[5 mm]
				\includegraphics[width=0.4\linewidth]{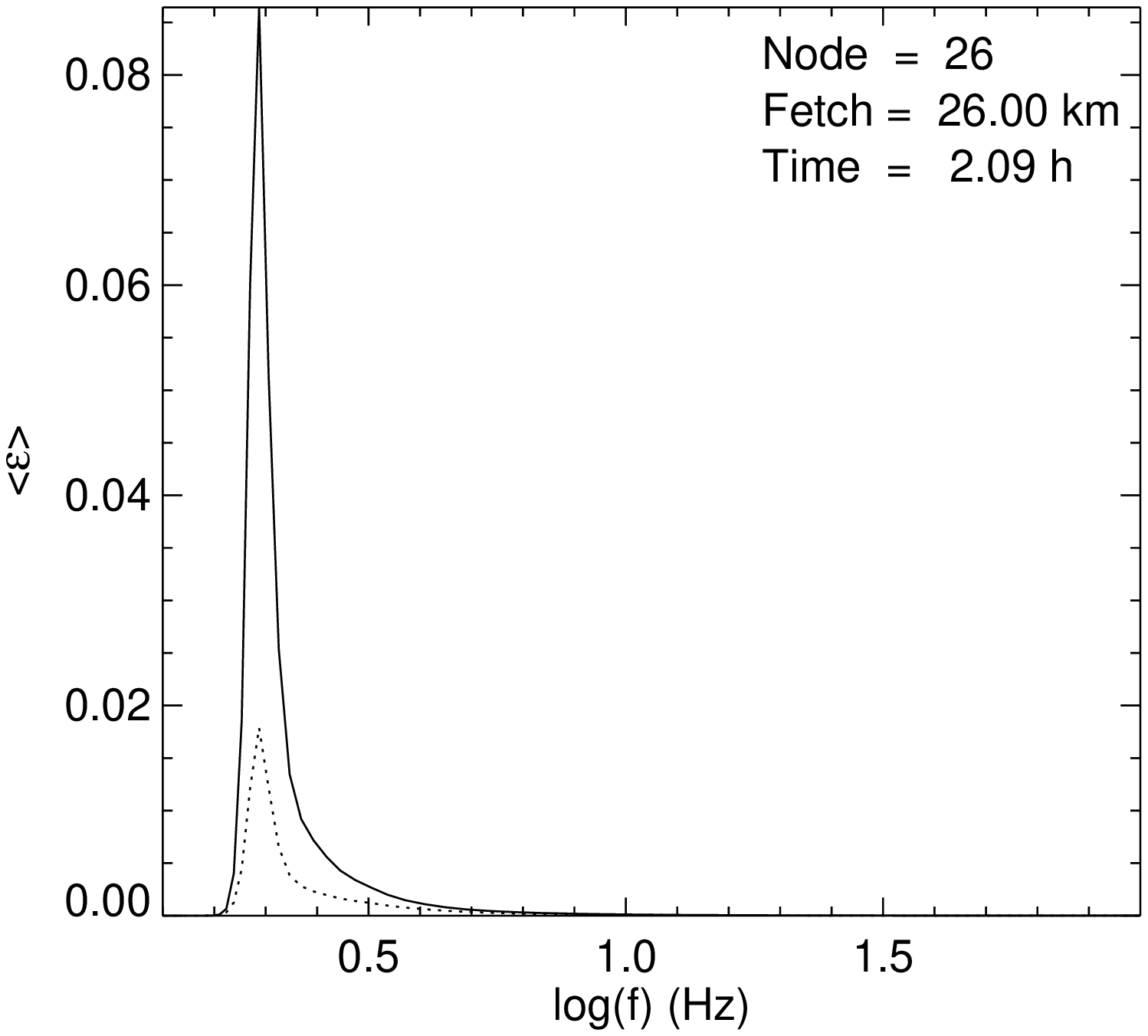} & \includegraphics[width=0.4\linewidth]{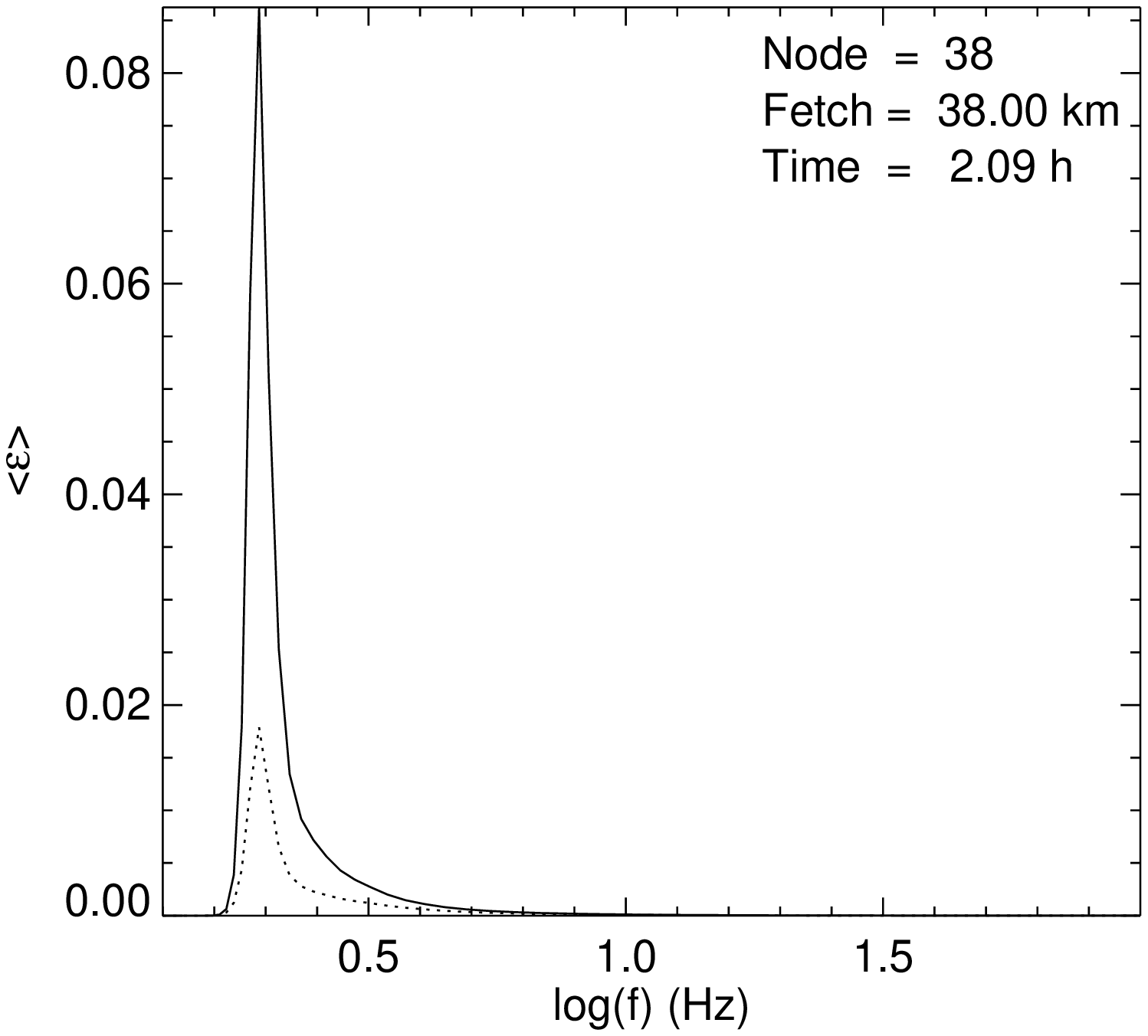} \\
			\end{tabular}
			\captionof{figure}{Angle averaged $<\varepsilon>_{\theta} = \frac{1}{2\pi}\int\limits_{0}^{2 \pi} \varepsilon (f,\theta,x,t) d\theta$ (dotted line) and along the wind $\varepsilon(f,\theta_{wind},x,t)$ (solid line) spectra for time $t=2$ h at fetch coordinates $x=$ 2, 14, 26 and 38 km.} \label{AngAvNormSpectra2}
		\end{center}
\end{table}

\begin{table}
	\centering
		\begin{center} 
			\begin{tabular}{c c}
				\includegraphics[width=0.4\linewidth]{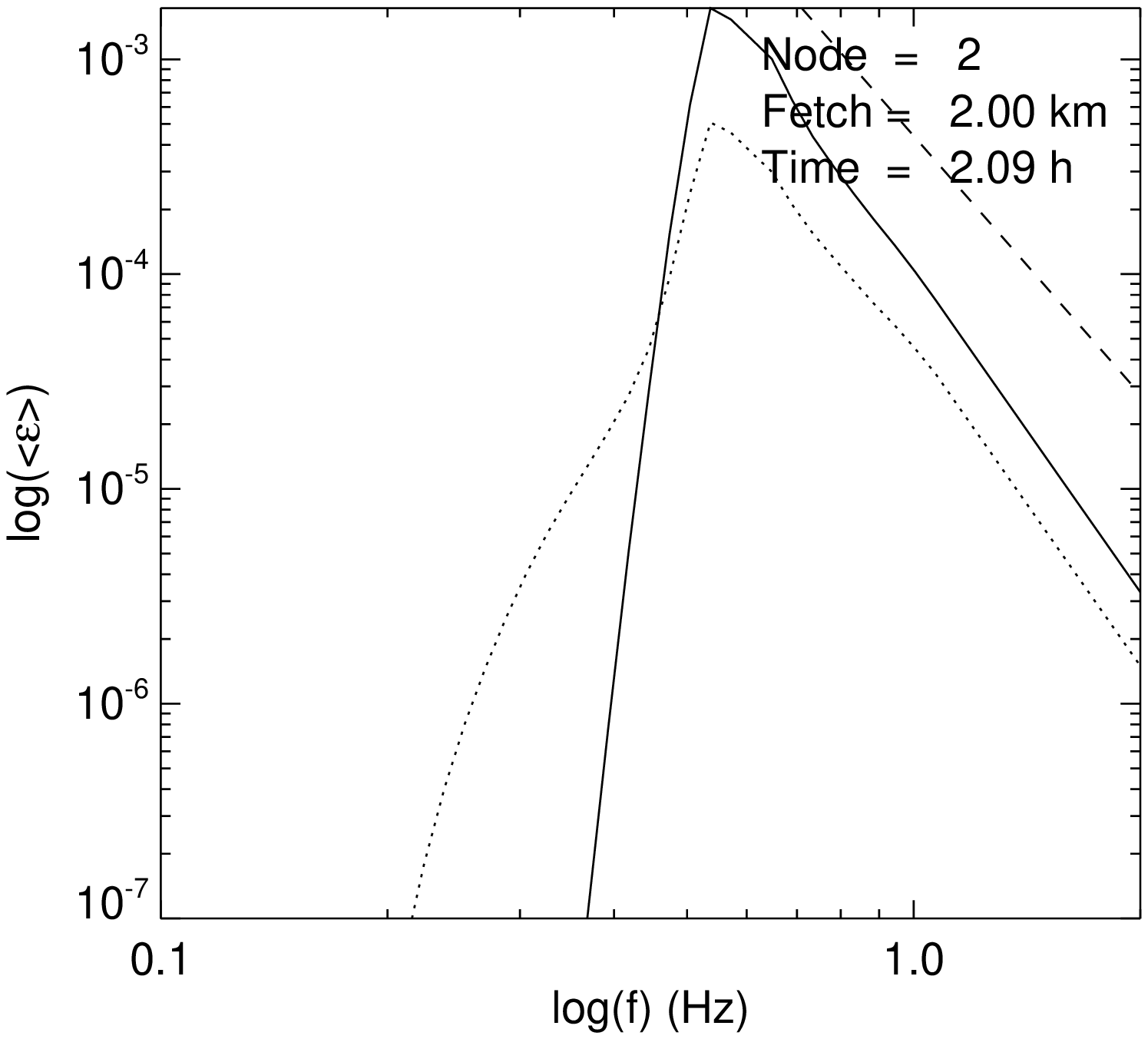} & \includegraphics[width=0.4\linewidth]{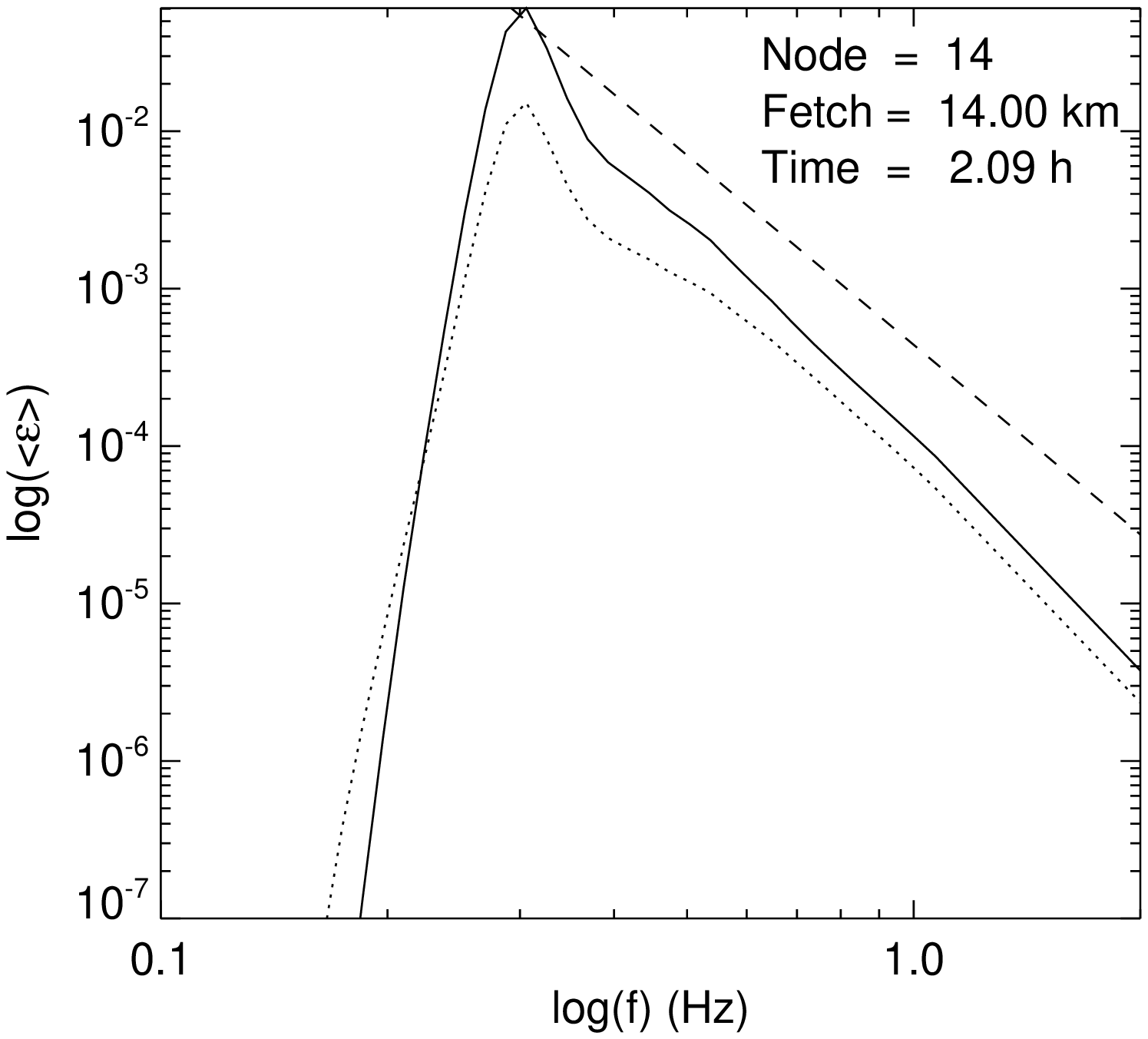} \\
[5 mm]
				\includegraphics[width=0.4\linewidth]{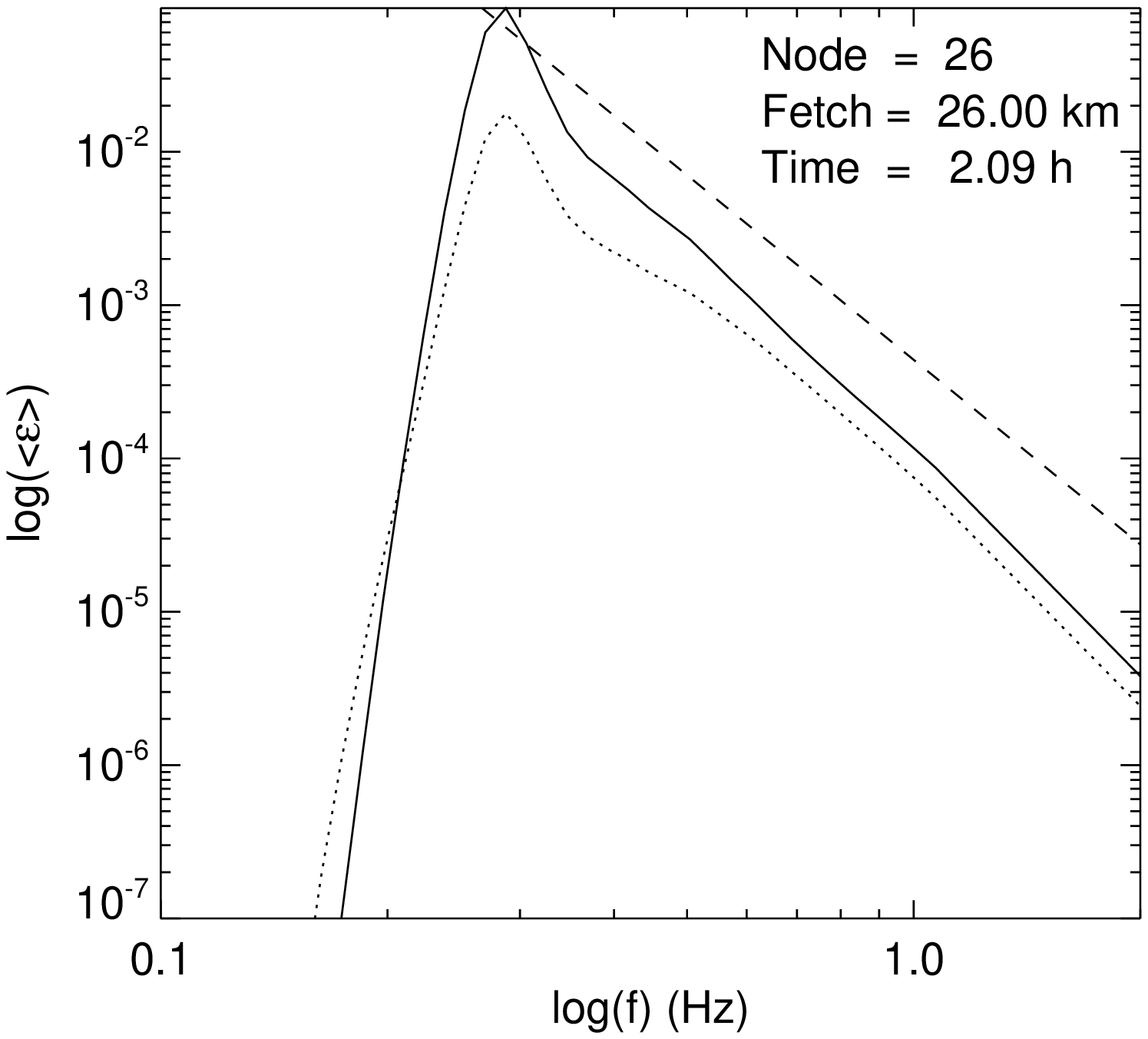} & \includegraphics[width=0.4\linewidth]{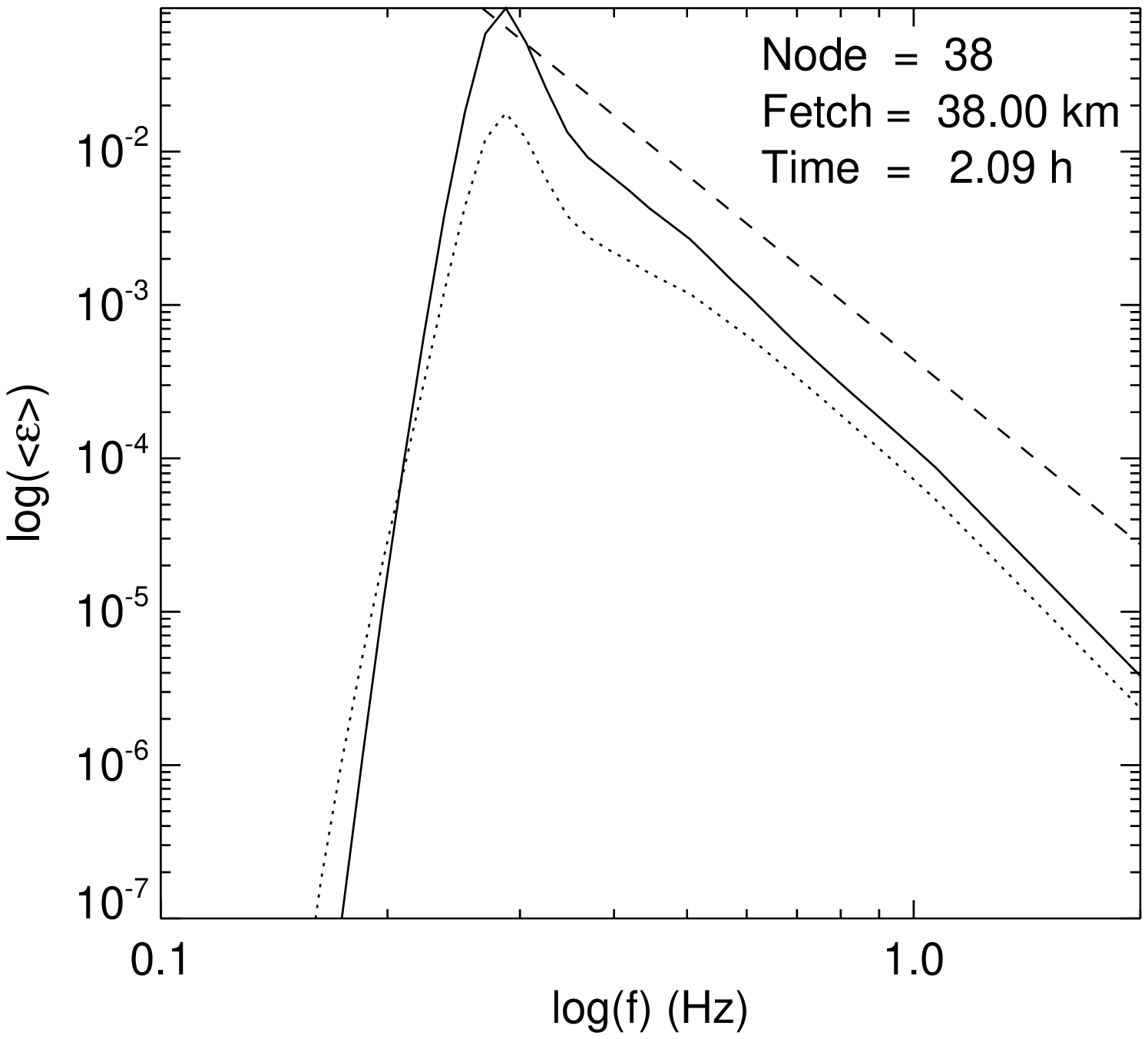} \\
			\end{tabular}
			\captionof{figure}{Decimal logarithm of the angle averaged $<\varepsilon>_{\theta} = \log ( \int\limits_{0}^{2 \pi} \varepsilon (f,\theta,x,t) d\theta$ ) (dotted line) and along the wind $\log ( \varepsilon(f,\theta_{wind},x,t) )$ (solid line) spectra for time $t=2$ h at fetch coordinates: $x=$ 2, 14, 26 and 38 km. Dashed line - KZ spectrum $\sim\omega^{-4}$.} \label{AngAvSpectra2}
		\end{center}
\end{table}


\begin{table}
\centering
	\begin{center} 
		\begin{tabular}{c c}
			\includegraphics[width=0.4\linewidth]{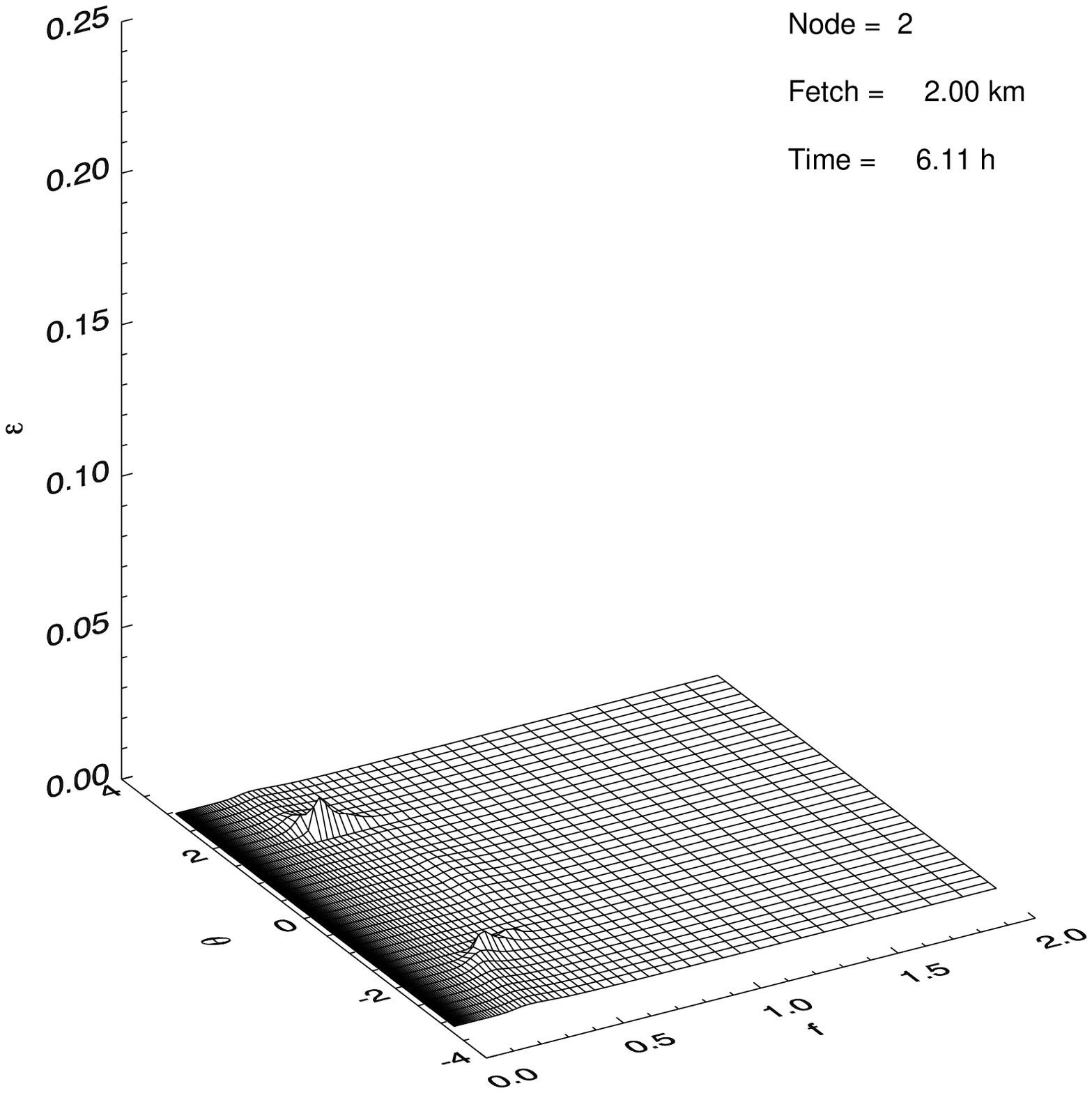} & \includegraphics[width=0.4\linewidth]{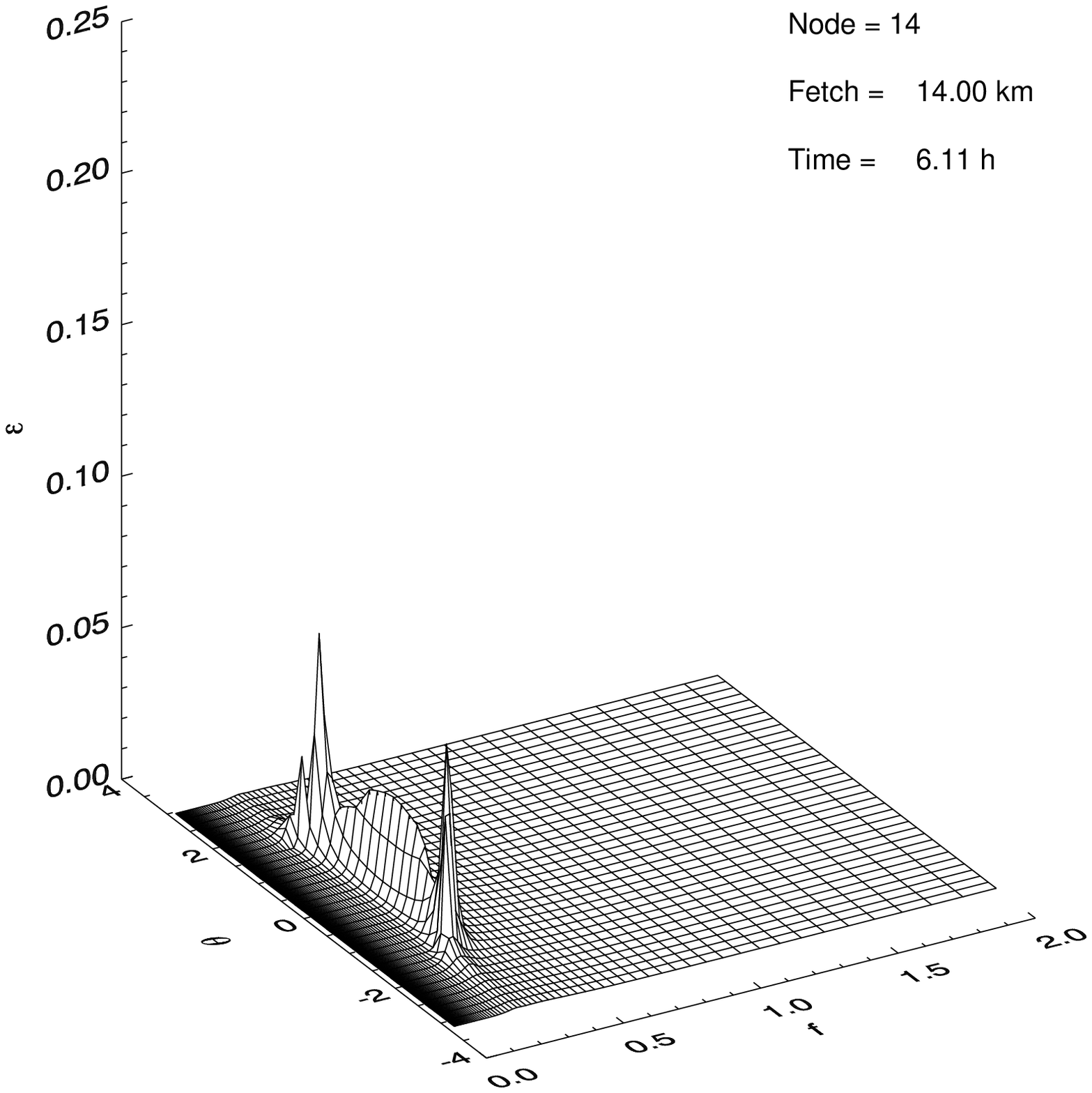}   \\ [5 mm]
			\includegraphics[width=0.4\linewidth]{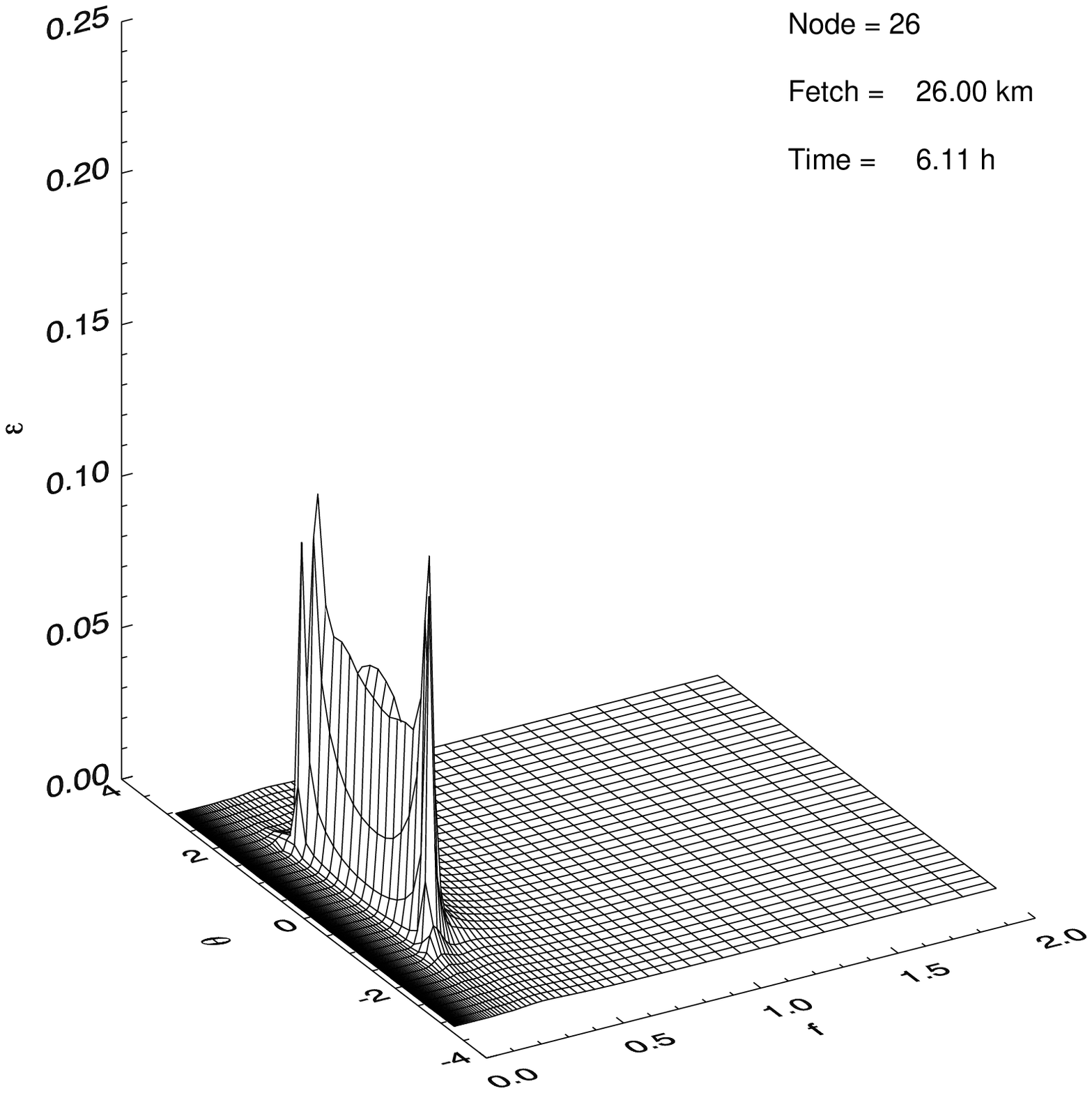} & \includegraphics[width=0.4\linewidth]{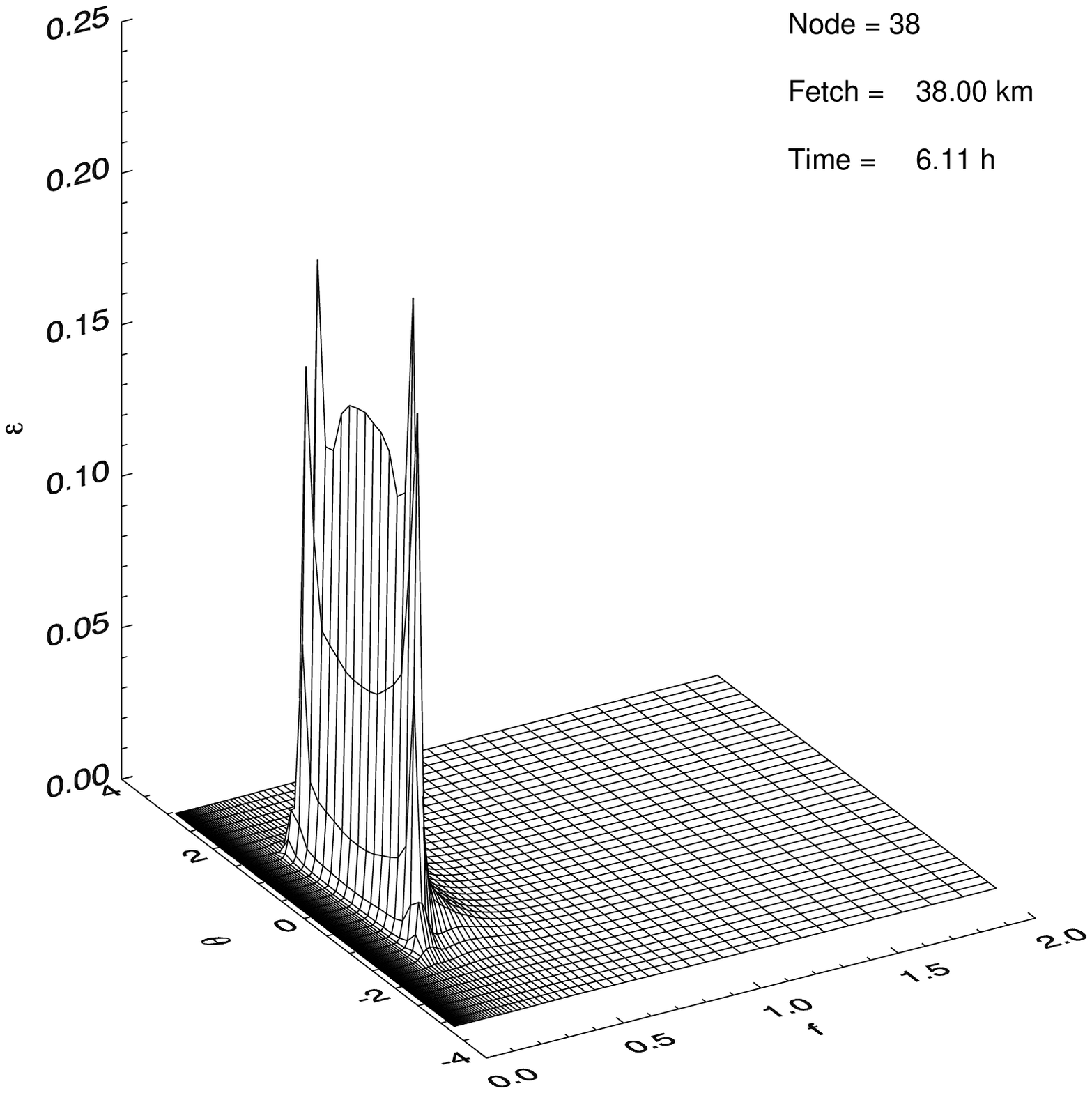} \\
		\end{tabular}
	\captionof{figure}{Energy spectrum $\varepsilon(f,\theta,x,t)$ as the function of the frequency $f$ and angle $\theta$ at the fetch coordinates $x=$ 2, 14, 26 and 38 km for time $t=6$ h.} \label{Spectrum3D6h}
	\end{center}
\end{table}

\begin{table}
\centering
	\begin{center} 
		\begin{tabular}{c c}
			\includegraphics[width=0.4\linewidth]{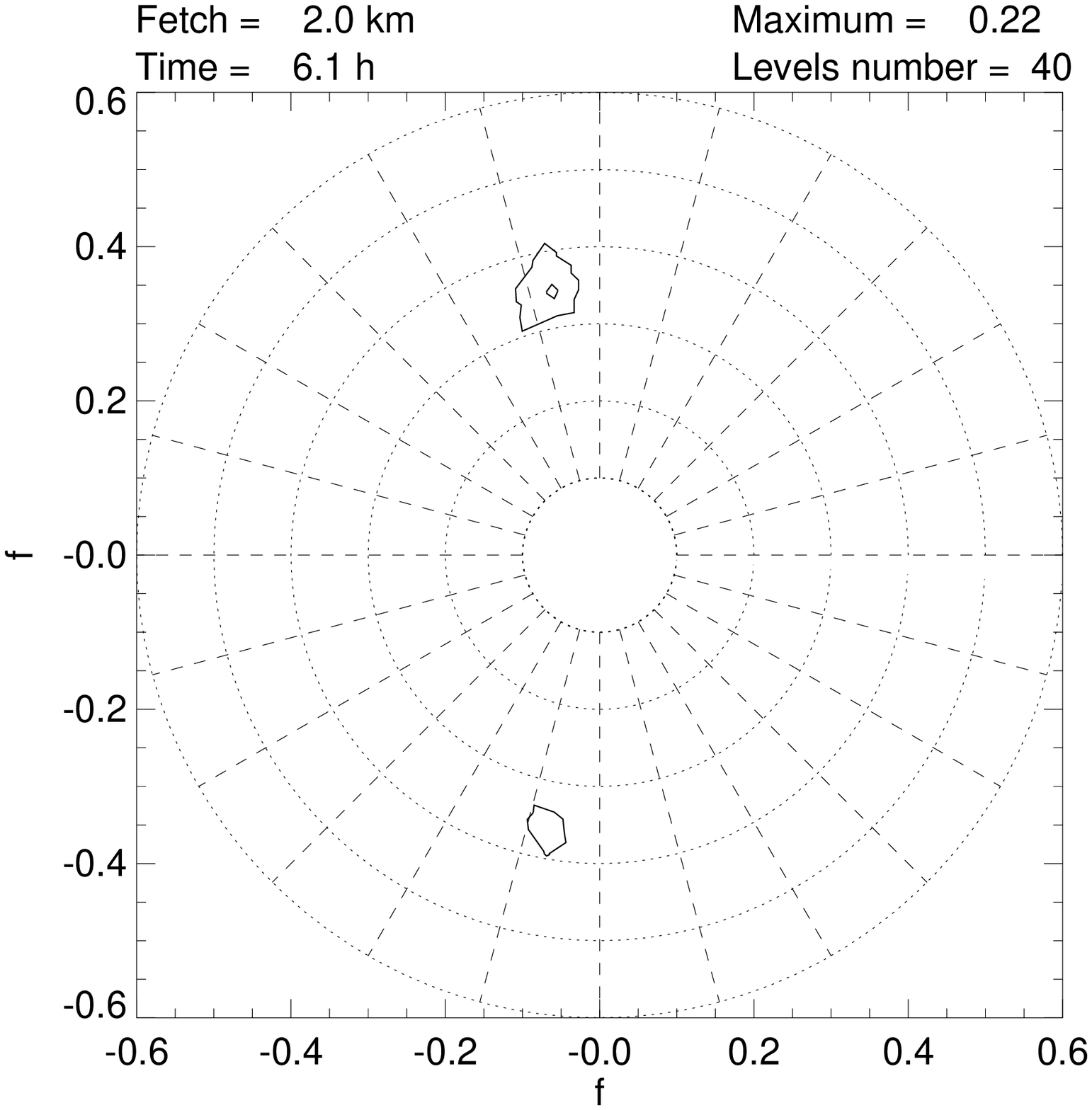} & \includegraphics[width=0.4\linewidth]{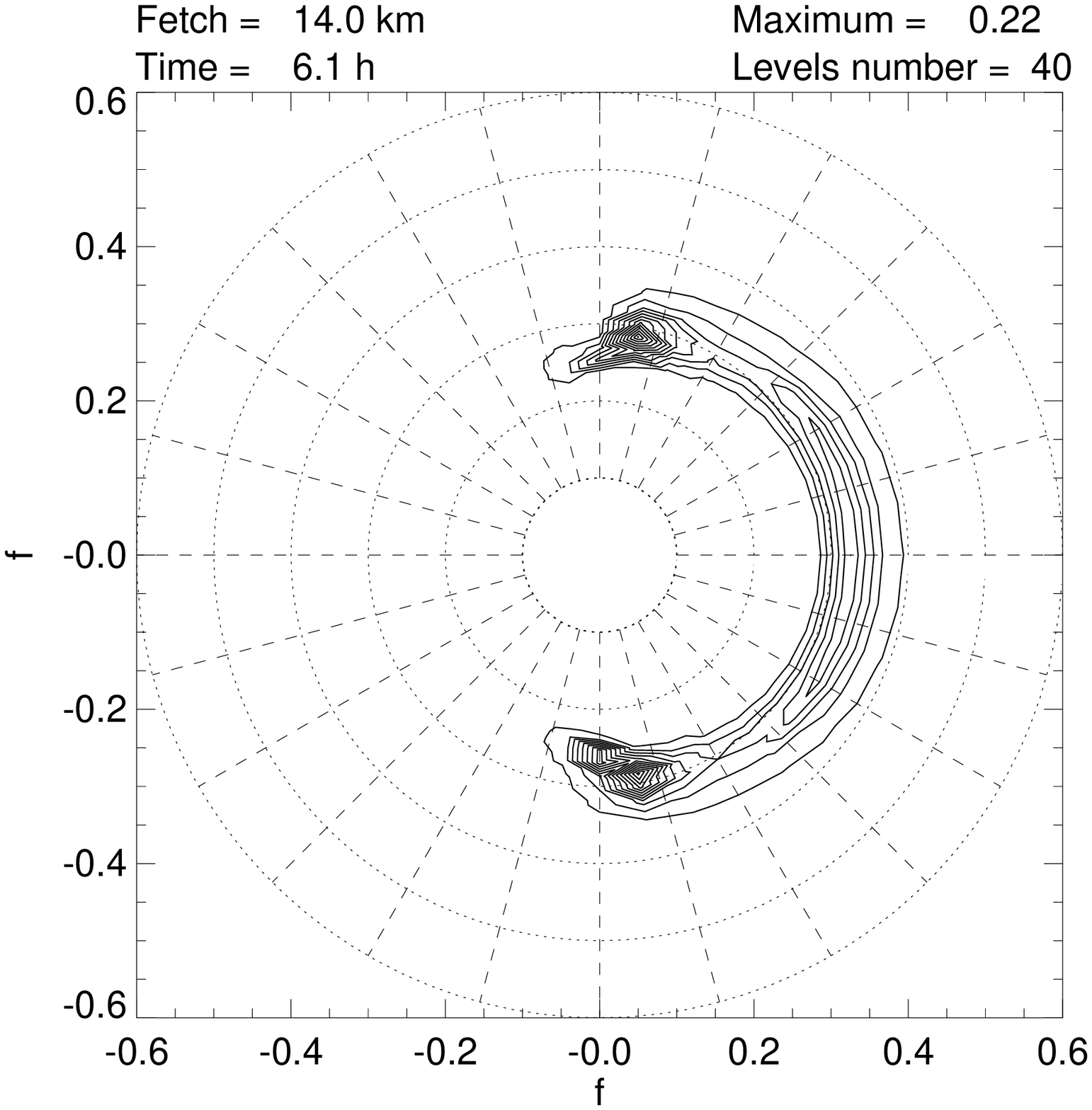}   \\ 
[5 mm]
			\includegraphics[width=0.4\linewidth]{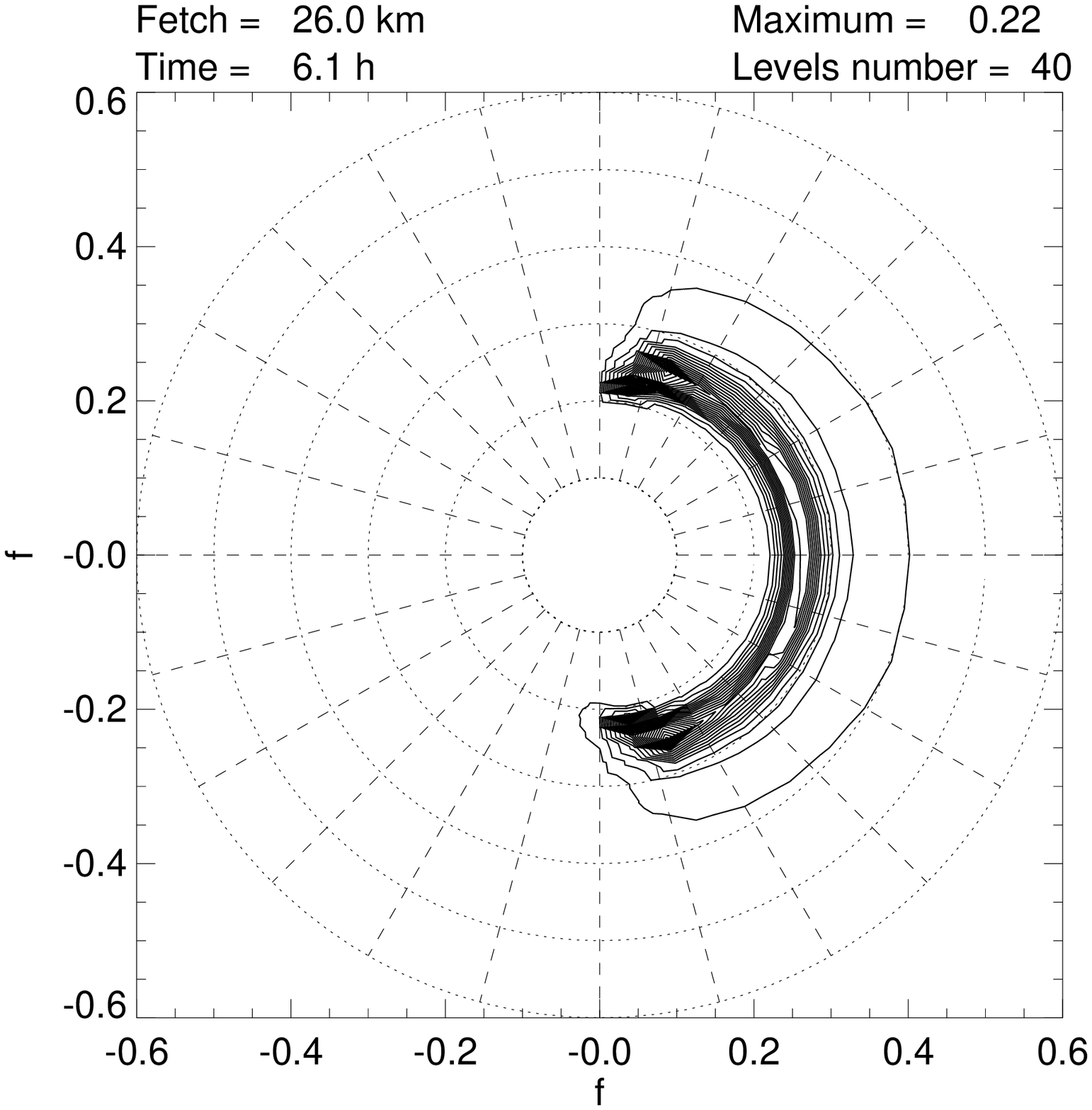} & \includegraphics[width=0.4\linewidth]{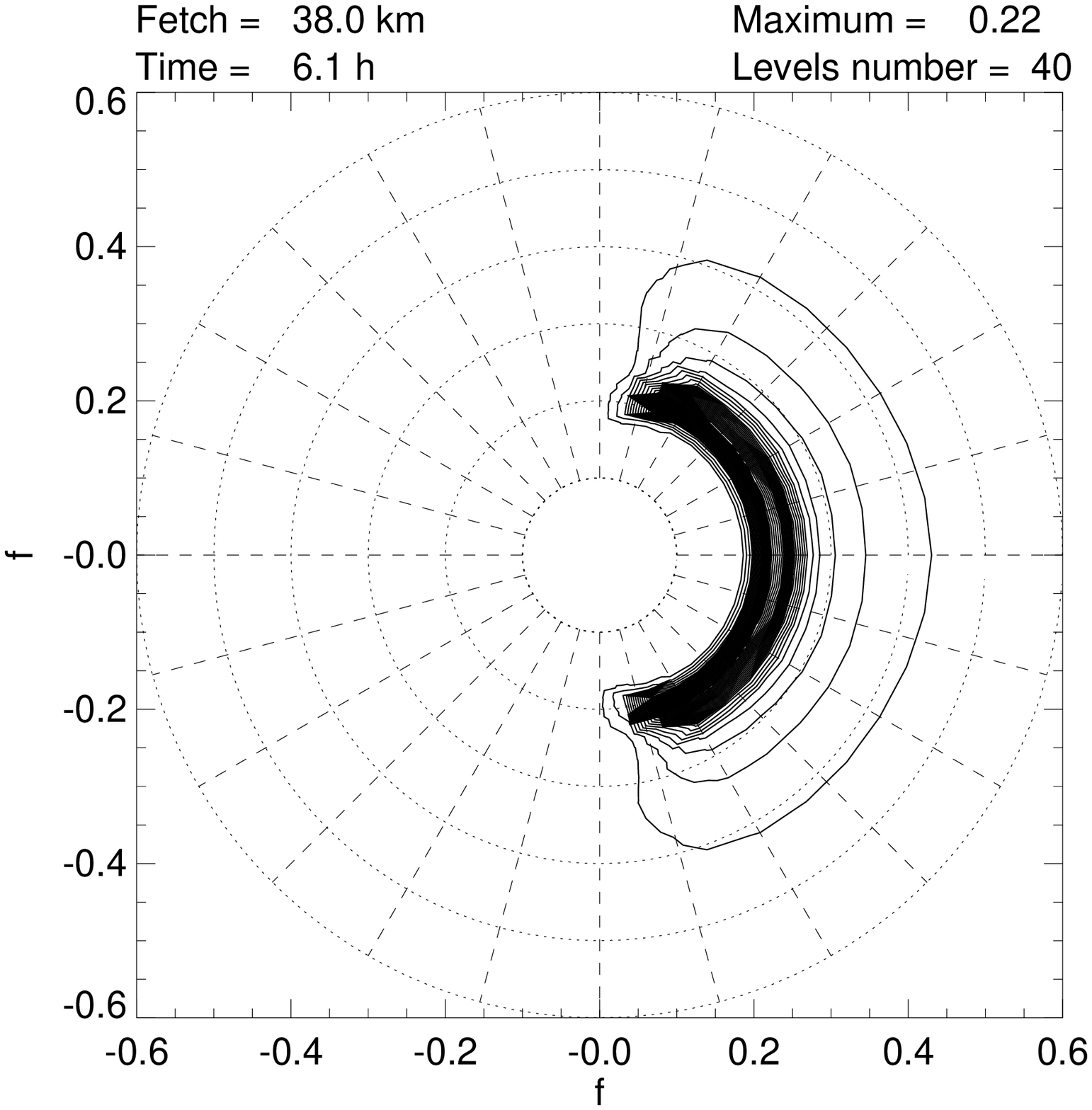} \\
		\end{tabular}
	\captionof{figure}{Energy spectrum $\varepsilon(f,\theta,x,t)$ as the function of the frequency $f$ and angle $\theta$ at the fetch coordinates $x=$ 2, 14, 26 and 38 km for time $t=6$ h.} \label{Polar6h}
	\end{center}
\end{table}

\begin{table}
	\centering
		\begin{center} 
			\begin{tabular}{c c}
				\includegraphics[width=0.4\linewidth]{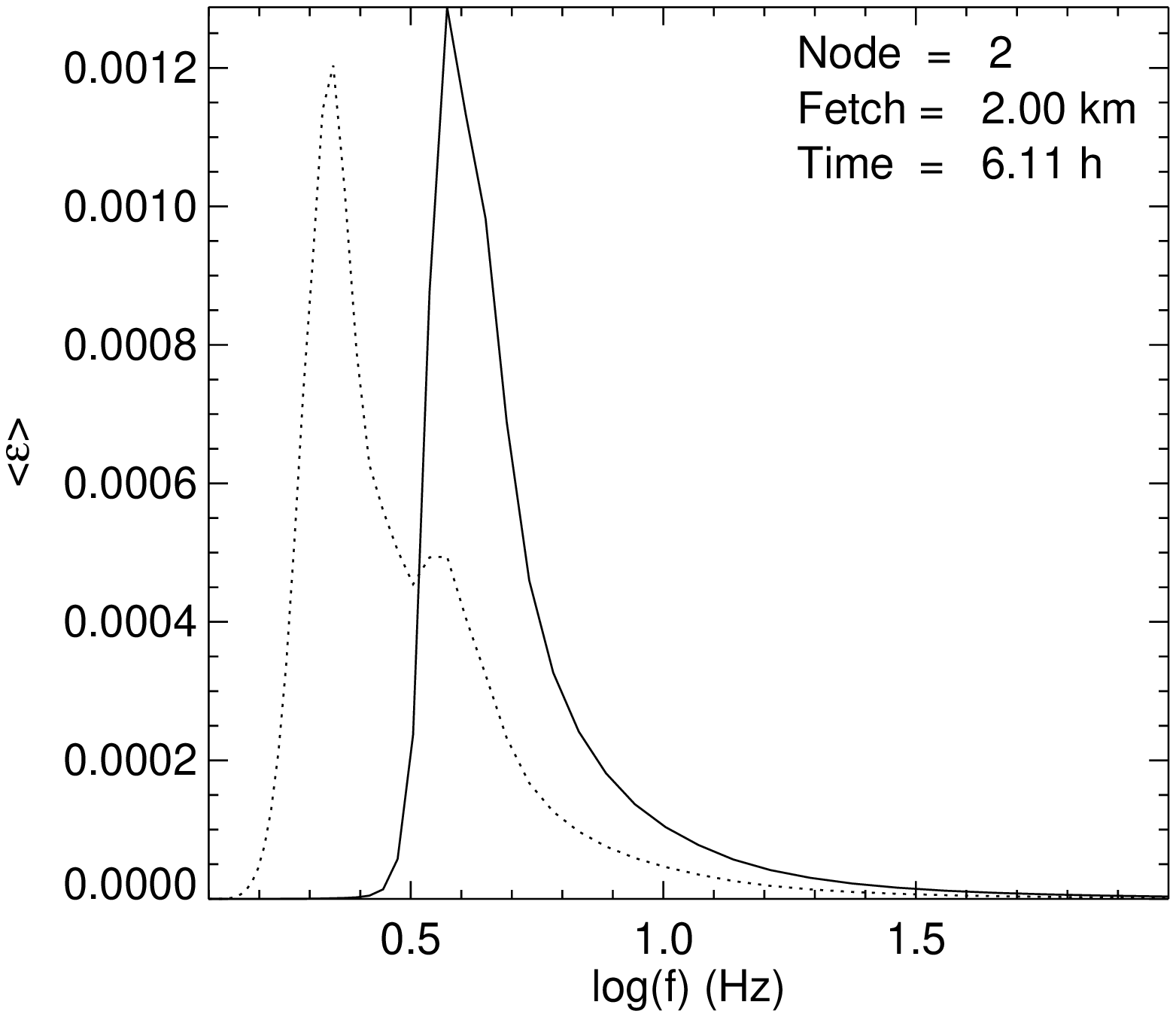} & \includegraphics[width=0.4\linewidth]{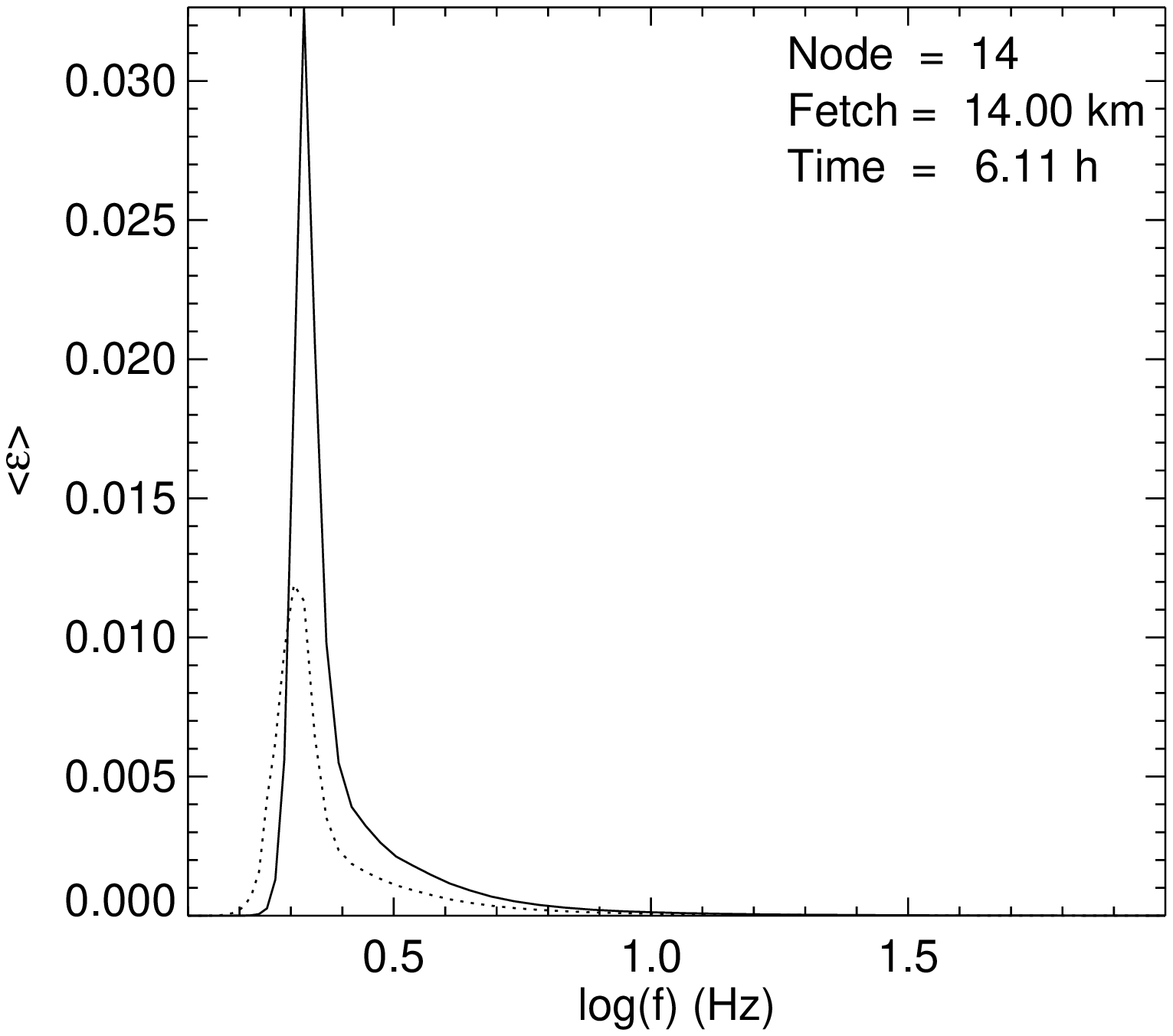} \\
[5 mm]
				\includegraphics[width=0.4\linewidth]{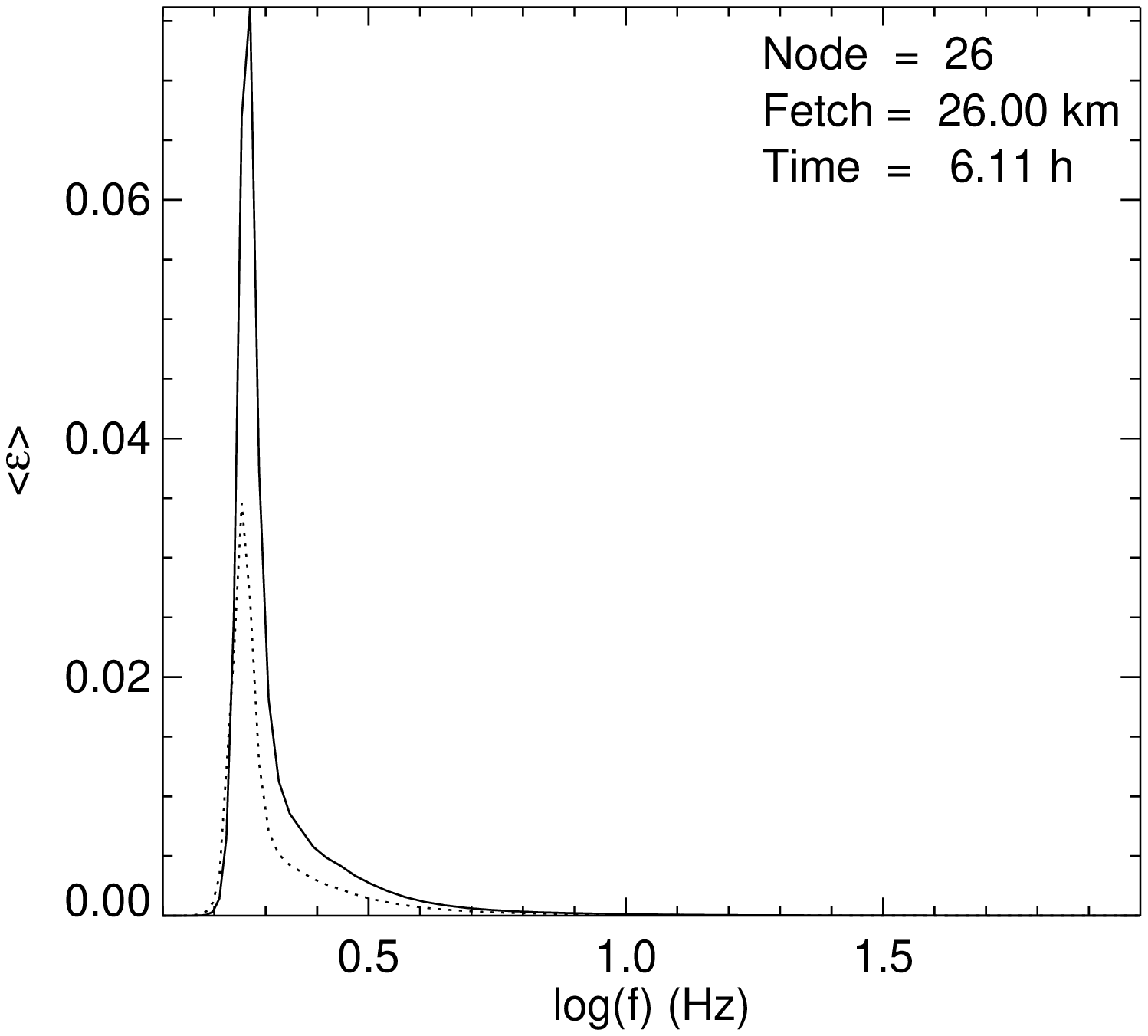} & \includegraphics[width=0.4\linewidth]{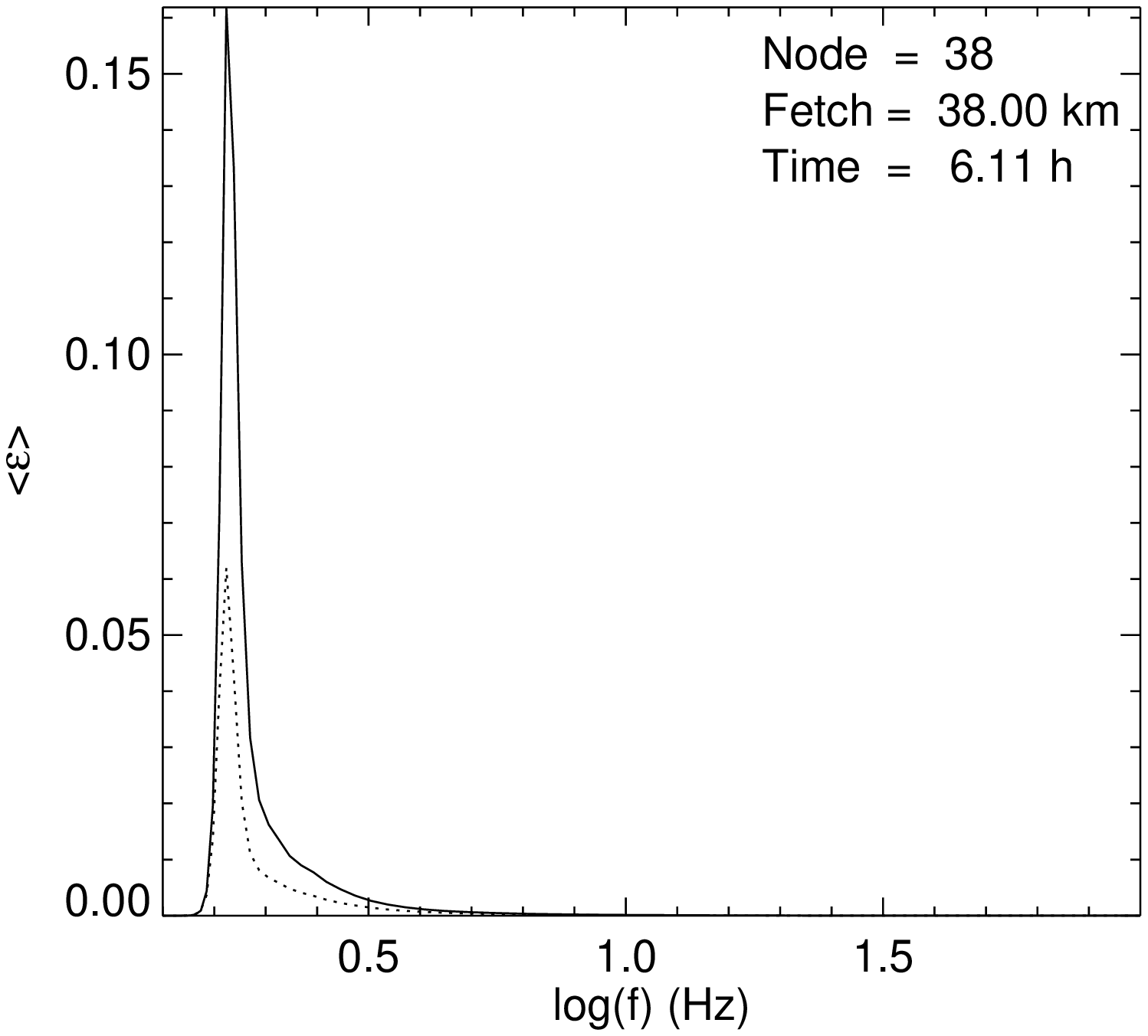} \\
			\end{tabular}
			\captionof{figure}{
Angle averaged $<\varepsilon>_{\theta} = \frac{1}{2\pi}\int\limits_{0}^{2 \pi} \varepsilon (f,\theta,x,t) d\theta$ (dotted line) and along the wind $\varepsilon(f,\theta_{wind},x,t)$ (solid line) spectra for time $t=6$ h at fetch coordinates $x=$ 2, 14, 26 and 38 km.} \label{AngAvNormSpectra6}
		\end{center}
\end{table}

\begin{table}
	\centering
		\begin{center} 
			\begin{tabular}{c c}
				\includegraphics[width=0.4\linewidth]{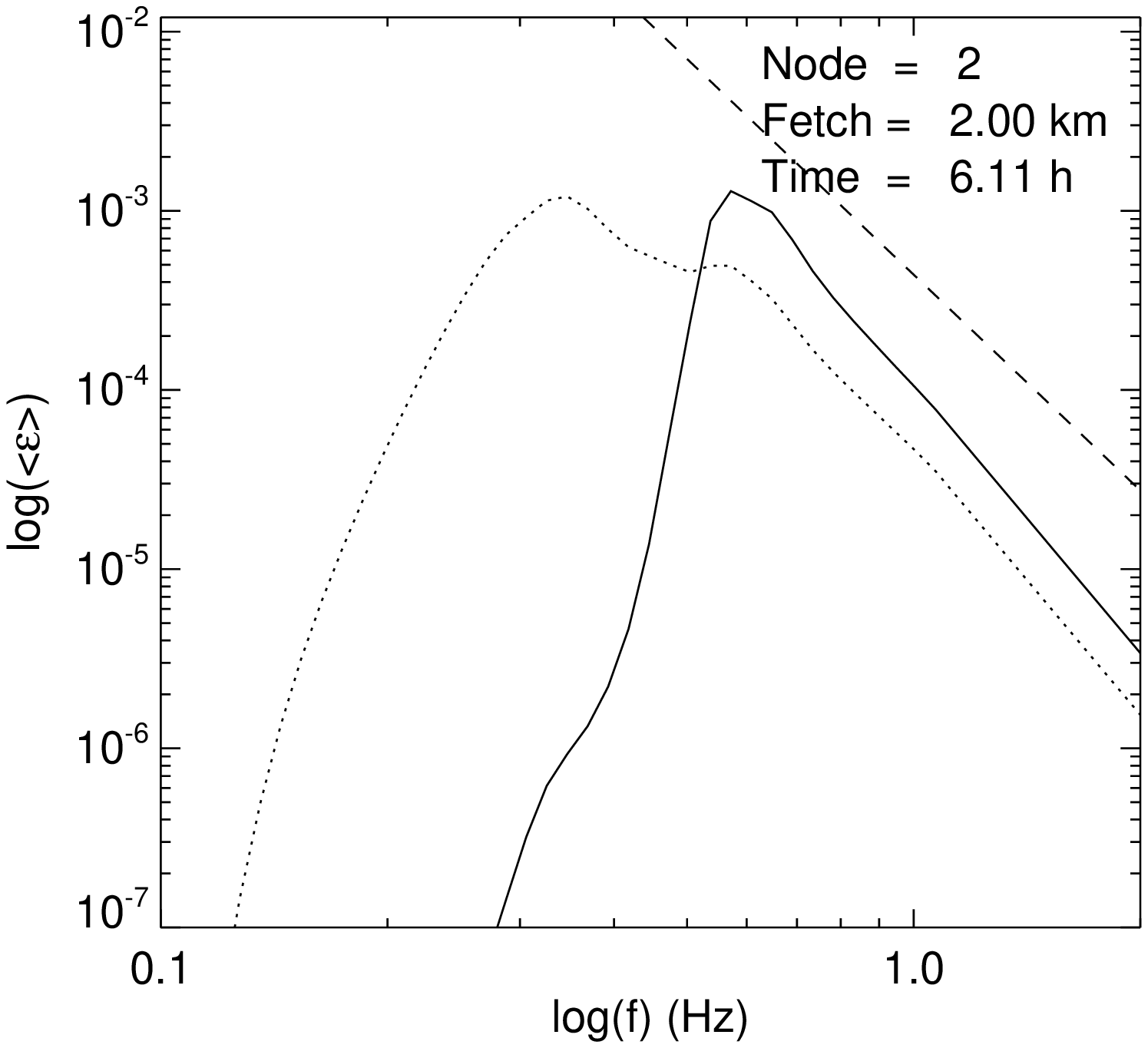} & \includegraphics[width=0.4\linewidth]{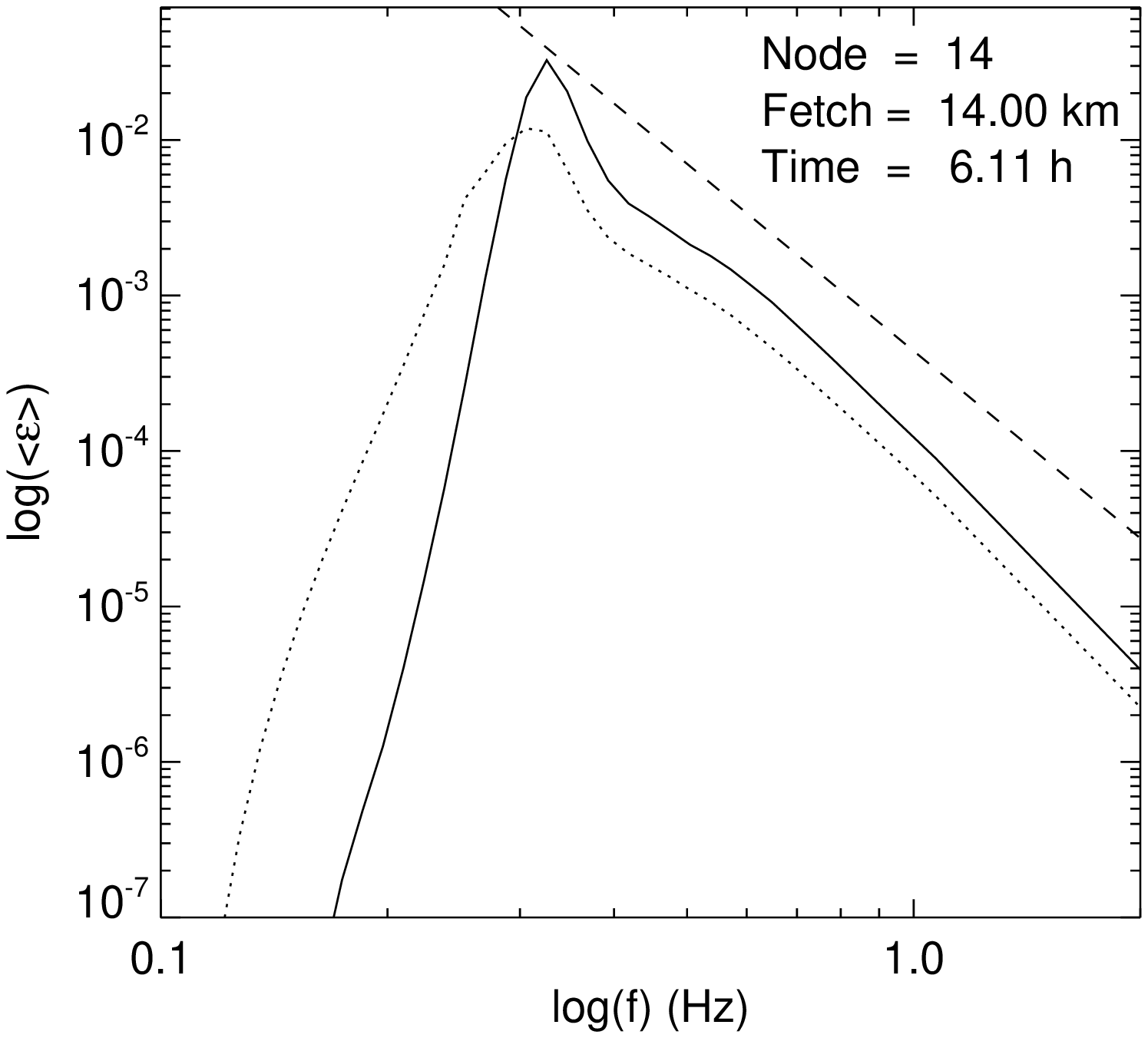} \\
[5 mm]
				\includegraphics[width=0.4\linewidth]{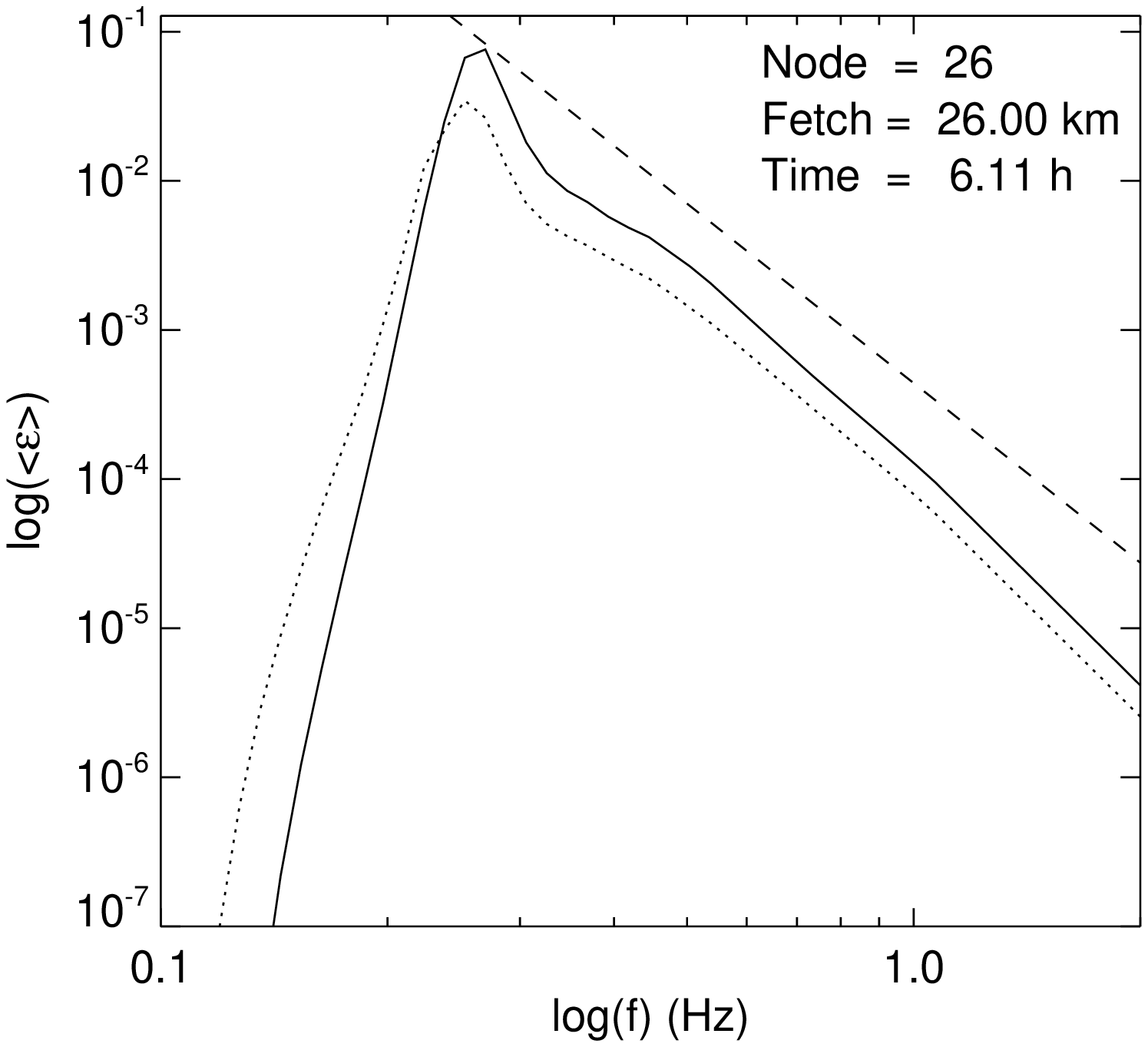} & \includegraphics[width=0.4\linewidth]{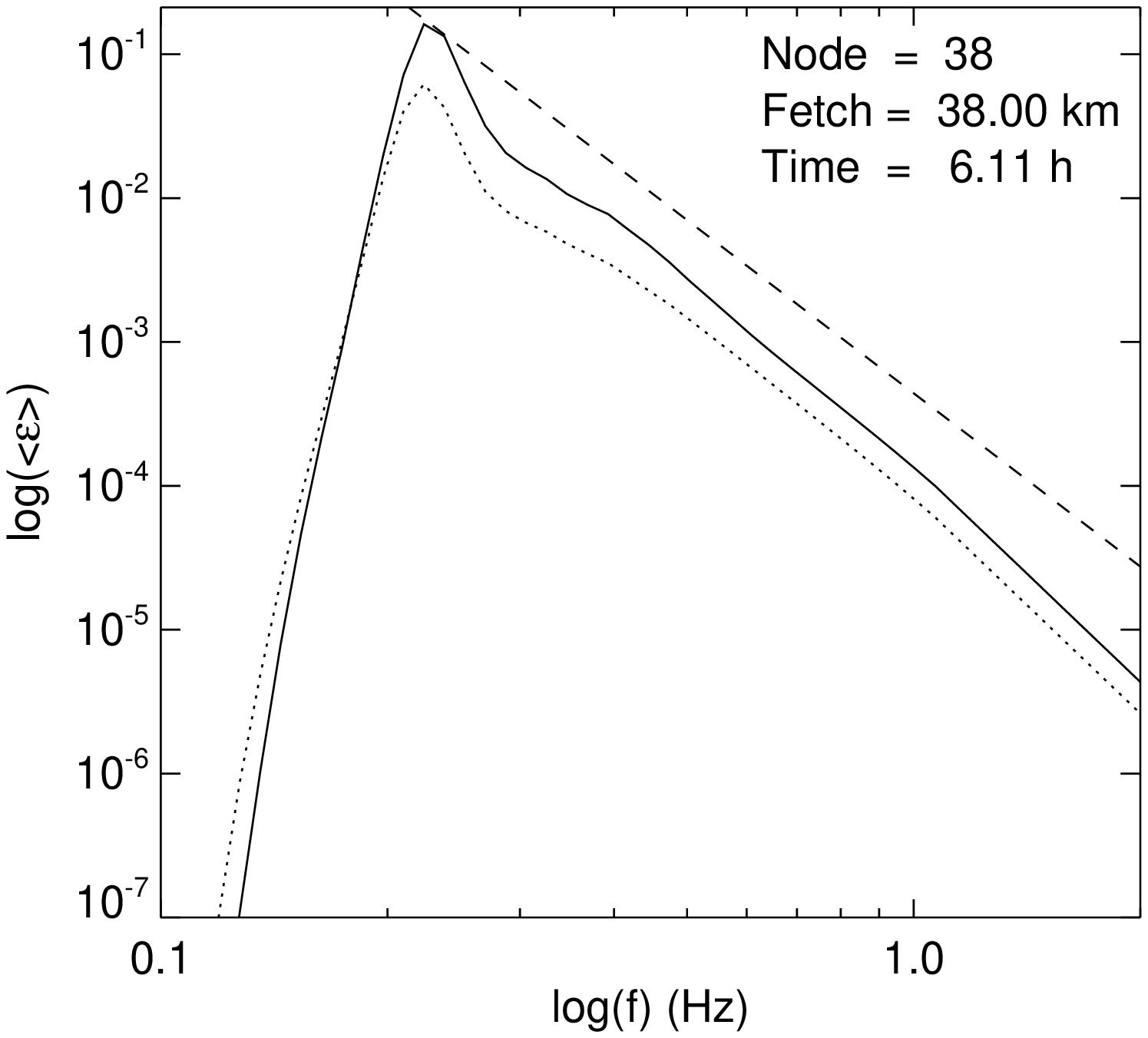} \\
			\end{tabular}
			\captionof{figure}{Decimal logarithm of angle averaged $<\varepsilon>_{\theta} = \log ( \int\limits_{0}^{2 \pi} \varepsilon (f,\theta,x,t) d\theta$ ) (dotted line) and along the wind $\log (\varepsilon(f,\theta_{wind},x,t) )$ (solid line) spectra for time $t=6$ h at fetch coordinates: $x=$ 2, 14, 26 and 38 km. Dashed line - KZ spectrum $\sim\omega^{-4}$.} \label{AngAvSpectra6}
		\end{center}
\end{table}

\begin{table}
\centering
	\begin{center} 
		\begin{tabular}{c c}
			\includegraphics[width=0.4\linewidth]{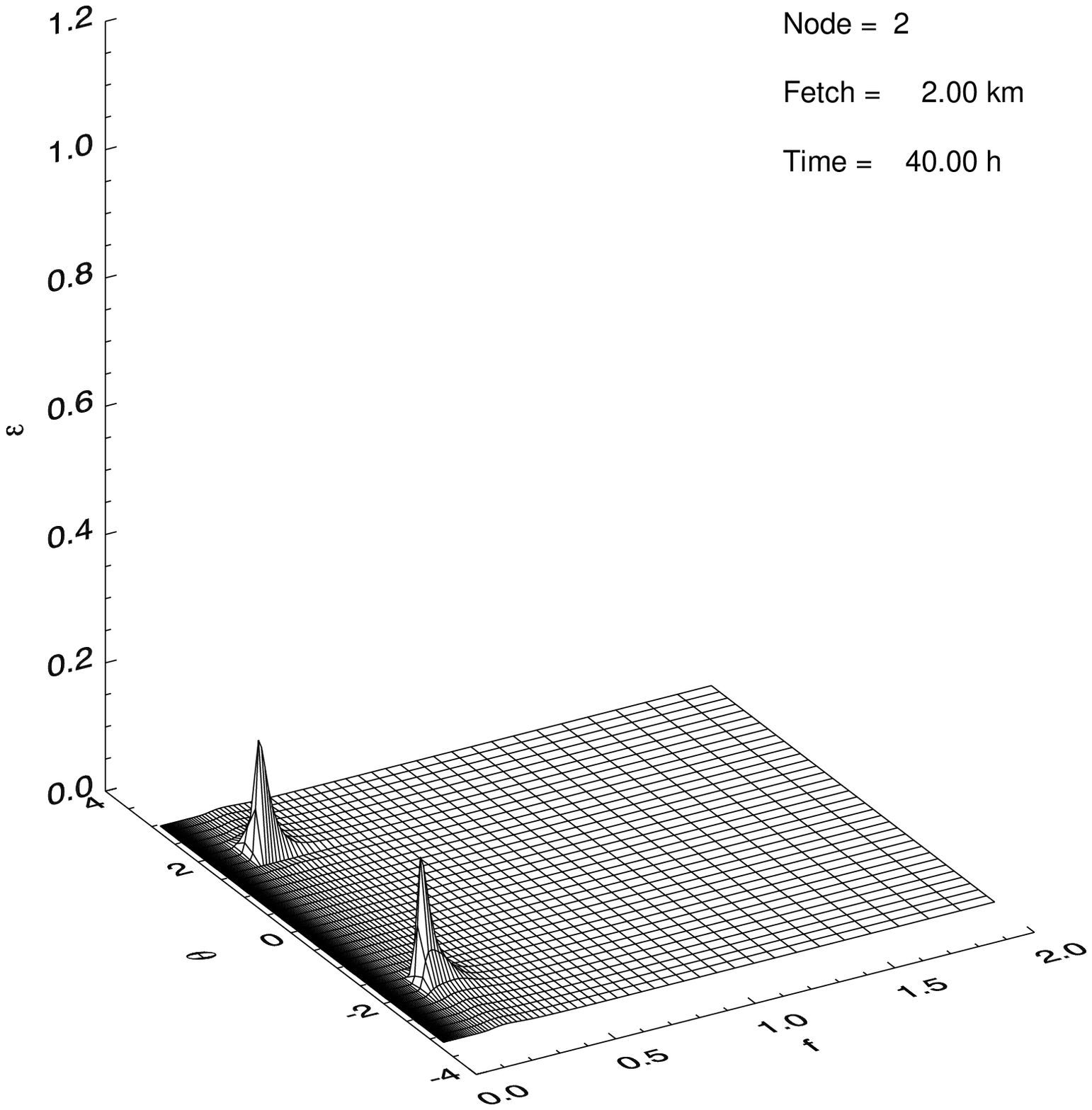} & \includegraphics[width=0.4\linewidth]{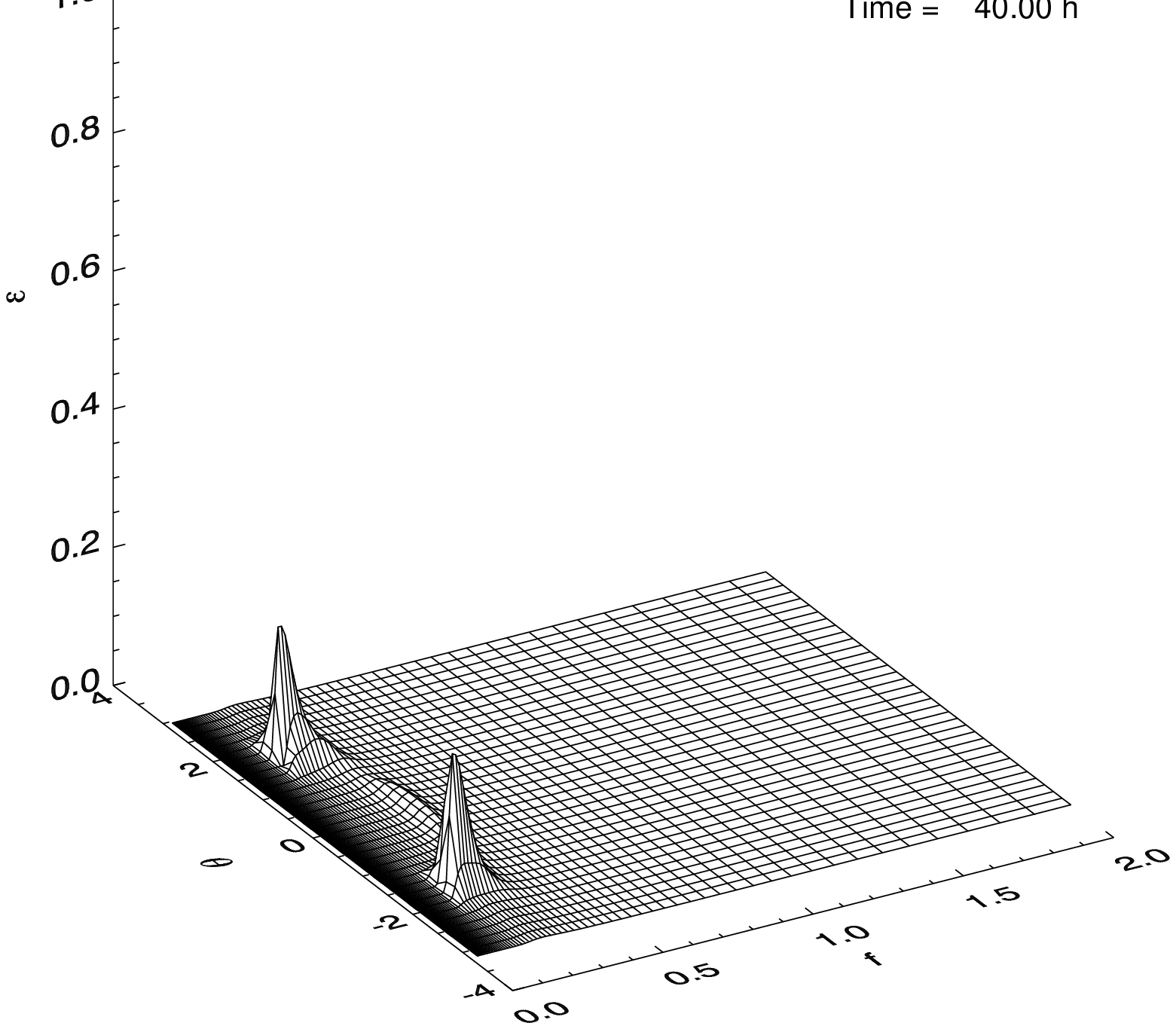}   \\ 
[5 mm]
			\includegraphics[width=0.4\linewidth]{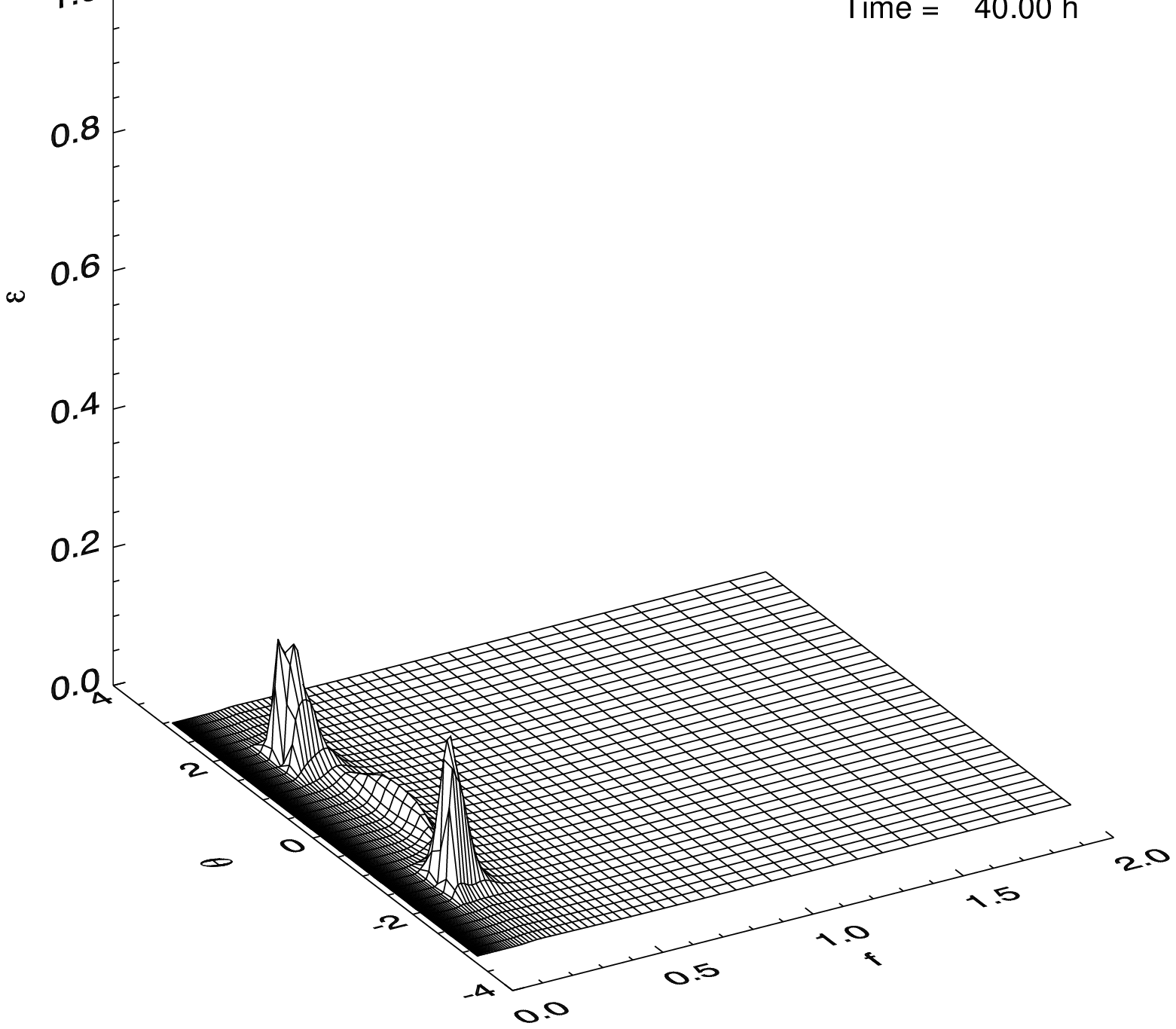} & \includegraphics[width=0.4\linewidth]{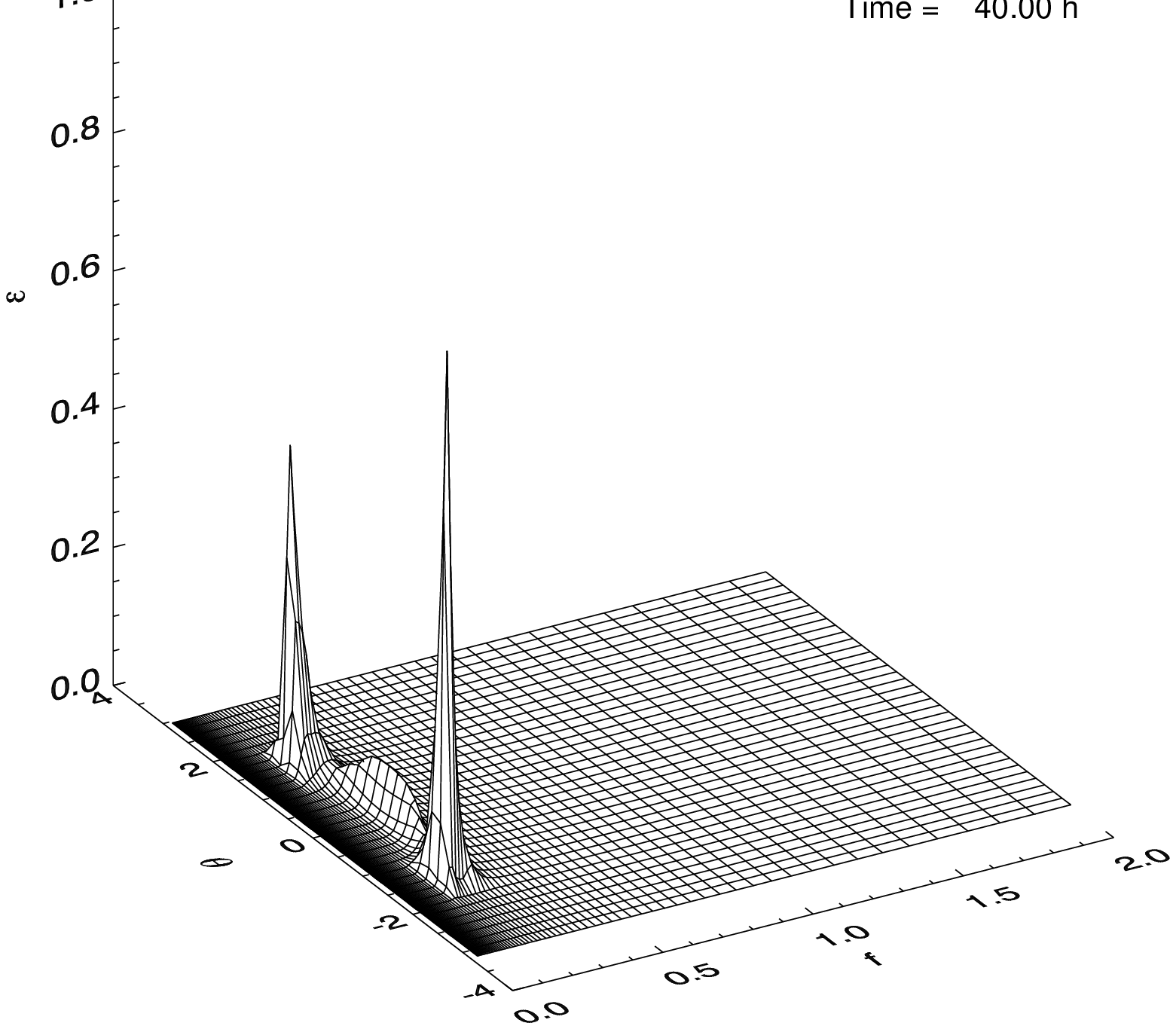} \\
		\end{tabular}
	\captionof{figure}{Energy spectrum $\varepsilon(f,\theta,x,t)$ as the function of the frequency $f$ and angle $\theta$ at the fetch coordinates $x=$ 2, 14, 26 and 38 km for time $t=40$ h.} \label{Spectrum3D40h}
	\end{center}
\end{table} 

\begin{table}
\centering
	\begin{center} 
		\begin{tabular}{c c}
			\includegraphics[width=0.4\linewidth]{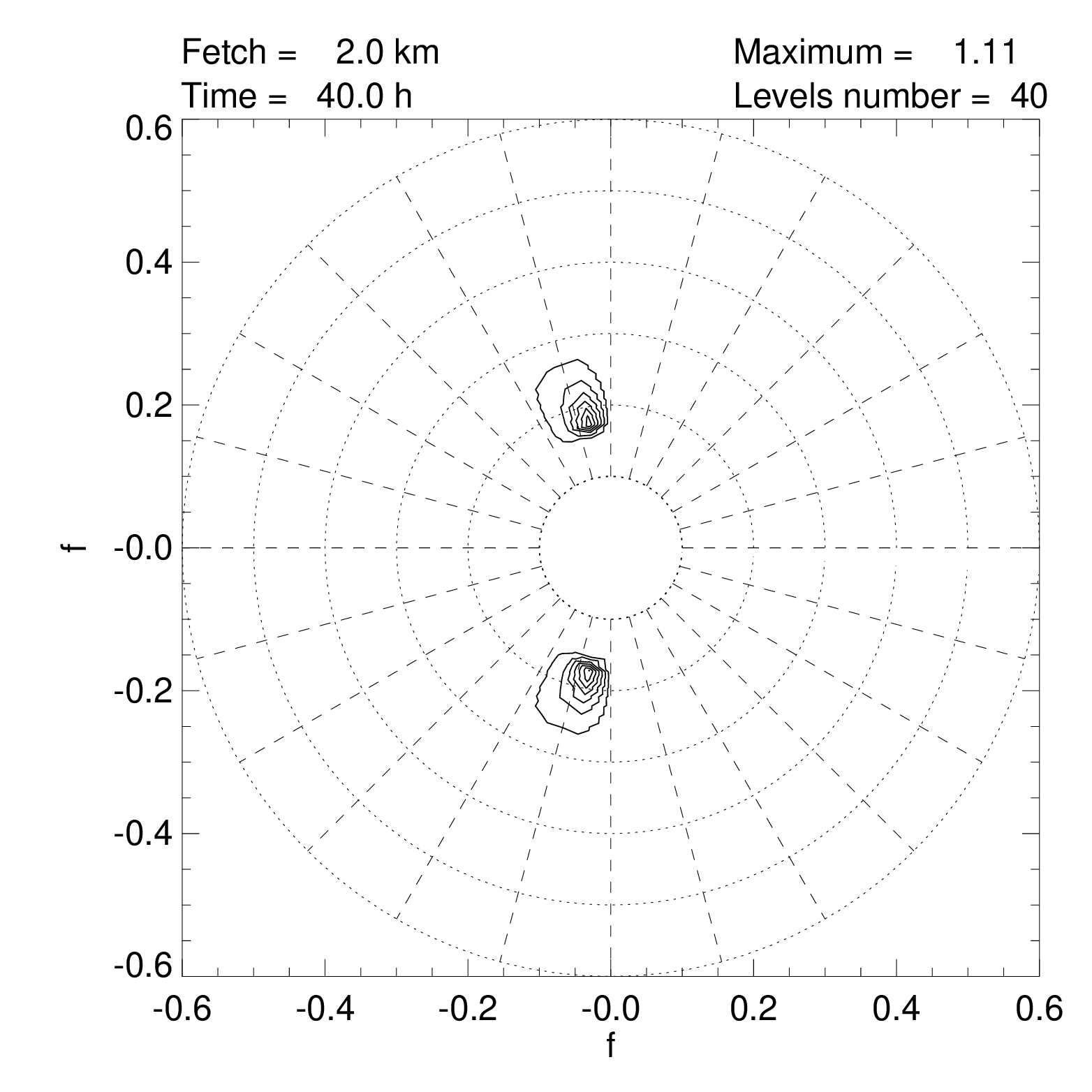} & \includegraphics[width=0.4\linewidth]{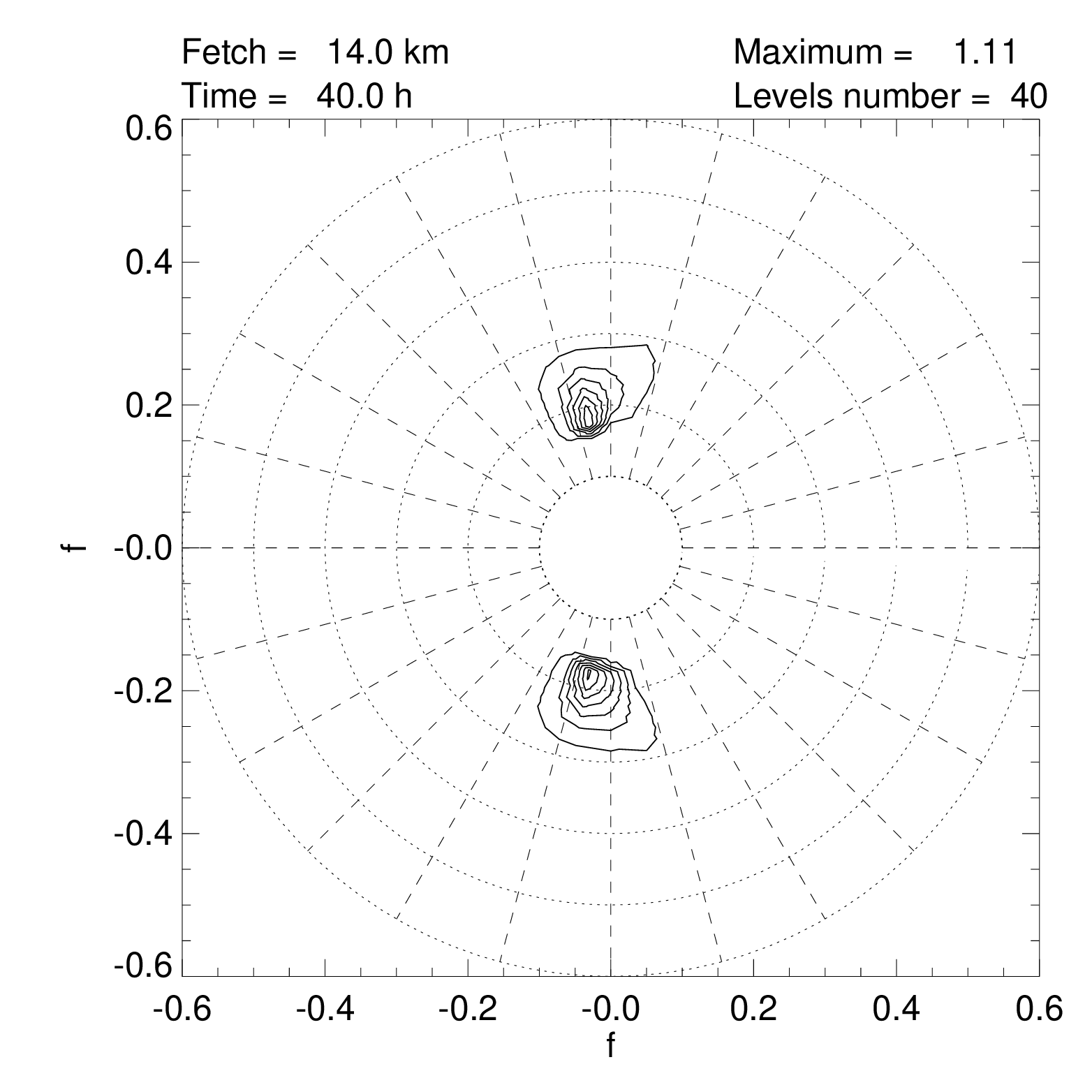}   \\
[5 mm]
			\includegraphics[width=0.4\linewidth]{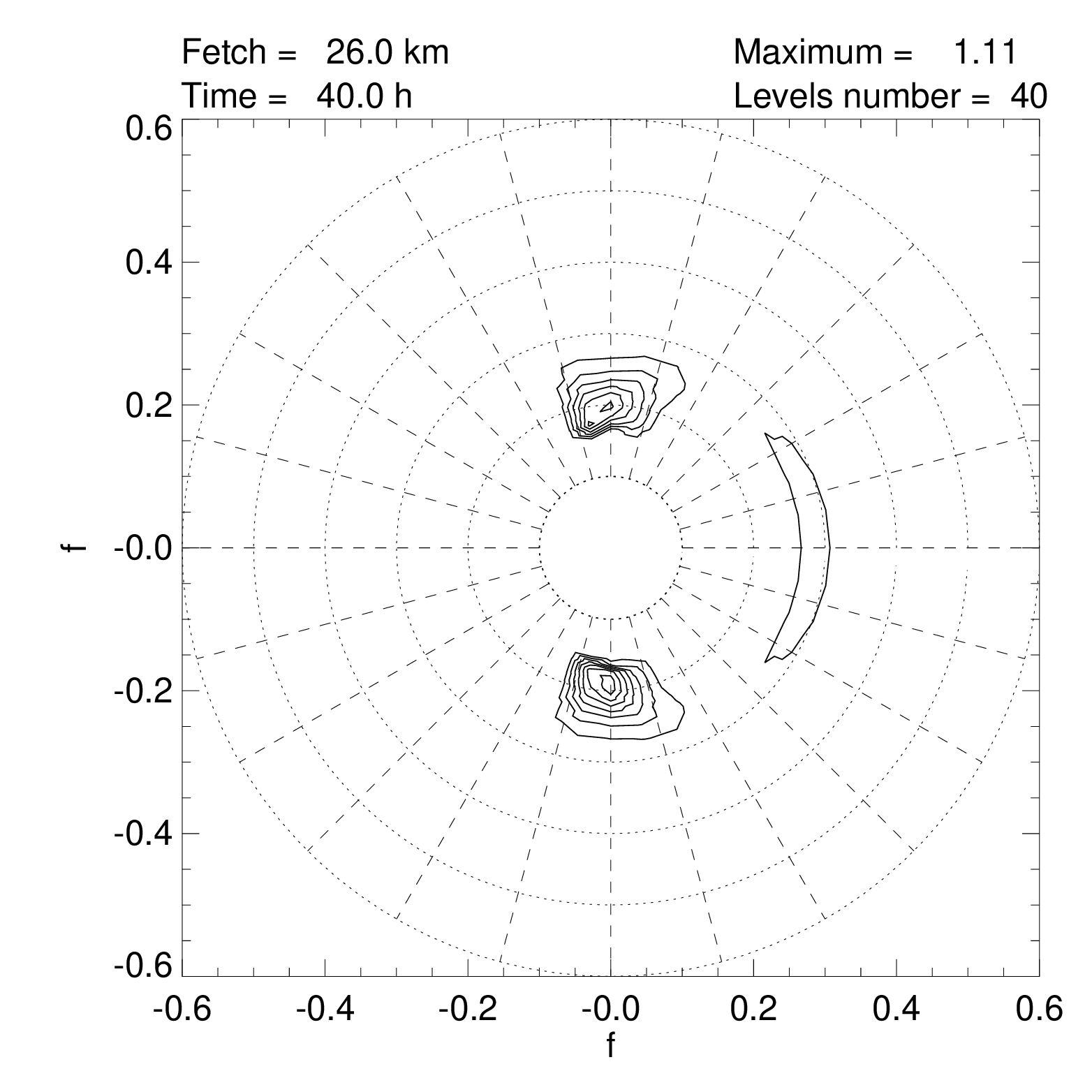} & \includegraphics[width=0.4\linewidth]{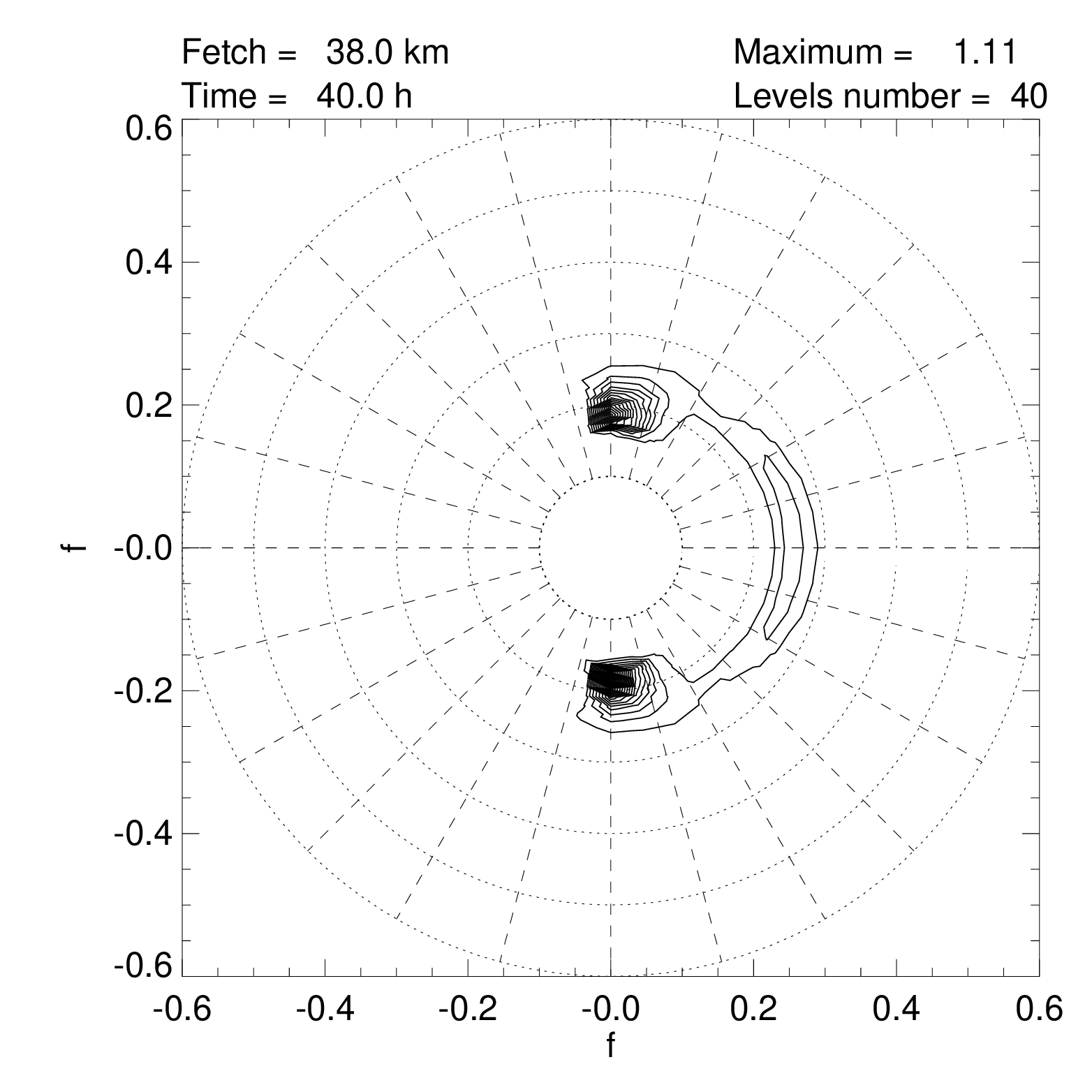} \\
		\end{tabular}
	\captionof{figure}{Energy spectrum $\varepsilon(f,\theta,x,t)$ as the function of the frequency $f$ and angle $\theta$ at the fetch coordinates $x=$ 2, 14, 26 and 38 km for time $t=40$ h.} \label{Polar40h}
	\end{center}
\end{table}

\begin{table}
	\centering
		\begin{center} 
			\begin{tabular}{c c}
				\includegraphics[width=0.4\linewidth]{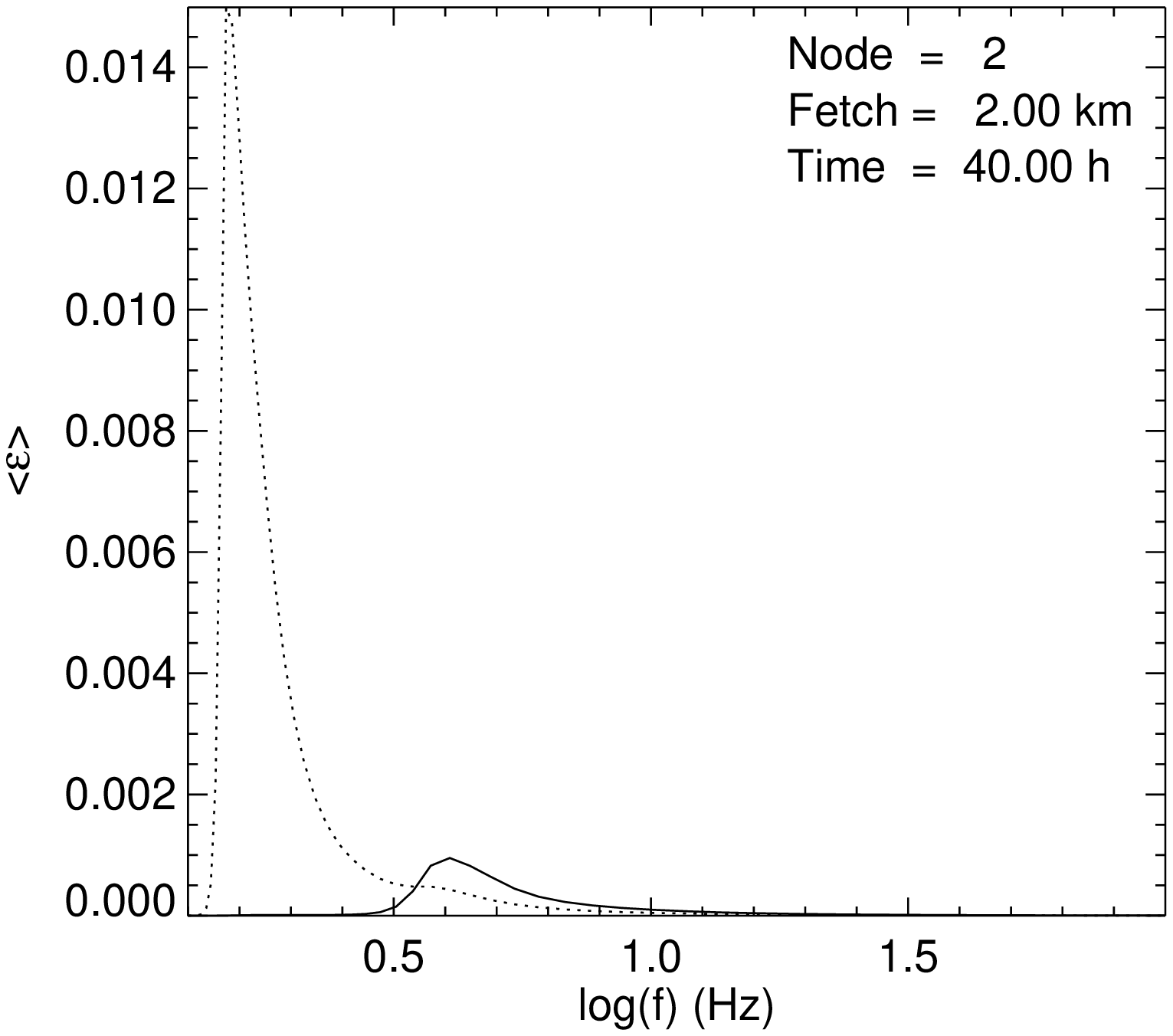} & \includegraphics[width=0.4\linewidth]{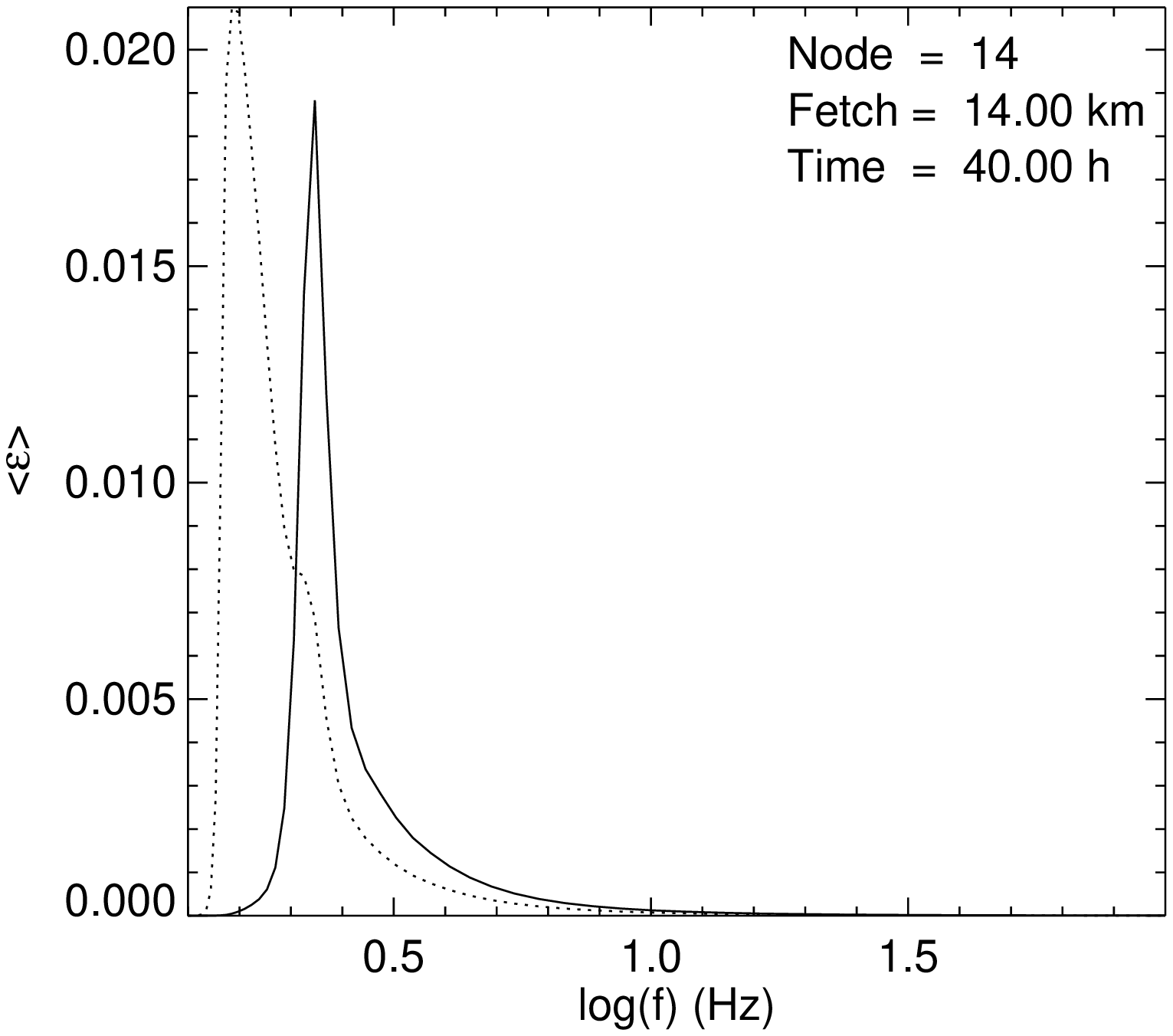} \\
[5 mm]
				\includegraphics[width=0.4\linewidth]{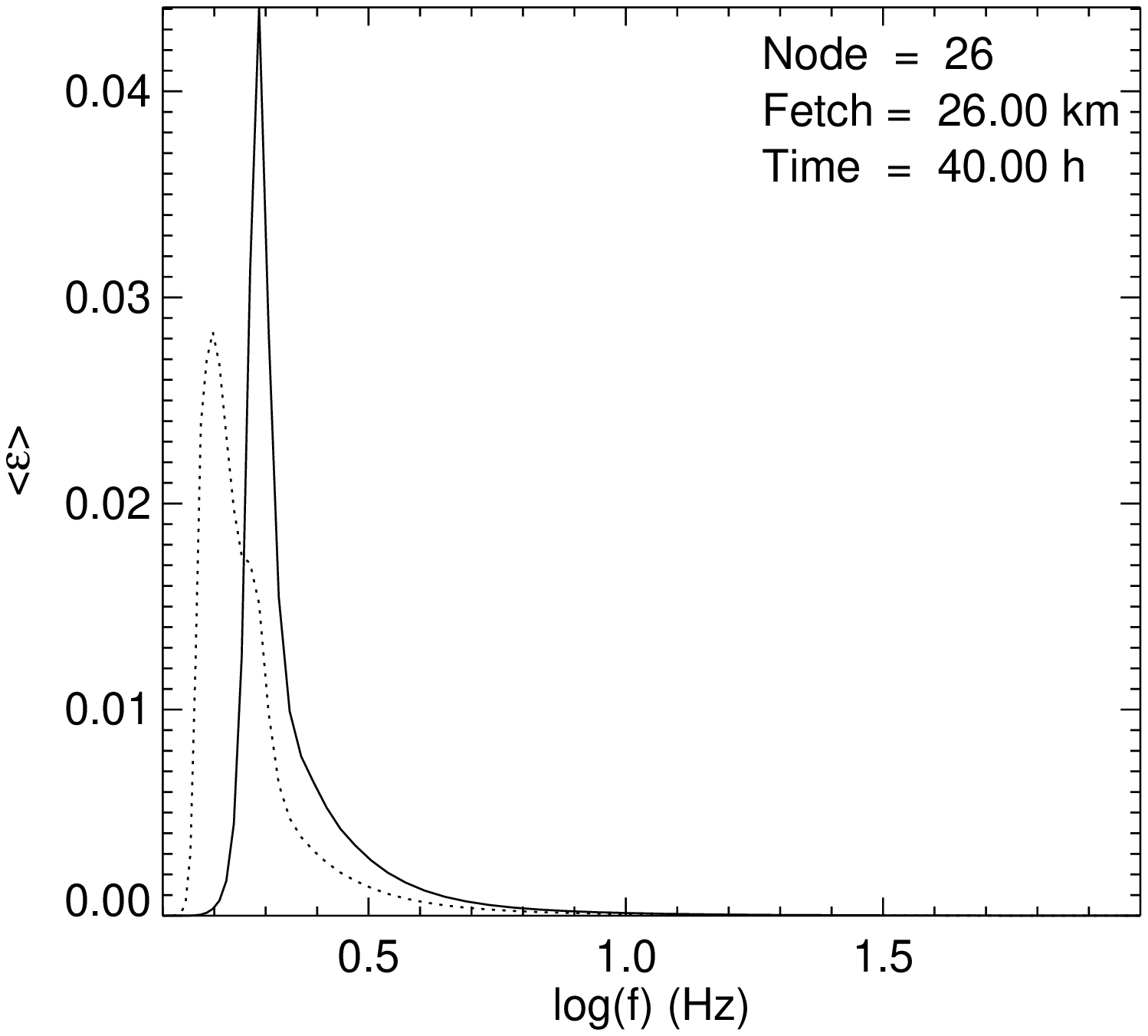} & \includegraphics[width=0.4\linewidth]{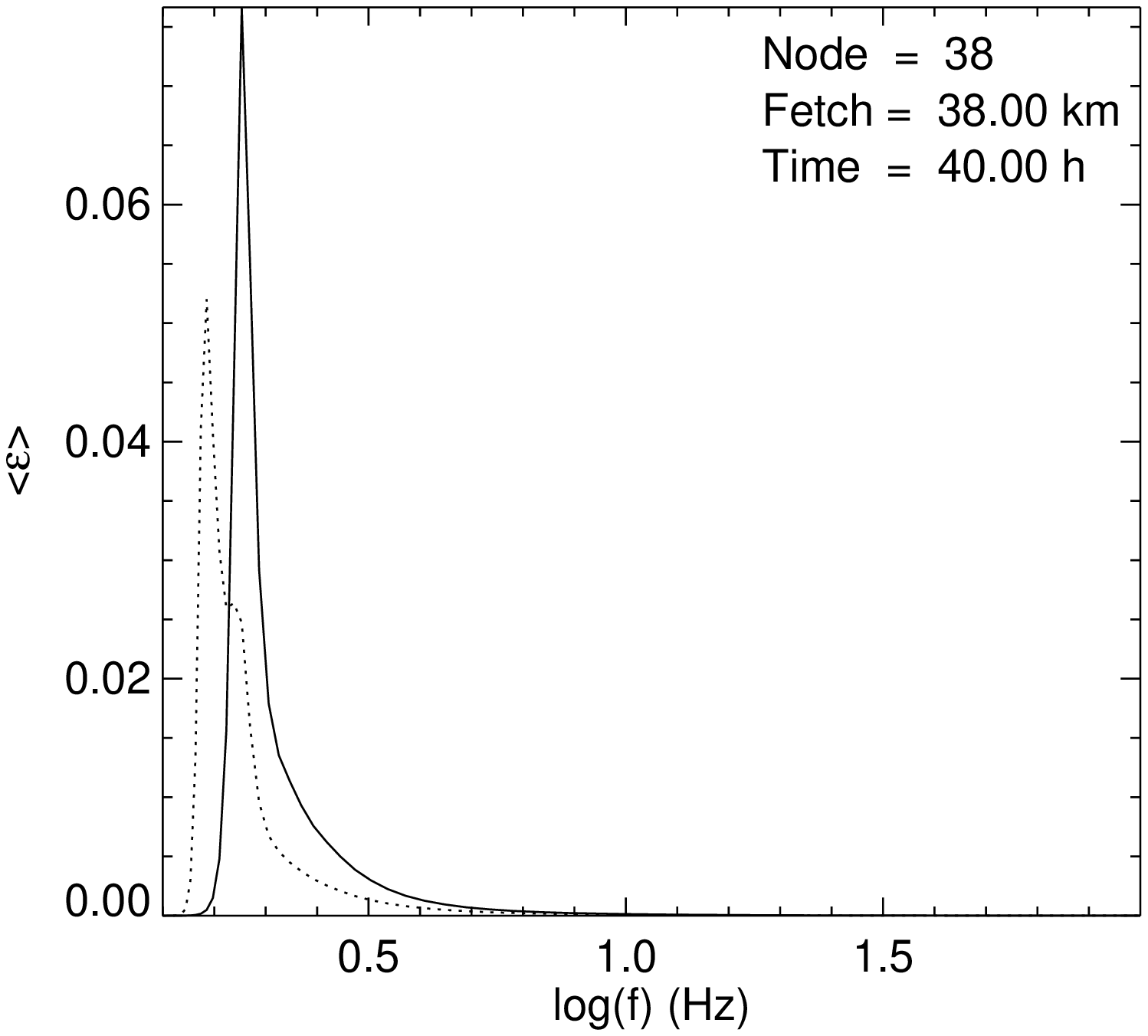} \\
			\end{tabular}
			\captionof{figure}{
Angle averaged $<\varepsilon>_{\theta} = \frac{1}{2\pi}\int\limits_{0}^{2 \pi} \varepsilon (f,\theta,x,t) d\theta$ (dotted line) and along the wind $\varepsilon(f,\theta_{wind},x,t)$ (solid line) spectra for time $t=40$ h at fetch coordinates $x=$ 2, 14, 26 and 38 km.} \label{AngAvNormSpectra40}
		\end{center}
\end{table}

\begin{table}
	\centering
		\begin{center} 
			\begin{tabular}{c c}
				\includegraphics[width=0.4\linewidth]{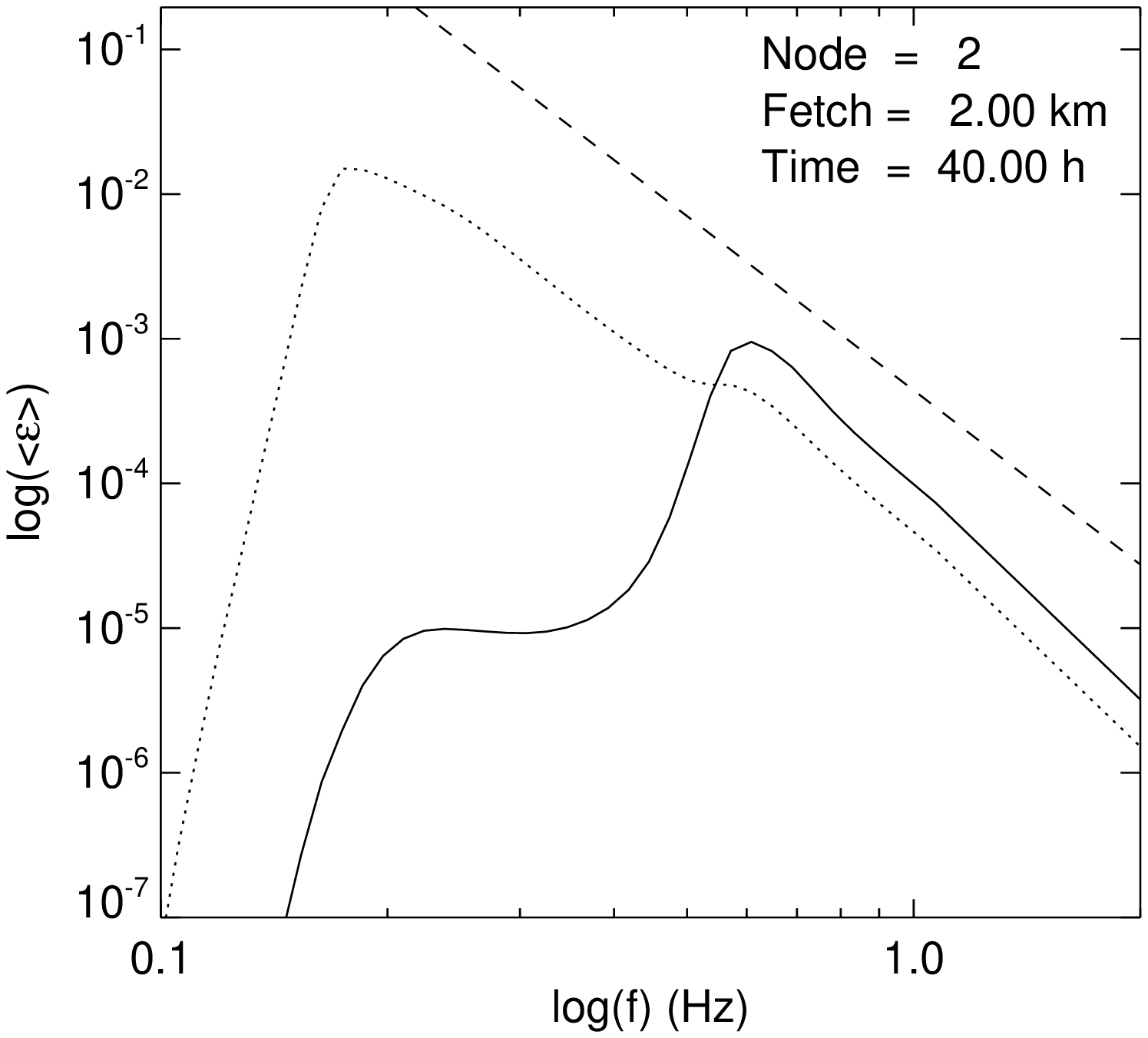} & \includegraphics[width=0.4\linewidth]{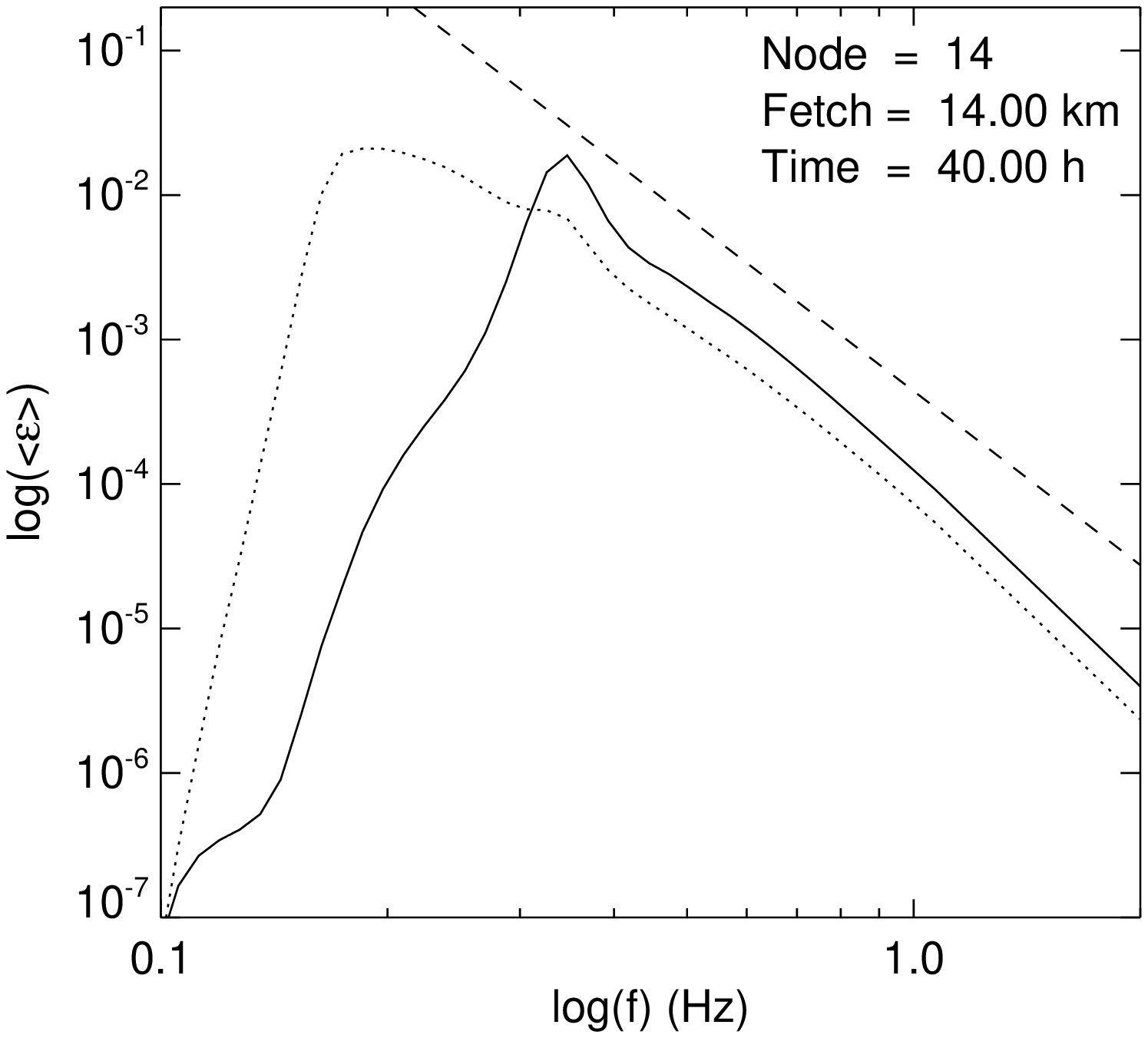} \\
[5 mm]
				\includegraphics[width=0.4\linewidth]{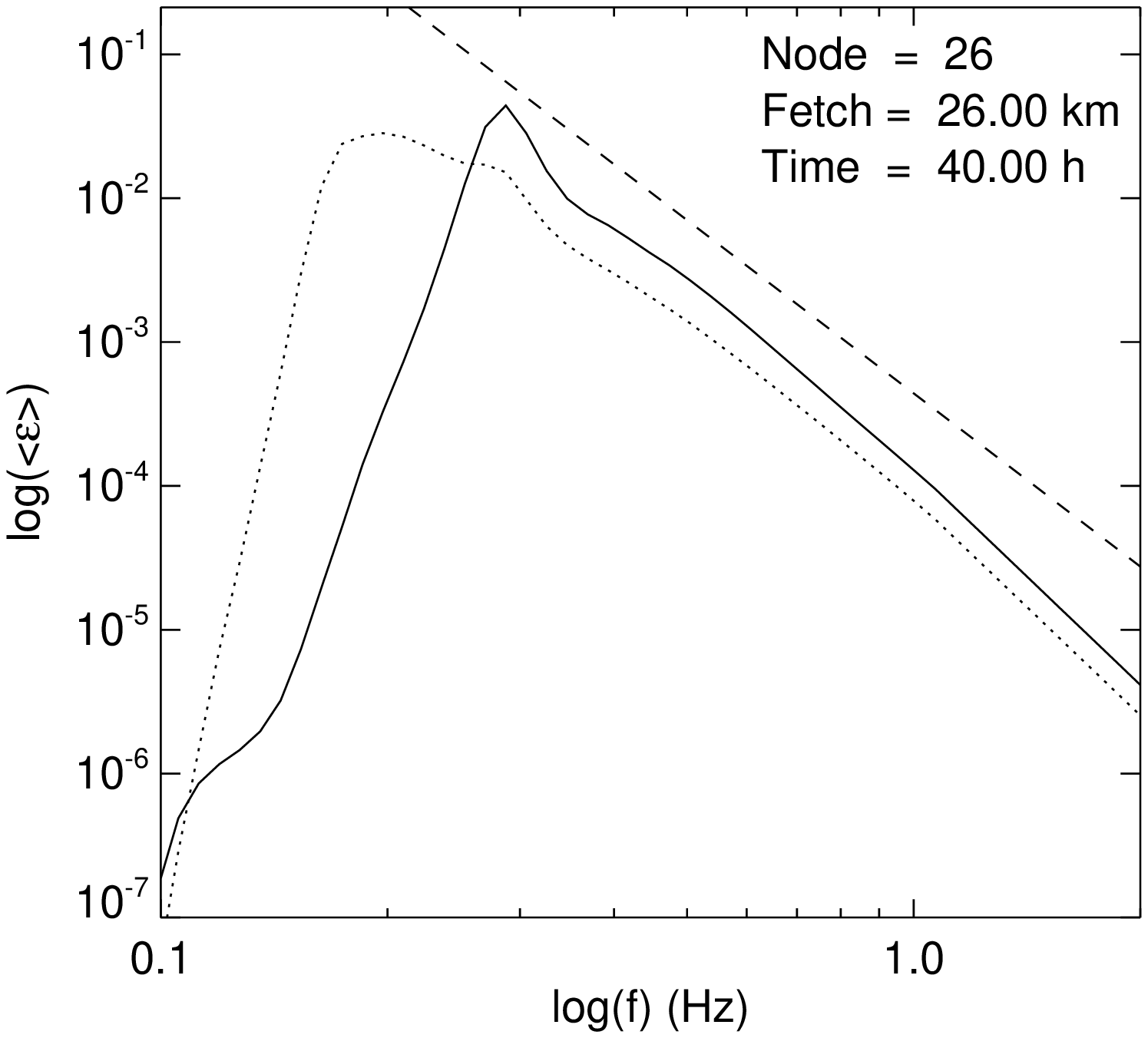} & \includegraphics[width=0.4\linewidth]{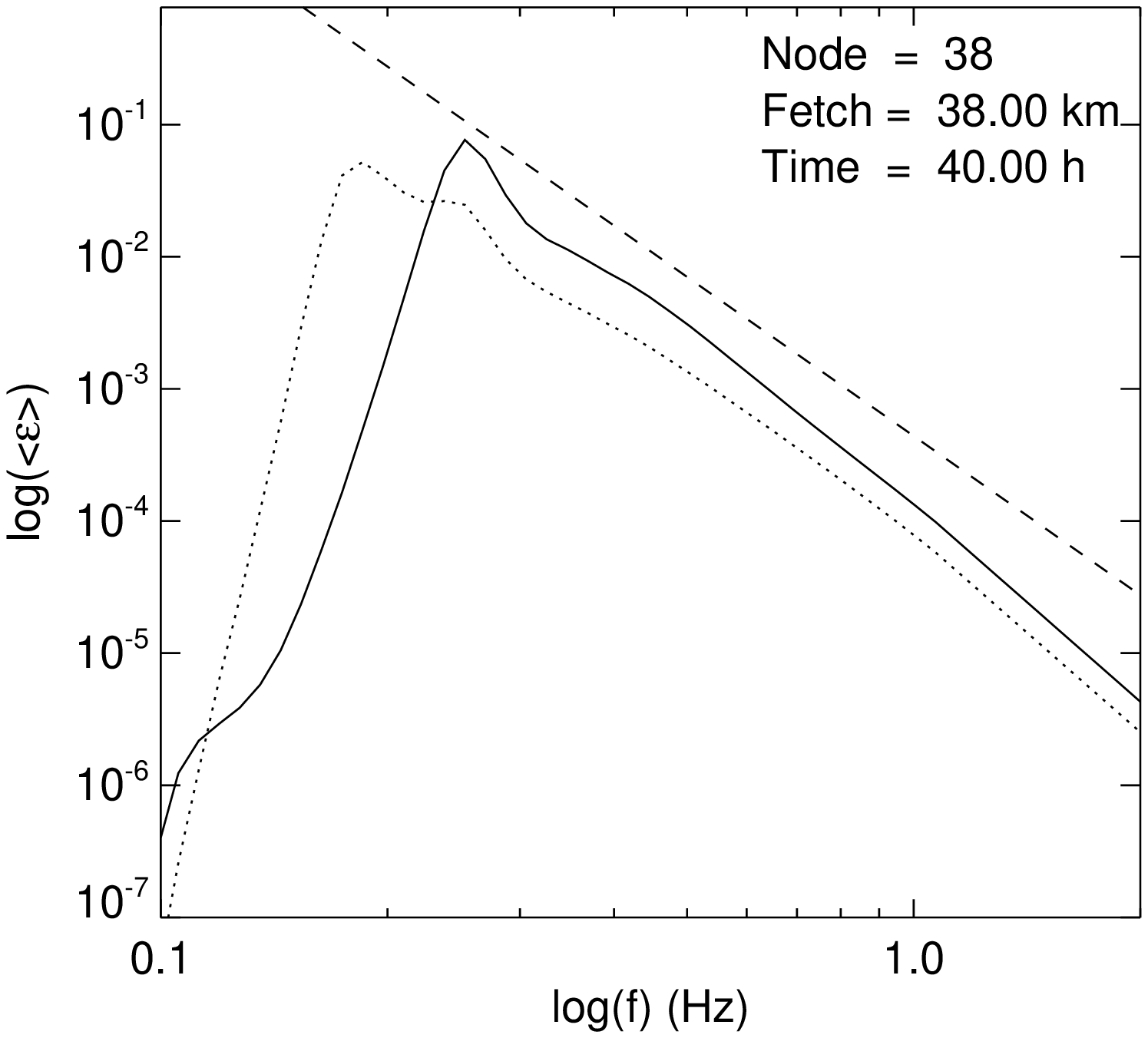} \\
			\end{tabular}
			\captionof{figure}{Decimal logarithm of the angle averaged $<\varepsilon>_{\theta} = \log ( \int\limits_{0}^{2 \pi} \varepsilon (f,\theta,x,t) d\theta$ ) (dotted line) and along the wind $\log ( \varepsilon(f,\theta_{wind},x,t) )$ (solid line) spectra for time $t=40$ h at fetch coordinates: $x=$ 2, 14, 26 and 38 km. Dashed line - KZ spectrum $\sim\omega^{-4}$.} \label{AngAvSpectra40}
		\end{center}
\end{table}

\begin{table}
\centering
	\begin{center} 
		\begin{tabular}{c c}
			\raisebox{0.01\height}{\includegraphics[scale=9.5]{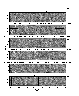}} &  \includegraphics[width=0.4\linewidth]{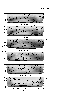}
		\end{tabular}
			\captionof{figure}{Left column: 3D surface topography of ocean waves along one of the four flight tracks at 6 different fetches (38.1, 31.5, 24.8, 18.2, 11.5 and 4.93 km from top to bottom). 
The wind is blowing from right to left in the coordinate system of each topographic image. \newline Right column: the corresponding 2D spectra calculated from the surface topographies shown in the left column. The wind direction is at $\theta_{wind} = 0^\circ$.} \label{Hwang}
	\end{center}
\end{table}

\end{document}